\documentclass[aps,pra,graphicx,10pt]{revtex4-2}
\usepackage{amssymb,amsbsy,times,fancyhdr,color}
\usepackage{amsmath}
\usepackage{mathrsfs,amsfonts}
\usepackage{latexsym,afterpage}
\usepackage{graphicx,subfigure,multirow}

\begin{document}


\title{Edge-wave phase-shifts versus normal-mode phase-tilts\\ in an Eady problem with a sloping boundary} 


\author{J. Mak}
\email[]{julian.c.l.mak@googlemail.com}
\affiliation{Department of Ocean Science, Hong Kong University of Science and Technology, Hong Kong}
\affiliation{Center for Ocean Research in Hong Kong and Macau, Hong Kong University of Science and Technology, Hong Kong}
\affiliation{National Oceanography Centre, Southampton, UK}


\author{N. Harnik}


\author{E. Heifetz}
\affiliation{Porter school of the Environment and Earth Sciences, Tel Aviv University,
 69978, Israel}
 

\author{G. Kumar}
\email[]{gautam.kmr10@zohomail.in}
\affiliation{Department of Ocean Science, Hong Kong University of Science and Technology, Hong Kong}
\affiliation{Center for Ocean Research in Hong Kong and Macau, Hong Kong University of Science and Technology, Hong Kong}
\affiliation{Department of Mathematics, Faculty of Science and Technology (IcfaiTech), ICFAI Foundation for Higher Education, Hyderabad 501203, Telangana, India}


\author{E. Q. Y. Ong}
\affiliation{Climate Change Research Centre, Australian Centre for Excellence in Antarctic Science, University of New South Wales, Sydney, NSW, Australia}
\affiliation{Australian Research Council Centre of Excellence for Climate Extremes, University of New South Wales, Sydney, NSW, Australia}






\begin{abstract}

One mechanistic interpretation of baroclinic instability is that of mutual
constructive interference of Rossby edge-waves. While the two edge-waves and
their relative phase-shifts are invoked as part of the mechanistic
interpretation, for example in relation to suppression of baroclinic instability
over slopes, the phase-tilts of the related normal modes are often presented
instead. Here we highlight the differences between edge-wave phase-shifts and
normal-mode phase-tilts, in the context of an Eady problem modified by the
presence of a sloping boundary. The resulting problem remains tractable
analytically, and the interacting Rossby edge-waves can be asymmetric, in
contrast to the standard Eady case. We argue and present evidence that the
normal-mode phase-tilt is potentially a misleading quantity to use, and
edge-wave phase-shifts should be the ones that are mechanistically relevant. We
also provide a clarification for the mechanistic rationalization for baroclinic
instability in the presence of slopes (such as suppression of growth rates) that
is valid over all parameter space, in contrast to previous attempts. We further
present evidence that there is a strong correlation between quantities diagnosed
from the GEOMETRIC framework with the edge-wave phase-shifts, but not the
normal-mode phase-tilts. The result is noteworthy in that the geometric
framework makes no explicit reference to the edge-wave structures in its
construction, but the correlation suggests that in problems where edge-wave
structures are not so well-defined or readily available, the GEOMETRIC framework
should still capture mechanistic and dynamical information. Some implications
for parameterization of baroclinic instability and relevant eddy-mean feedbacks
are discussed. For completeness, we also provide an explicit demonstration that
the linear instability problem of the present modified Eady problem is
parity-time symmetric, and speculate on some suggestive links between
parity-time symmetry, shear instability, and the edge-wave interaction
mechanism.


\end{abstract}


\maketitle 



\section{Introduction}

Baroclinic dynamics and its turbulence are ubiquitous features in rotating
stratified systems, playing a key role in systems such as the Earth's ocean and
atmosphere as well as other geophysical/astrophysical systems, for the
associated transport of buoyancy and impacts the overturning circulation
\cite[e.g.,][]{Vallis-GFD, Lovelace-et-al99, KaspiFlierl07,
Hughes-et-al-Tachocline, Teed-et-al10, PolichtchoukCho12, GilmanDikpati14,
Gilman15, Read-et-al20, YellinBergovoy-et-al21}. Understanding the mechanisms
and conditions for instability, its transition to turbulence and its eventual
saturation is of interest in understanding and modeling of the evolution in the
relevant rotating stratified systems.

It is not too controversial to say the subject of baroclinic instability is
rather well-understood at least in the hydrodynamic regime, where the linear
instability phase of idealized models have analytical solutions
\cite[e.g.,][]{Charney47, Eady49, Phillips56, Green60}, and general stability
theorems may be derived \cite[e.g.,][]{CharneyStern62, Pedlosky64a, Pedlosky64b,
Shepherd90}. The nonlinear phase can also be tackled, mostly by numerical means
\cite[e.g.,][]{Shepherd88, Shepherd89, Thorncroft-et-al93, LarichevHeld95,
SpallChapman98, ThompsonYoung07, Esler08a, BachmanFoxKemper13, Bachman-et-al17,
ChangHeld22}. Often of interest in those cases are the associated statistics
such as meridional eddy buoyancy fluxes (baroclinic instability usually leads to
poleward eddy buoyancy flux in order for reduction of available potential energy
\cite[e.g.,][]{Vallis-GFD}), which plays a role in the eddy-mean interaction in
the relevant rotating stratified systems, and informs on the parameterization in
numerical general circulation models. A link that has been of particular
interest is that of \emph{quasi-linear control}, i.e., to what degree does the
linear instability characteristics have an imprint on the nonlinear dynamics.
While one could argue that the processes that are being parameterized are
inherent manifestations of the nonlinear dynamics, and there is no strong reason
that the relevant linear analysis should play any role, the fact remains that
there does appear a relation between the two \cite[e.g.,][]{Green70, Stone72,
Killworth97, Killworth98, Eden11}. A piece of work of relevance here is the
GEOMETRIC framework of \cite{Marshall-et-al12, MaddisonMarshall13} (see also
\cite{Hoskins-et-al83, WatermanHoskins13}), which has highlighted a link between
the associated eddy fluxes in terms of geometric quantities associated with eddy
variance ellipses (such as anisotropy factors and angles) and the linear
instability properties. The scalings provided from the GEOMETRIC framework has
found particular skill in the parameterization of eddy buoyancy fluxes in
numerical ocean general circulation models \cite[e.g.,][]{Mak-et-al18,
Mak-et-al22, Mak-et-al23}.

In the present case we are interested in the instability characteristics of
baroclinic instability in the presence of a slope, where `slope' is broadly
interpreted to mean a slope as a physical boundary (e.g. topography in the
atmosphere and/or ocean), or motion in the presence of an impermeable surface
arising from the relevant fluid properties (e.g., adiabatic flow above/below a
sloping isentrope in planetary atmospheres, magnetic field effects in the solar
tachocline above the radiative zone). Analogous investigations of classical
baroclinic instability in the presence of weak slopes in the linear and
nonlinear regime exist, and is of particular relevance in the field of
oceanography. The presence of continental slopes is generally seen to suppress
eddy buoyancy fluxes over the slope regions \cite[e.g.,][]{BlumsackGierasch72,
Mechoso80, Isachsen11, Brink12, ChenKamenkovich13, Isachsen15, Pedlosky16,
Manucharyan-et-al17, Hetland17, TrodahlIsachsen18, WangStewart18,
ManucharyanIsachsen19, Chen-et-al20, Tanaka21, WeiWang21, Wei-et-al22,
Wei-et-al24, NummelinIsachsen24}, with consequences for the material exchange
between the shelf and open ocean environment. One possible contributing factor
for the observed suppression over slope regions is that the linear instability
is itself suppressed and/or less efficient over regions with slopes
\cite[e.g.,][]{BlumsackGierasch72, Mechoso80, Isachsen11, Brink12,
ChenKamenkovich13, Pedlosky16, Chen-et-al20, Tanaka21}. In relation to the
GEOMETRIC parameterization, the work of \cite{Wei-et-al22} has found, by
diagnoses, that a tuning parameter $\alpha$ normally interpreted as a baroclinic
eddy efficiency for the feedback onto the mean state \cite[e.g.,][]{Mak-et-al18}
is suppressed over the slope regions. A slope suppressed $\alpha$ has been
found to lead to improvements in idealized prognostic calculations
\cite{Wei-et-al24}, providing additional evidence that there should be reduced
eddy feedback over slopes. The links between the observed suppression of this
$\alpha$ parameter and possible links with linear stability analysis are to be
clarified, and is one of the goals of the present work.

\begin{figure}
  \includegraphics[width=\textwidth]{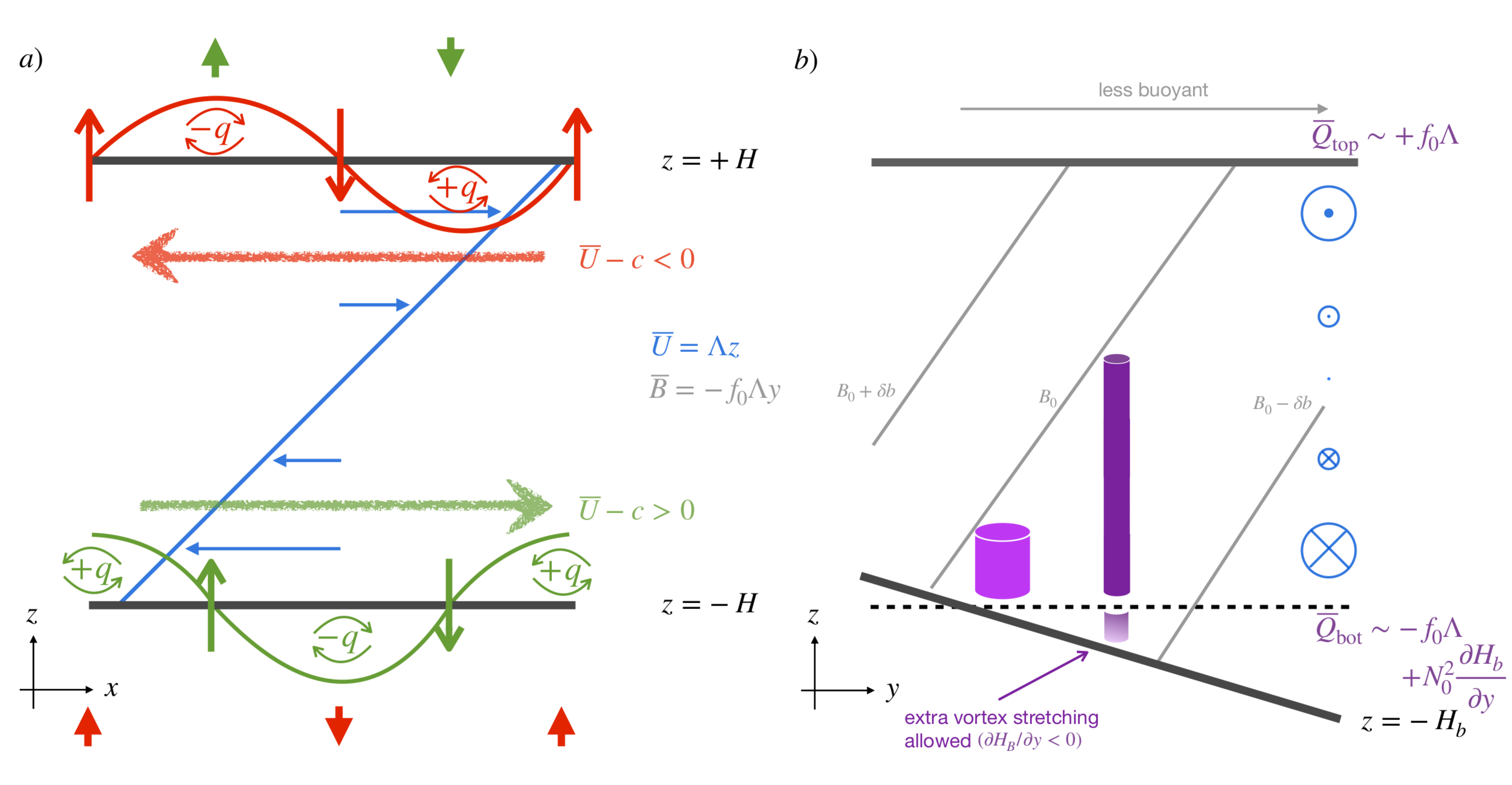}
  \caption{Physical schematic for the current modified Eady problem. ($a$)
  Cross-section of the set up, showing the basic state linear flow, with Rossby
  edge-waves drawn on at the vertical boundaries in an unstable configuration
  for a case where the interaction is symmetric (cf., a standard Eady problem,
  $\delta=0$ here); see text for a description of the counter-propagating Rossby
  waves mechanism. ($b$) Head on view, showing the basic state linear flow going
  in and coming out of the page, and the associated buoyancy profile via thermal
  wind shear relation. The $\delta < 0$ case (topography and basic state
  buoyancy profile at opposite orientations) is illustrated, where the
  contribution from the slope \emph{reinforces} the bottom PV gradient by
  allowing extra stretching of a fluid column.}
  \label{fig:setup}
\end{figure}

Why exactly are baroclinic instabilities suppressed over slopes? For this, we
note first that, in the absence of slopes, a kinematic/mechanistic
interpretation of baroclinic instability is normally given in terms of a pair of
counter-propagating Rossby edge-waves \cite[e.g.,][]{Bretherton66a,
Hoskins-et-al85, Heifetz-et-al04a} (although the concept of instability arising
from a constructive interference of edge-waves appears to hold for general shear
instabilities \cite[e.g.,][]{Heifetz-et-al99, HarnikHeifetz07,
Rabinovich-et-al11, Heifetz-et-al15, HeifetzGuha19}). If the edge-waves are the
building blocks of the instability, then knowing how the edge-waves interact and
form the instability should help with parameterization efforts. The mechanism is
illustrated in Fig.~\ref{fig:setup}$a$ for the classical Eady problem
\cite{Eady49}, and proceeds as follows:

\begin{enumerate}
  \item Rossby waves are supported on potential vorticity (PV) gradients, and in
  the standard Eady set up, PV is only non-zero and localized at the upper and
  lower boundaries, associated with buoyancy anomalies
  \cite[e.g.][]{Vallis-GFD}, hence Rossby edge-waves;
  \item the Rossby edge-waves carry PV anomalies, and the self-induced PV
  anomalies have associated with it a velocity (via PV inversion, cf.
  \cite{Hoskins-et-al85}), such that the edge-waves are
  \emph{counter}-propagating against the background mean flow;
  \item the waves interact with each other, and depending on the phase-shift of
  the edge-waves, can lead to mutual amplification of the wave displacement (the
  theoretical optimum being quarter of a wavelength or $\pi/2$ out of phase), as
  well as some hindering/helping of the other wave's propagation;
  \item the counter-propagation against the mean flow and the mutual interaction
  between the waves can lead to a phase-locked configuration, and if the phase
  shift is also conducive for displacement amplification, then we have (modal)
  instability.
\end{enumerate}

How is the counter-propagating Rossby wave mechanism modified by the presence of
slopes? In Fig.~\ref{fig:setup}$b$ we show a case where the slope and the
isopycnal configuration are in opposite orientations (the $\delta<0$ regime
later), adapted from the work of \cite{Chen-et-al20}. In this set up, the vortex
tube when moved in the meridional direction $y$ is allowed to stretch more in
the vertical direction $z$, and so the vortex tube strengthens more than it
would relative to the flat case. Then it is perhaps clear what is going to
happen is this setting: the bottom edge-wave is modified because the slope (if
small) reinforces the value of the PV value at the bottom boundary, increasing
the bottom PV gradient, and thus changing the characteristic of the bottom
edge-wave. The work of \cite{Isachsen11} alludes to this but stops short of
elaborating on the details. The work of \cite{Chen-et-al20} goes slightly
further by arguing that, in the setting as in Fig.~\ref{fig:setup}$b$, the
bottom Rossby edge-wave would propagate \emph{faster} (because the associated PV
gradient is enhanced), the associated doppler shifted velocity
$\overline{U}_{\rm bot}-c > 0$ must be smaller in value (here $\overline{U}$
denotes the basic state velocity, and suffix `bot' and `top' have the expected
meanings). In order to maintain phase-locking, we need $\overline{U}_{\rm
top}-c$ to match the $\overline{U}_{\rm bot}-c > 0$, and that can only happen if
$c$ \emph{decreases} in value, which for Rossby waves means selecting a higher
wavenumber, i.e. shorter waves, providing an explanation for the observed
instability bandwidth (see Fig~\ref{fig:growth_rates_cr_l00}$a$ in the
$\delta<0$ regime).

We argue here the explanation provided by \cite{Chen-et-al20} is incomplete. A
first issue is that the argument solely based on phase-speed matching does not
extend to the converse case ($\delta > 0$): the instability does not
monotonically go to the longest wavelength as the slope is increased in the
other orientation. Second is that the argument as is does not provide an
explanation for the growth rate behavior over parameter space (at least not
explicitly). One of our aims of this work is to provide a more complete argument
and support it with further analysis: essentially, \emph{the mutual interaction
matters}. We further clarify a point regarding the usage of the term
`phase-shift'. The work of \cite{Chen-et-al20} (and also in standard textbooks
such as \cite{Vallis-GFD}) invoke `phase-shifts' when talking about the
instability mechanism associated with Rossby edge-waves in terms of PV
signatures, but demonstrate it with a phase-\emph{tilt} of the
\emph{streamfunction} eigenfunction, where the phase-tilt coincides with that of
the anticipated optimum of $\pi/2$ (quarter of a wavelength; (see for example
Fig.~1$d$ of \cite{Chen-et-al20}, and essentially
Fig.~\ref{fig:edge_vs_eigen_shift_l00}$c$ in the present work). We provide
arguments that this can be misleading: it is the Rossby edge-wave basis in the
PV signature and their phase-shifts that are relevant for the mechanistic
interpretation, and those phase-shifts generically do not coincide with the
phase-\emph{tilts} diagnosed from the streamfunction eigenfunction. We show here
that while the diagnosed phase-tilt in the streamfunction \emph{happens} to be
the theoretical optimum $\pi/2$ for the most unstable mode, the associated PV
edge-wave phase-\emph{shifts} are essentially \emph{never} at $\pi/2$ over the
whole parameter space, the latter because, again, the \emph{mutual interaction
matters}.

While we show that the quantities diagnosed from the streamfunction
eigenfunctions directly bear little resemblance to the quantities associated
with the PV edge-wave basis, we report here that applying the GEOMETRIC
framework to the modified Eady problem and performing analysis in terms of
appropriate combinations of the normal-mode eigenfunctions, the resulting
quantities, in particular the vertical eddy tilt, \emph{do} correspond well to
the quantities diagnosed from the PV edge-wave basis, even though the GEOMETRIC
framework makes no explicit reference to the edge-waves. Within the GEOMETRIC
framework and in the context of linear theory, the suppression of the
instability and eddy efficiency as invoked in parameterization of baroclinic
eddies is attributed principally to changes in the \emph{buoyancy anisotropy} of
the eddies. The link between GEOMETRIC and PV edge-waves may be of further
interest: in cases where the edge-wave basis are not well-defined, it may in
fact be useful to perform an analysis in the GEOMETRIC framework instead.

In Sec.~\ref{sec:eady} we formulate and provide an overview of the instability
characteristics of the modified Eady problem, and make precise our arguments on
why we think the existing descriptions of the instability mechanism in the
modified Eady problem is incomplete. We additionally highlight that the modified
Eady problem here is in fact parity-time ($\mathcal{PT}$) symmetric
\cite[e.g.,][]{Bender-PT, Qin-et-al19, Zhang-et-al20, David-et-al22}, which has
consequences for the solution spectrum, but defer the expanded details and
related discussion to Appendix \ref{app:A}. In Sec.~\ref{sec:edgewave} we
provide a rephrasing of the modified Eady problem in terms of an edge-wave basis
\cite[cf.,][]{DaviesBishop94, Heifetz-et-al04a}, where the presence of the slope
(related to a topographic PV signature) manifests as an independent adjustable
parameter that controls the degree of asymmetry between upper and lower
edge-waves (the standard Eady problem being the case with symmetric
interaction). We provide an internally consistent physical rationalization of
the instability mechanism, with explicit references to the phase-shift and
asymmetry in the wave amplitude ratios that is valid over all parameter space.
We further clarify the issue of `phase-shifts', by comparing results from the
edge-wave basis and a standard analysis of the streamfunction eigenfunction. In
Sec.~\ref{sec:geometric} we demonstrate links between the quantities of interest
from the GEOMETRIC framework and the edge-wave analysis, highlighting a link
between the eddy angles with edge-wave phase-shifts
\cite[cf.,][]{Tamarin-et-al16}, as well as providing an analysis for what
contributes to the suppression to the eddy efficiency parameter $\alpha$ that is
prescribed in parameterizations. We summarize our results in
Sec.~\ref{sec:conclude}, and discuss some implications of our results for
parameterization of baroclinic processes.


\section{Overview of the modified Eady problem}\label{sec:eady}

The physical set up is as illustrated in Fig.~\ref{fig:setup}, for the Northern
Hemisphere with Coriolis parameter $f_0>0$. We start with the quasi-geostrophic
(QG) equations \cite[e.g.][]{Vallis-GFD} formulated on a $f$-plane with
potential vorticity (PV) advection in the interior, and QG buoyancy advection on
the vertical surfaces, i.e.,
\begin{equation}
  \frac{\mathrm{D}q}{\mathrm{D}t}=0, \quad z \in(H, -H_b), \qquad \qquad
  \frac{\mathrm{D}b}{\mathrm{D}t}=0, \quad z = H, -H_b,
\end{equation}
where the domain of interest is between $z=H$ and $z=-H_b(y)$, $H_b(y)$
represents the bottom slope, $\mathrm{D}/\mathrm{D}t = \partial/\partial t +
\boldsymbol{u}\cdot\nabla$ is the material derivative, $\boldsymbol{u}$ is the
geostrophic velocity with associated streamfunction $\psi$, $\nabla$ the
horizontal gradient operator, and $(x,y,z)$ denotes the zonal (East-West),
meridional (North-South) and vertical co-ordinate. The PV $q$ and QG buoyancy
$b$ are defined as
\begin{equation*}
  q = \nabla^2\psi + \frac{\partial}{\partial z}\frac{f_0}{N_0^2} b, \qquad b = f_0 \frac{\partial\psi}{\partial z},
\end{equation*}
where $N_0^2=\partial\overline{B}/\partial z$ is the buoyancy frequency
associated with the prescribed background stratification, where $\overline{B}$
is the basic state buoyancy profile to be prescribed with the basic state
velocity $\overline{U}$. Contributions from the small slope will arise through
the buoyancy equation in the advective term via $w=\boldsymbol{u}\cdot\nabla
H_b$, arising from the no-normal flow condition on the bottom boundary.

We make an assumption that $\partial H_b(y) / \partial y$ is small (more
precisely, that $(N_0/f_0)\partial H_b / \partial y$ is of order Rossby number
\cite[e.g.,][]{Isachsen11, Chen-et-al20}), and that $\partial H_b(y) / \partial
y$ can be approximated by a small constant contribution only in the boundary
condition at $z=-H$. Then, linearizing against the basic state $\boldsymbol{u} =
\overline{U}\boldsymbol{e}_x = \Lambda z$ (and so $\overline{B} = -f_0\Lambda y$
by thermal wind shear relation), the governing linear equations are
\begin{subequations}
\begin{align}
  &\left(\frac{\partial}{\partial t} + \Lambda z\frac{\partial}{\partial x}\right)\left(\nabla^2 \psi + \frac{f_0^2}{N_0^2} \frac{\partial^2\psi}{\partial z^2}\right)=0, \qquad &z \in(-H, H),\\
  &\left(\frac{\partial}{\partial t} + \Lambda H \frac{\partial}{\partial x}\right)\frac{\partial\psi}{\partial z} - \Lambda\frac{\partial\psi}{\partial x} = 0, \qquad &z = H,\\
  &\left(\frac{\partial}{\partial t} - \Lambda H \frac{\partial}{\partial x}\right)\frac{\partial\psi}{\partial z} - \left(\Lambda - \frac{N_0^2}{f_0}\frac{\partial H_b}{\partial y}\right)\frac{\partial\psi}{\partial x} = 0, \qquad &z = -H.
\end{align}
\end{subequations}
Non-dimensionalizing by the horizontal length-scale $L$, vertical length-scale
$H$, and time-scale $T = L / U = L / (\Lambda H)$, we have
\begin{subequations}\label{eq:lin-nondim}
\begin{align}
  &\left(\frac{\partial}{\partial t} + z\frac{\partial}{\partial x}\right)\left(\nabla^2 \psi + F^2 \frac{\partial^2\psi}{\partial z^2}\right)=0, \qquad &z \in(-1, 1),\\
  &\left(\frac{\partial}{\partial t} + \frac{\partial}{\partial x}\right)\frac{\partial\psi}{\partial z} - \frac{\partial\psi}{\partial x} = 0, \qquad &z = 1,\\
  &\left(\frac{\partial}{\partial t} - \frac{\partial}{\partial x}\right)\frac{\partial\psi}{\partial z} - (1 - \delta)\frac{\partial\psi}{\partial x} = 0, \qquad &z = -1,
\end{align}
\end{subequations}
where $F^2 = (f_0 L / NH)^2$ and is related to the inverse of the Burger number.
A key non-dimensional parameter in the present system is (in terms of
dimensional variables)
\begin{equation}
  \delta = \frac{\partial H_b}{\partial y}\left(\frac{N_0^2}{f_0 \Lambda}\right) = \left.\frac{\partial H_b}{\partial y} \right/ \frac{-\partial \overline{B} / \partial y}{\partial \overline{B} / \partial z} = \frac{\partial H_b/\partial y}{s},
\end{equation}
i.e. the parameter $\delta$ relates to the orientation of intersection between
the background isopycnal slopes $s$ with the bottom slope (a $\delta<0$ case is
illustrated in Fig.~\ref{fig:setup}$b$); this parameter is related to the
$\alpha_T$ parameter in \cite{Pedlosky16}. With that, the case $\delta < 0$ and
$\delta > 0$ are sometimes known as retrograde or prograde configurations,
although we will not be using that terminology here. One could relate the
$\delta$ parameter here to a topographic $\beta$ term, but we refer to reader to
the work of \cite{Isachsen11} for that since we do not invoke that term in this
work.

With appropriate horizontal boundary choices and conditions (periodic in zonal
$x$ direction, periodic or appropriate no-normal flow boundary conditions in
meridional $y$ direction), we consider solutions of the form
\begin{equation}\label{eq:eigenfunc}
  \psi(x,y,z,t) = \tilde{\psi}(z)\exp[\mathrm{i}(kx - \omega t)]g(y),
\end{equation}
where $\tilde{\psi}$ is a vertical structure function in the streamfunction,
$g(y)$ is an appropriate eigenfunction of the Laplacian operator so that
$\partial^2 g / \partial y^2 = -l^2 g$ (e.g., combinations of $\sin ly$ and
$\cos ly$ as appropriate), $\mathrm{i} = \sqrt{-1}$, $(k,l)$ are the zonal and
meridional wavenumbers, $\omega = kc = k(c_r + \mathrm{i}c_i)$ is the angular
frequency, $c$ is the (complex) phase-speed; we have modal instability if $c_i >
0$. The modified Eady problem has zero PV signature in the interior, so the
vertical structure function satisfies
\begin{equation}\label{fig:vert_struc}
  \tilde{\psi}(z) = a \cosh \mu z + b \sinh \mu z, \qquad \mu^2 = F^2(k^2 + l^2).
\end{equation}
The constants $a$ and $b$ are fixed by the compatibility condition resulting
from the vertical boundary conditions. Making the shorthand $C = \cosh \mu$ and
$S = \sinh \mu$, after some algebraic manipulation, the dispersion relation is
given by
\begin{equation}\label{eq:c-equ}
  0 = c^2 + \frac{\delta}{2\mu} \left(\frac{C}{S} + \frac{S}{C}\right)c + \frac{\delta^2}{4\mu^2} - \left(\frac{1-\delta/2}{\mu} - \frac{C}{S}\right)\left(\frac{1-\delta/2}{\mu} - \frac{S}{C}\right).
\end{equation}

\begin{figure}
  \includegraphics[width=\textwidth]{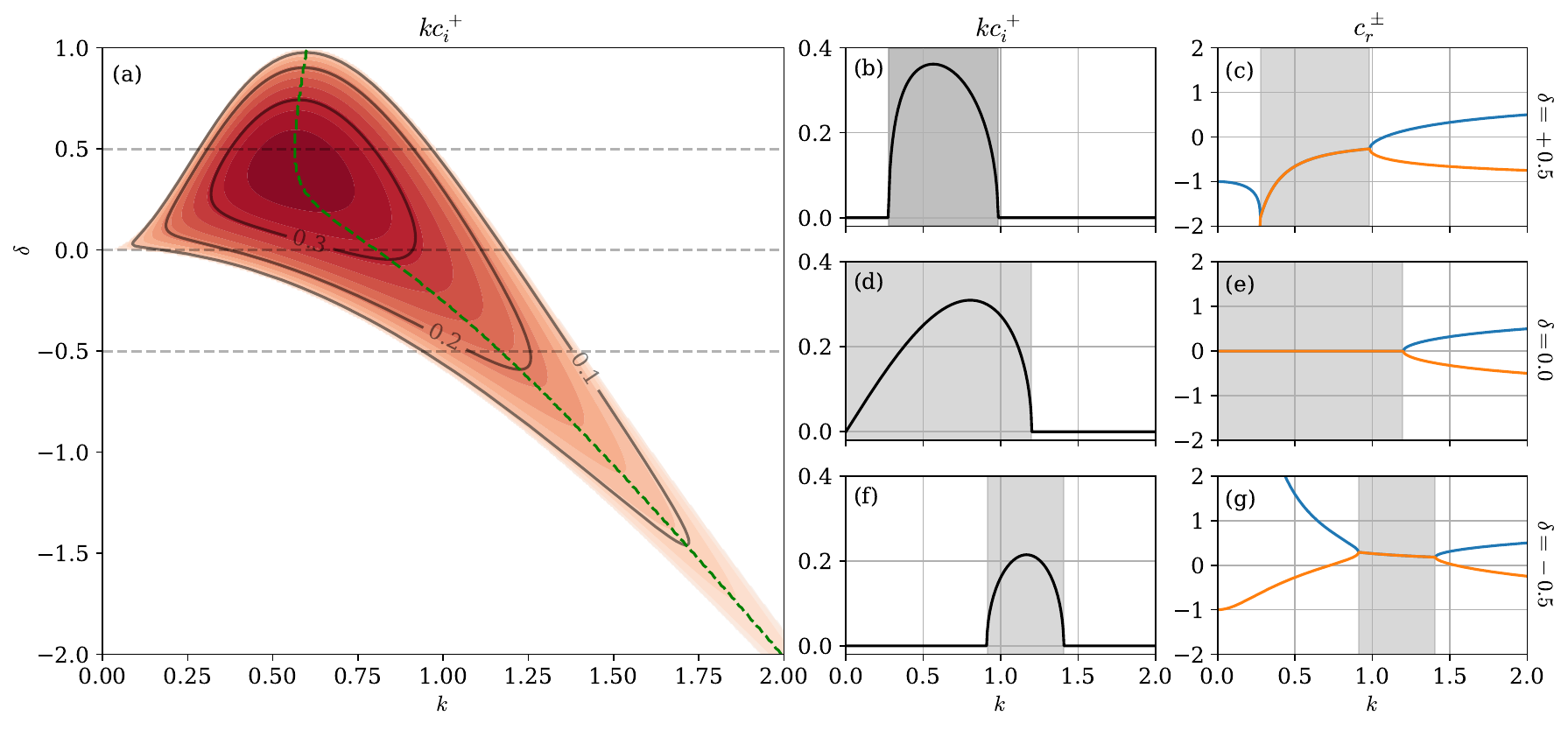}
  \caption{General instability characteristics for the case of $l=0$. ($a$)
  Growth rate as a function of the non-dimensional zonal wavenumber $k$ and
  $\delta$ parameter, with darker shadings denoting higher values, sample
  contours of growth rates, and the green dashed line denoting
  $k_{\max}(\delta)$ where the growth rate is maximized. Also shown are the
  growth rates $kc^+_i$ and phase-speed $c_r^\pm$ for ($b,c$) $\delta = +0.5$,
  ($d,e$) the standard Eady problem $\delta = 0.0$, and ($f,g$) $\delta = -0.5$.
  The shaded regions in panels $b$-$g$ denote the regions where there the growth
  rates are non-zero.}
  \label{fig:growth_rates_cr_l00}
\end{figure}

Computing for $c$ analytically or numerically, and denoting the solutions of the
plus and minus branch as $c^{\pm}$ for ease of discussion, plots of $c^\pm_r$
and growth rate of the instability $kc_i^\pm$ as a function of wavenumber $k$
and $\delta$ can be constructed, and a sample of this for the case $l=0$ is
shown in Fig.~\ref{fig:growth_rates_cr_l00}; we will focus mostly on the $l=0$
case in this work since the gravest meridional mode seems to be associated with
the largest growth rates at fixed $\delta$. Note that for our computations shown
for the remainder of this work that we take $F=1$ for simplicity. Additionally,
compared to more standard non-dimensional formulations
\cite[e.g.,][]{Vallis-GFD, Isachsen11, Pedlosky16, Chen-et-al20}, our values of
$k$ and $c$ are smaller and larger by a factor of 2 respectively, but $kc_i$
remain the same.

We make an observation that Eq.~\eqref{eq:lin-nondim} is invariant under the
transformation
\begin{equation}
  \mathcal{P}: (x,y,z) \mapsto (-x,-y,-z), \qquad \mathcal{T}: (t, \psi) \mapsto (-t, -\psi),
\end{equation}
so the system is parity-time ($\mathcal{PT}$) symmetric, where parity refers to
a flipping or mirroring of space, while time symmetry refers to reversal of time
\cite{Bender-PT, Qin-et-al19, Zhang-et-al20, David-et-al22}. Note that we regard
$\delta$ as a given and prescribed parameter of the system, so $\delta$ does not
transform under $\mathcal{P}$. The fact that the equations are $\mathcal{PT}$
symmetric implies the solution spectrum has certain properties (e.g.,
Fig.~\ref{fig:growth_rates_cr_l00}$b$-$g$), and has very suggestive links for
shear instability as well as the edge-wave interpretation for shear
instabilities. The details are somewhat of a digression, and we refer the reader
to Appendix \ref{app:A} for details.

There are a few features that may be observed from
Fig.~\ref{fig:growth_rates_cr_l00}:
\begin{enumerate}
  \item when $\delta = 0$ and there is instability, $c^\pm_r = 0$;
  \item when $\delta \neq 0$ and there is instability, $c^+_r = c^-_r$, and if
  $\delta > 0$, $c^\pm_r < 0$ (and vice-versa for $\delta < 0$); 
  \item there is no instability for $\delta \geq 1$, but instability seems to
  persist for $\delta < 0$;
  \item relative to the standard $\delta = 0$ case, the most unstable wavelength
  $k_{\rm max}$ decreases somewhat for $\delta > 0$, but persistently increases
  for $\delta < 0$, with the unstable bandwidth shifting to larger wavenumbers.
\end{enumerate}
A test of the wave interaction interpretation would be whether we are able to
rationalize the changes in the instability characteristics via changes to the
properties of these edge-waves that is valid over the whole parameter space.

The first point simply arises from our choice of problem formulation, where our
domain goes from $z=\pm1$ and our velocity profile is anti-symmetric about $z=0$
(cf. $c^\pm_r=(\overline{U}_{\rm top} + \overline{U}_{\rm bot}) / 2=0.5$ with
the more standard formulations and choice of non-dimensionalization in other
works \cite[e.g.,][]{Vallis-GFD, Isachsen11, Pedlosky16, Chen-et-al20}). The
second point relates to the first point, and together with the third point can
be rationalized as follows. Noting first that, with counter-propagation of
Rossby edge-waves against the mean flow, the top wave intrinsically propagates
to the left, while the bottom wave would propagate to the right. Noting that
$\partial H_b / \partial y \sim \delta$, the presence of the slope is to alter
the background PV gradient on which the the edge-waves are propagating (cf.
Fig.~\ref{fig:setup}$b$, via PV stretching), and in the present setting, $\delta
> 0$ \emph{counter}acts while $\delta < 0$ \emph{reinforces} the PV gradient
provided by the background state; mathematically this is through the $(1 -
\delta)$ term in Eq.~\eqref{eq:lin-nondim}. For $\delta > 0$, the bottom wave is
\emph{weakened} and propagates \emph{slower} for a fixed wavenumber, and one
might expect the instability has more of the characteristic of the top wave
propagating \emph{left}, resulting in an unstable mode with $c^\pm_r <
c^\pm_r(\delta = 0) = 0$. On the other hand, for $\delta < 0$, the bottom wave
is \emph{strengthened} and propagates \emph{faster} for a fixed wavenumber, the
bottom wave dominates and resulting in an unstable mode with $c^\pm_r >
c^\pm_r(\delta = 0) = 0$. These observations can be verified with standard
formulations, where we have instead $c^\pm_r \lessgtr (\overline{U}_{\max} +
\overline{U}_{\min}) / 2 = c^\pm_r(\delta = 0) = 0$. The third point is also
consistent with the mechanistic picture: there is no instability for $\delta
\geq 1$, because counter-propagation is then no longer possible, but for $\delta
< 0$ counter-propagation appears to always be possible, and could in principle
persist at increasing wavenumbers, albeit over a decreasing bandwidth. Some of
these points were already noted by the work of \cite{Pedlosky16, Chen-et-al20}.

The fourth point however is not covered by the explanation given in
\cite{Pedlosky16, Chen-et-al20} as such (although it was neither work's main
focus). Those works argue that for phase-locking the edge-wave phase-speeds must
match, and therefore look for conditions where the phase-speeds match. While
that argument functions well for the $\delta<0$ regime, it (1) fails for the
opposite case of $\delta > 0$, where the most unstable wavenumber do not
uniformly go to longer waves or smaller wavenumbers (green dashed line of
Fig.~\ref{fig:growth_rates_cr_l00}$a$), and (2) does not explain the changes in
the strength of instability over parameter space. The reason we will argue for
is simply that \emph{mutual edge-wave interaction matters}, is part of the
solution and central to the counter-propagating Rossby wave mechanism, and
cannot simply be ignored, as is done when considering simply phase-speeds can
match. The associated mutual interaction leads to extra helping/hindering of the
wave propagation that affects phase-locking, and strength of interaction affects
the growth rates.


\section{Clarifying the instability mechanism in terms of Rossby edge-waves}\label{sec:edgewave}


\subsection{edge-wave formulation in phase-amplitude variables}

To quantify the impact of slopes and its modification to the background PV on
the wave interaction mechanism, here we consider expressing the problems
explicitly in terms of Rossby or PV edge-waves and its interaction, rather than
in the streamfunction eigenfunction. The streamfunction eigenfunction
$\tilde{\psi}(z)$ as it stands is in general a tilted structure in space that
could be regarded as a superposition of edge-wave structures, and the problem
here is in the definition of an appropriate basis of PV edge-wave functions.
While there is a general approach for constructing the PV edge-waves in terms of
wave activity variables such as pseudomomentum and pseudoenergy
\cite[e.g.,][]{Held85, Shepherd90, Heifetz-et-al04a, Heifetz-et-al09,
HeifetzGuha19}, we do not need that amount of complexity here since the
edge-wave locations are well defined for the present problem. For elucidation
purposes we will derive the structure and governing equations explicitly.

Consider expressing the streamfunction eigenfunction $\tilde{\psi}(z)$ in terms
of a linear superposition of untilted structures focused at the top and bottom
boundary (subscript $T$ and $B$ respectively) such that
\begin{equation}
  \tilde{\psi} = \tilde{\psi}_T + \tilde{\psi}_B, \qquad \tilde{q} = \tilde{q}_T + \tilde{q}_B,
\end{equation}
where tilde denotes functions that are $z$ only. Assuming modal solutions as in
\eqref{eq:eigenfunc}, $\tilde{q}$ and $\tilde{\psi}$ are related via
\begin{equation}\label{eq:helmholtz}
  \tilde{q} = -\mu^2 \tilde{\psi} + \frac{\partial^2\tilde{\psi}}{\partial z^2},
\end{equation}
subject to the boundary conditions that
\begin{equation}\label{eq:bcs}
  \left.\frac{\partial \tilde{\psi}_T}{\partial z}\right|_{z=-1} = 0, \qquad \left.\frac{\partial \tilde{\psi}_B}{\partial z}\right|_{z=+1} = 0.
\end{equation}
Denoting $\hat{\delta}$ to be the Dirac $\delta$-distribution, if we take
(abusing mathematical rigor somewhat) 
\begin{equation}
  \tilde{q}_B = \hat{q}_B(t)\hat{\delta}(z+1), \qquad \tilde{q}_T = \hat{q}_T(t)\hat{\delta}(z-1),
\end{equation}
i.e. PV anomaly of an edge-wave is non-zero only at the associated locations,
then either by manually constructing a solution \cite[cf.][]{DaviesBishop94}, or
by noting that we are in effect looking for the Green's function associated with
the one-dimensional Helmholtz operator (in Eq.~\ref{eq:helmholtz}) subject to
homogeneous Neumann conditions (in Eq.~\ref{eq:bcs}), for which solutions are
documented (e.g., online Green's function libraries, with appropriate changes of
variable), or otherwise, the relevant solutions are
\begin{equation}\label{eq:psi_wave}
  \tilde{\psi}_B = -\hat{q}_B\frac{\cosh\mu(1-z)}{\mu\sinh2\mu}, \qquad \tilde{\psi}_T = -\hat{q}_T\frac{\cosh\mu(1+z)}{\mu\sinh2\mu}.
\end{equation}
Note that, with \eqref{eq:psi_wave},
\begin{equation}
  \tilde{b} = \frac{\partial\tilde{\psi}}{\partial z} = \begin{cases}-\hat{q}_T, & z=+1, \\ +\hat{q}_B, & z=-1,\end{cases}
\end{equation}
demonstrating the explicit relation between buoyancy and PV anomalies, and that
the top edge-wave induces no PV anomaly at the location of the other edge-wave.

Taking $\hat{q}_T = T\mathrm{e}^{\mathrm{i}\epsilon_T}$ and $\hat{q}_B =
B\mathrm{e}^{\mathrm{i}\epsilon_B}$, substituting \eqref{eq:psi_wave} into the
governing equations (\ref{eq:lin-nondim}$b,c$) and considering the real and
imaginary parts lead to
\begin{subequations}\label{eq:amp_phase}
\begin{align}
  \frac{1}{T}\frac{\partial T}{\partial t} &= +\frac{k}{\mu \sinh 2\mu}\frac{B}{T}\sin\Delta\epsilon,\\
  \frac{1}{B}\frac{\partial B}{\partial t} &= -\frac{k(1-\delta)}{\mu \sinh 2\mu}\frac{T}{B}\sin\Delta\epsilon,\\
  -\frac{1}{k}\frac{\partial \epsilon_T}{\partial t} &= +\left[1 - \frac{1}{\mu\sinh2\mu}\left(\cosh2\mu + \frac{B}{T}\cos\Delta\epsilon\right)\right],\\
  -\frac{1}{k}\frac{\partial \epsilon_B}{\partial t} &= -\left[1 - \frac{(1-\delta)}{\mu\sinh2\mu}\left(\cosh2\mu - \frac{T}{B}\cos\Delta\epsilon\right)\right],
\end{align}
\end{subequations}
where we define $\Delta\epsilon = \epsilon_T - \epsilon_B$ as the phase-shift of
the edge-wave in terms of PV signature; $\Delta\epsilon > 0$ means the top wave
has a PV signature that is \emph{lagging behind} the bottom wave PV signature
(cf. Fig.~\ref{fig:setup}$a$). The set of equations are cast in a form that is
more similar to Eq. (14) of \cite{Heifetz-et-al99} for the dimensional
formulation of the Rayleigh shear profile problem, but is equivalent to Eq. (7)
of \cite{DaviesBishop94}, who consider the phase-shift in terms of the buoyancy
variable instead. Note that since $\tilde{\psi}\sim -\tilde{q}$, the phase-shift
applies also to the streamfunction; contrast this to $\tilde{v} = \mathrm{i} k
\tilde{\psi}$, which would be shifted by $\pi/2$, and $\tilde{b}$ which would be
shifted instead by $\pi$. Here, $k / (\mu\sinh2\mu)$ plays the role of the
vertical interaction (cf. $\mathrm{e}^{-k}$ in \cite{Heifetz-et-al99} for the
Rayleigh profile in the barotropic setting). Taking the amplitudes $T$ and $B$
as positive without loss of generality, we note that we need
$\Delta\epsilon\in(0, \pi)$ for growth of edge-waves, which is consistent with
what we know about baroclinic instability: an unstable mode has the PV,
streamfunction and meridional flow patterns leaning \emph{against} the shear
(top signal lagging bottom signal; see Fig.~\ref{fig:edge_vs_eigen_shift_l00}
for example), while the buoyancy pattern leans \emph{into} the shear
corresponding to a shift by $\pi$ \cite[e.g.,][]{Vallis-GFD}.

While there are four independent variables, the equations depend only on the
amplitude ratios and the phase difference, and could be considered a
two-dimensional dynamical system. Following the notation of
\cite{Heifetz-et-al99}, we define the amplitude ratio as $\tan\gamma = T/B$.
Noting then various trigonometric identities such as
\begin{equation*}
  \frac{B^2 - T^2}{B^2 + T^2} = \cos2\gamma, \qquad \frac{B^2 + T^2}{2BT} = \frac{1}{2\sin2\gamma},
\end{equation*}
\eqref{eq:amp_phase} takes the form
\begin{subequations}\label{eq:amp_phase_ratios}
\begin{align}
    \frac{\partial \gamma}{\partial t} &= \frac{k}{\mu \sinh2\mu} \sin\Delta \epsilon (\cos 2\gamma + \delta \sin^2\gamma),\\
    \frac{\partial \Delta\epsilon}{\partial t} &= \frac{2k}{\mu \sinh2\mu} \left[\left(1-\frac{\delta}{2}\right)\cosh2\mu - \mu\sinh2\mu + \left(\frac{1}{\sin 2\gamma} - \frac{\delta}{2}\tan\gamma\right)\cos\Delta\epsilon \right].
\end{align}
\end{subequations}
The set of equations should be compared with Eq. (15) of \cite{Heifetz-et-al99},
noting the difference in the interaction function ($k /(\mu\sinh2\mu$) vs.
$\mathrm{e}^{-k}$), arising from the differing physics between the systems being
considered, encapsulated in the different Green's functions of the associated
system.

\begin{figure}
  \includegraphics[width=0.8\textwidth]{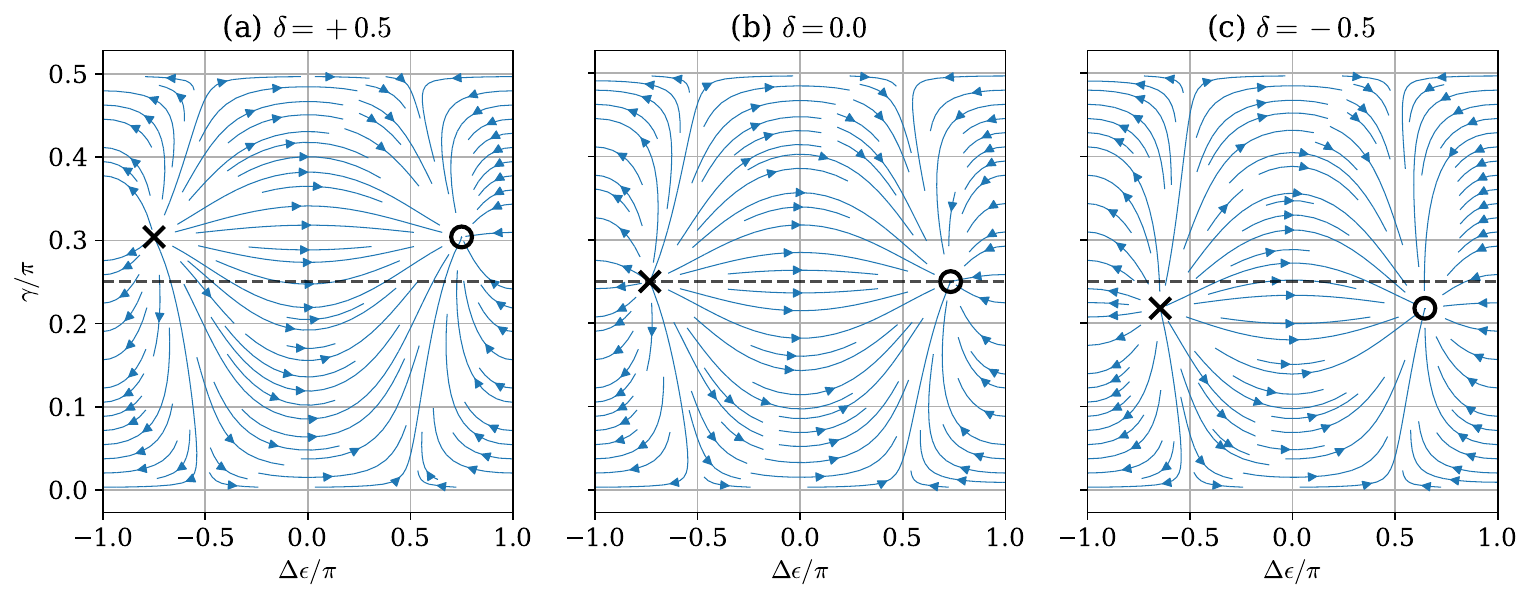}
  \caption{Phase portrait and the fixed points of the dynamical system (stable
  and unstable mode marked on as a circle and cross respectively), for three
  choices of $\delta$, at wavenumber ($k = k_{\max}(\delta)$, $l=0$).}
  \label{fig:phase_portrait_l00}
\end{figure}

\begin{figure}
  \includegraphics[width=0.9\textwidth]{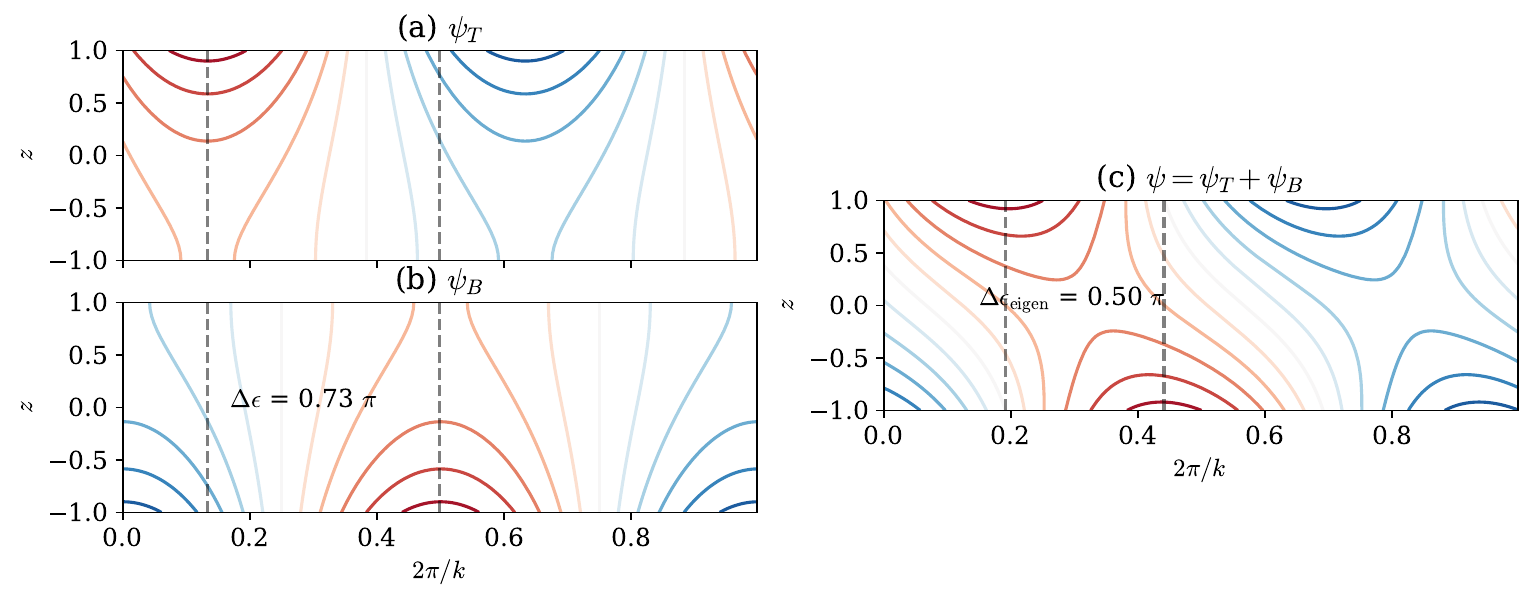}
  \caption{Edge-waves components and the combination demonstrated for the most
  unstable mode of the standard Eady problem ($\delta=0$). ($a,b$) Associated
  edge-wave structure $\psi_{T,B}$ (referenced by $B=1$ and $\epsilon_B=0$)
  given in \eqref{eq:psi_wave} of the attracting fixed point in
  Fig.~\ref{fig:phase_portrait_l00}$b$, with $\Delta\epsilon \neq \pi/2$. ($c$)
  The reconstructed normal-mode $\psi = \psi_T + \psi_B$ that has a phase-tilt
  $\Delta\epsilon_{\rm eigen} = \pi/2$.}
  \label{fig:edge_vs_eigen_shift_l00}
\end{figure}

A dynamical system could be described in terms of phase portraits
\cite[e.g.,][]{Strogatz-Dynamical}, and phase portraits associated with the most
unstable wavenumber $k$ (with $l=0$) for sample choices of $\delta$ are shown in
Fig.~\ref{fig:phase_portrait_l00}. Note that the stable and unstable equilibrium
points of the dynamical system (repellers and attractors) are associated with
the unstable and stable normal-modes respectively in the unstable bandwidth
\cite[cf.][]{Heifetz-et-al99, Heifetz-et-al04a}. As the stability boundaries are
passed there is a bifurcation, and the equilibrium points become centers located
at $\gamma=0, \pi/2$, and $\Delta\epsilon = \pm\pi$ (not shown), associated with
neutral and freely propagating edge-waves with no change in amplitude or phase.
The non-equilibrium points have been argued to correspond to non-modal growth
\cite[e.g.,][]{DaviesBishop94, Heifetz-et-al99}, and the phase portraits
indicate the regime transition in terms of edge-waves as the non-modal
instabilities develop, but we leave this for the interested reader to pursue.

Focusing on the unstable modes (the attractors), we note that the associated PV
phase-shifts $\Delta\epsilon$ is \emph{not} $\pi/2$, even for the $\delta=0$
case, which seems to contradict what is generally documented about the Eady
problem having a phase-shift of $\pi/2$ \cite[e.g.,][]{Vallis-GFD}. There is in
fact no discrepancy: taking the $\delta=0$ case as an example (i.e., standard
Eady problem), we compute the values of $\gamma$ and $\Delta\epsilon$ associated
with the stable equilibrium point and construct the edge-wave couplets as well
as their sum, and these are shown in Fig.~\ref{fig:edge_vs_eigen_shift_l00}
(recalling $\hat{q} = B\mathrm{e}^{\mathrm{i}\epsilon_B}$, we take the reference
to be $B=1$ and $\epsilon_B = 0$) . While $\Delta\epsilon \neq \pi/2$, their
combination does lead to a phase-\emph{tilt} in the \emph{streamfunction
eigenfunction} $\Delta\epsilon_{\rm eigen} = \pi/2$, and the reconstructed
solution can be shown to coincide exactly with the one obtained from the more
standard normal-mode analysis. A non-optimal shift is realized simply because
the mutual interaction from the interaction function (which is wavenumber
dependent) is also part of the solution, and provides extra hindering of the
wave propagation required for phase-locking \cite[e.g.,][]{Heifetz-et-al99,
Heifetz-et-al04a, Tamarin-et-al16, Heifetz-et-al15}. Considerations based simply
on phase-shift (or phase-tilt) and/or phase-speed matching is incomplete,
because it is ignoring the interaction component.

The statement in the above paragraph holds true for different values of $k$ and
$\delta$ (not shown). For synchronized growth of edge-waves associated with
stable equilibrium points of \eqref{eq:amp_phase_ratios}, we have
$\Delta\epsilon\in(0, \pi)$ and $\gamma \in (0, \pi/2)$, so that the growth rate
can be inferred from \eqref{eq:amp_phase}, given by
\begin{equation}\label{eq:sync_growth}
  \sigma = \left|\frac{k}{\mu\sinh2\mu}\frac{1}{\tan\gamma}\sin\Delta\epsilon\right| = \left|\frac{k(1-\delta)}{\mu\sinh2\mu}\tan\gamma\sin\Delta\epsilon\right|,
\end{equation}
which numerically coincides with that of Fig.~\ref{fig:growth_rates_cr_l00}$a$
(not shown). The phase-speeds of the normal-modes $c_r$ can also in principle
be reconstructed in principle from the edge-wave basis via consideration of the
self- and induced-propagation by the edge-waves \cite[cf.,][]{Heifetz-et-al04a,
HeifetzGuha19}. The edge-wave formulation here encompasses the standard
formulation of the modified Eady problem, which is not surprising given it is
really a reformulation of the same problem, but expressing it in a basis that
allows for a mechanistic interpretation to be drawn from.

\begin{figure}
  \includegraphics[width=0.6\textwidth]{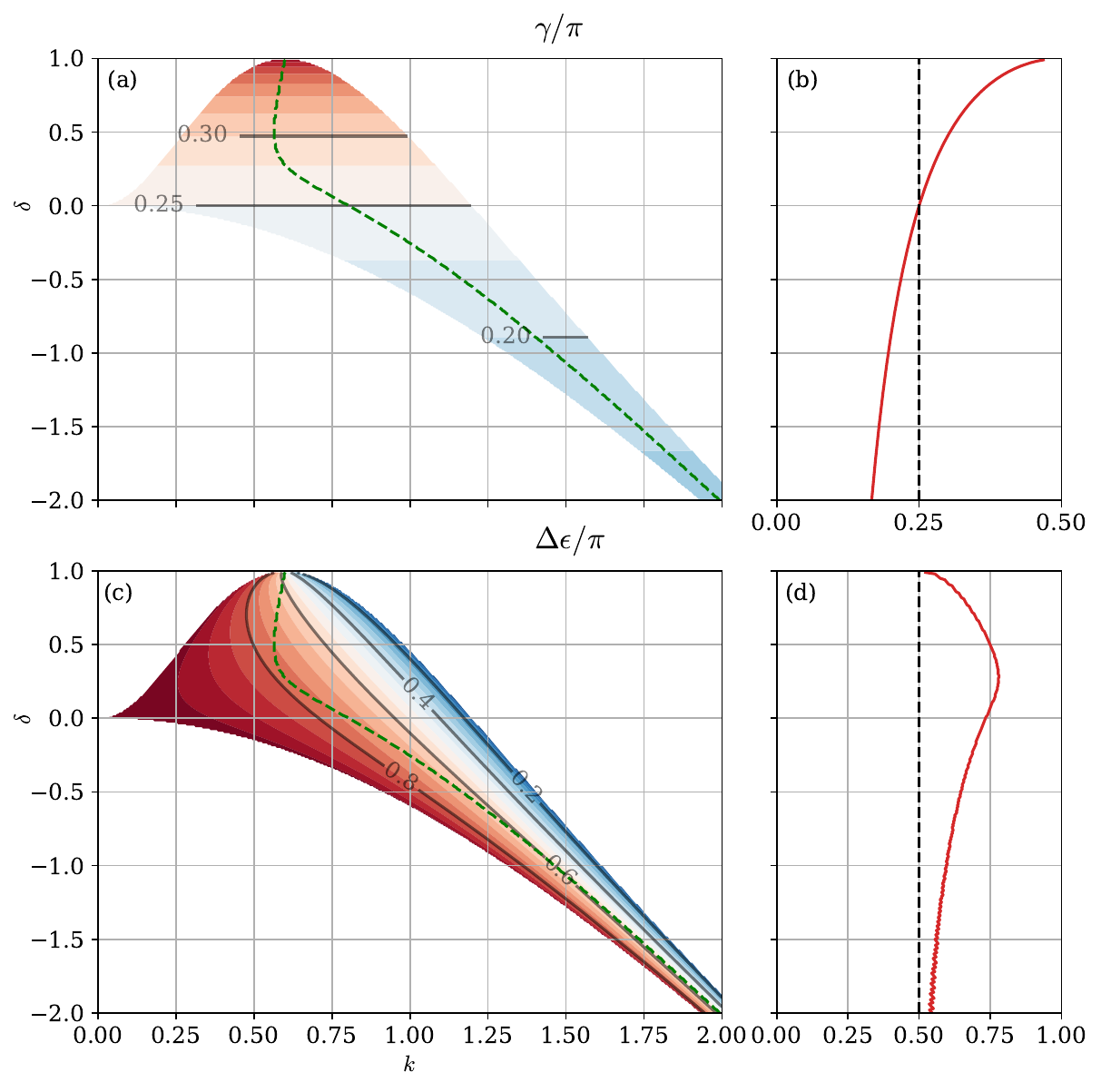}
  \caption{($a$) Normalized amplitude ratio $\gamma$ in multiples of $\pi$ as a
  function of $k$ and $\delta$, and ($b$) the same quantity along the line of
  $k_{\max}(\delta)$ (red line, with $\gamma_{\rm eigen}$ as faint green line);
  note $0.25$ here denotes equal amplitude at top and bottom. ($c$) The
  normalized edge-wave phase-shift $\Delta\epsilon$ in multiples of $\pi$ as a
  function of $k$ and $\delta$, and ($d$) the same quantity along the line of
  $k_{\max}(\delta)$ (red line); note $+0.5$ here denotes that the top edge-wave
  \emph{lags behind} the bottom edge-wave by a quarter wavelength. The line of
  $k_{\max}(\delta)$ has been marked on panel ($a,c$) as the green dashed line.}
  \label{fig:amp_phase_from_edge_wave}
\end{figure}

We show in Fig.~\ref{fig:amp_phase_from_edge_wave} the amplitude ratios $\gamma$
and edge-wave phase-shift $\Delta\epsilon$ of equilibrium points of
\eqref{eq:amp_phase_ratios} associated with unstable modes. Starting first with
the amplitude ratio $\gamma$, it is clear that $\gamma$ only depend on $\delta$.
The analytic expression for $\gamma$ can be obtained by noting that, with
synchronized growth and growth rate given by \eqref{eq:sync_growth}, we must
have (with appropriate normalization and/or shifts in the phase)
\begin{equation}\label{eq:amp_ratio}
  \left|\frac{1}{\tan\gamma}\right| = \left|\frac{B}{T}\right| = \sqrt{1-\delta},
\end{equation}
which is $k$ independent. Notice that: (1) $|B/T|$ is ill-defined for $\delta >
1$, corresponding to the case where there is no instability (since there is no
counter-propagation possible for the bottom wave as the bottom background PV
gradient has switched signs); (2) $|B| = 0$ for $\delta=1$ for physically sound
solutions (coinciding with vanishing PV gradient at the bottom); (3) $|T| > |B|$
for $\delta\in(0, 1)$, i.e. weaker bottom wave; (4) $|T| = |B|$ for $\delta =
0$, and there is no asymmetry in the standard Eady case; (5) $|B| > |T|$ for
$\delta < 0$, and there is \emph{always} instability possible for $\delta < 0$.
The observations are consistent with our physical expectations highlighted in
the previous sections.

The behavior of $\Delta\epsilon$ in Fig.~\ref{fig:amp_phase_from_edge_wave}$c$
is in line with kinematic arguments from edge-waves that help or hinder each
other's propagation, with implications for phase-locking. We generally need
$\Delta\epsilon\in(0, \pi)$ for constructive interference. At fixed $\delta$,
long waves propagate faster (since these are Rossby edge-waves), and the
edge-waves need to \emph{hinder} each other to maintain phase-locking, which for
instability requires $\Delta\epsilon\in(\pi/2, \pi)$. The converse holds for
shorter waves, requiring $\Delta\epsilon\in(0, \pi/2)$
\cite[e.g.,][]{Heifetz-et-al04a}.


\subsection{Physical rationalization of the interacting edge-wave process}\label{sec:rationalise}

Our interpretation for the role of topographic PV on the Counter-propagation
Rossby Wave mechanism is then as follows. Starting with the $\delta<0$ case, the
bottom wave is strengthened and propagate faster, but also leads to a stronger
induced velocity at the top wave. The strong interaction leads to a stronger
\emph{hindering} effect for fixed wavenumber, which will in general lead to a
sub-optimal phase-shift configuration. One way to drive the configuration
towards a more optimal configuration would be to go towards \emph{shorter}
waves, which offsets the increased interaction introduced by a stronger bottom
wave from the fact that $\delta < 0$ (since the interaction function goes like
$k/(\mu\sinh2\mu)$ for the present Eady system). This also then explains the
reduction in the growth rate as we shift towards shorter waves
(Fig.~\ref{fig:growth_rates_cr_l00}$a$): while the efficiency could be changed
via the phase-shift, the interaction strength is \emph{decreased}. It would seem
that, in the $\delta<0$ setting, it is \emph{always} possible to compensate for
the increase in interaction from $\delta\to-\infty$ by reducing the interaction
function via increases in $k$, albeit over an increasingly narrow bandwidth of
wavenumbers. The phase-speeds are positive
(Fig.~\ref{fig:growth_rates_cr_l00}$g$) because the bottom edge-wave propagating
to the right (or eastwards) dominates over the top edge-wave.

For the $1 > \delta > 0$ regime, the bottom wave is weakened (from
\eqref{eq:amp_ratio}) and leads to a weaker hindering effect in general. What
this means is that the top wave is now propagating too fast, and this effect
would have to be offset by increasing the interaction function via decreasing in
the wavenumber, i.e., going to longer wavelengths. However, unlike the
$\delta<0$ case, reduction in $k$ leads an increase in edge-wave propagation
since we are dealing with Rossby edge-waves, and beyond a certain point it is
simply not possible for the bottom wave's induced velocity and the background
flow to hold the top wave into a phase-locked position, and instability is no
longer possible. As $\delta\nearrow1$, the PV gradient vanishes,
counter-propagation is no longer possible, and no phase-locking can be achieved.
Note that $\delta \geq 1$ coincides with the non-satisfaction of the
Charney--Stern condition that it is necessary for the background PV gradient to
change sign in the domain in order for instability, which had previously been
interpreted as a condition required for counter-propagation
\cite[e.g.,][]{Heifetz-et-al04a, Heifetz-et-al09, HeifetzGuha17}. The
phase-speeds are negative (Fig.~\ref{fig:growth_rates_cr_l00}$c$) because the
top edge-wave propagating to the left (or wastwards) dominates over the bottom
edge-wave.


\subsection{Analysis in terms of the instability normal-mode}\label{sec:eigenfunc}

\begin{figure}
  \includegraphics[width=0.6\textwidth]{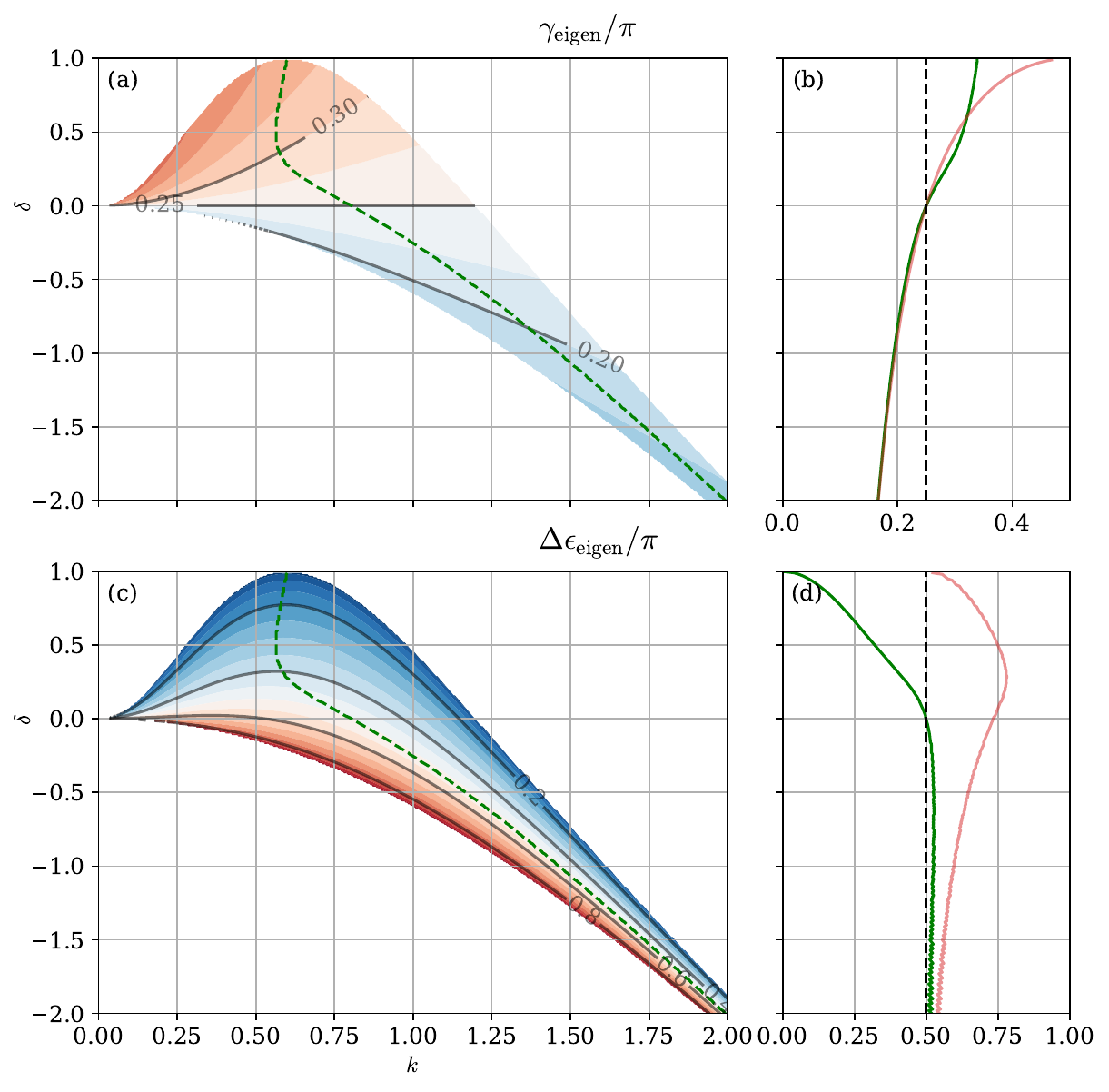}
  \caption{Analogue of Fig.~\ref{fig:amp_phase_from_edge_wave} but for
  quantities diagnosed from the normal-mode directly. ($a$) Normalized
  amplitude ratio $\gamma_{\rm eigen}$ in units of $\pi$ as a function of $k$
  and $\delta$, and ($b$) the same quantity along the line of $k_{\max}(\delta)$
  (green line, with $\gamma$ from Fig.~\ref{fig:amp_phase_from_edge_wave}$b$ as
  faint red line); note $0.25$ here denotes equal amplitude at top and bottom.
  ($c$) The normalized phase-tilt in the streamfunction $\Delta\epsilon_{\rm
  eigen}$ in units of $\pi$ as a function of $k$ and $\delta$, and ($d$) the
  same quantity along the line of $k_{\max}(\delta)$ (green line, with
  $\Delta\epsilon$ from Fig.~\ref{fig:amp_phase_from_edge_wave}$d$ as faint red
  line); note $+0.5$ here denotes that the top edge-wave \emph{lags behind} the
  bottom edge-wave by a quarter wavelength. The line of $k_{\max}(\delta)$
  marked on as green has been marked on panel ($a,c$) as the green dashed line.}
  \label{fig:amp_phase_from_Phi}
\end{figure}

Here we provide an analogous analysis to demonstrate how different the results
are if the instability streamfunction eigenfunction from \eqref{eq:eigenfunc} is
utilized instead. Focusing on unstable modes, given a value of $c$, we can
obtain the coefficients $a$ and $b$ for the vertical structure function
$\tilde{\psi}(z)$ from \eqref{fig:vert_struc}. Given $\tilde{\psi}(z) =
\tilde{\psi}_r + \mathrm{i}\tilde{\psi}_i$, we can compute for a (normalized)
amplitude and phase of the eigenfunction via \cite[e.g.,][]{Vallis-GFD,
Chen-et-al20}
\begin{equation}
  |\tilde{\psi}(z)|^2 = \tilde{\psi}_r^2(z) + \tilde{\psi}_i^2(z), \qquad \epsilon(z) = \arctan\frac{\tilde{\psi}_i(z)}{\tilde{\psi}_r(z)}.
\end{equation}
Analogous to our edge-wave analysis in Sec.~\ref{sec:edgewave}, we introduce the
quantities
\begin{equation}
  \tan\gamma_{\rm eigen} = \frac{|\tilde{\psi}(z=1)|}{|\tilde{\psi}(z=-1)|}, \qquad \Delta\epsilon_{\rm eigen} = \epsilon(z=1) - \epsilon(z=-1)
\end{equation}
as a measure of the amplitude ratio and phase-\emph{tilt} (rather than
phase-shift) between the streamfunction eigenfunction at the top and bottom of
the domain respectively. Fig.~\ref{fig:amp_phase_from_Phi} shows the amplitude
ratio $\gamma_{\rm eigen}$ and phase-tilt $\Delta\epsilon_{\rm eigen}$ as
measured through the eigenfunction $\tilde{\psi}(z)$.

Starting first with the amplitude ratio, we note that the standard Eady case
with $\delta=0$ has $\gamma_{\rm eigen} = \pi/4$ throughout the unstable
bandwidth, i.e. the top and bottom of the normal-mode have equal amplitude,
and in fact the normal-mode is symmetric about $z=0$ for $\delta=0$ (not
shown; \cite[cf.,][]{Nakamura93b}). For $\delta > 0$, $\gamma_{\rm eigen} >
\pi/4$, i.e. the normal-mode amplitude at the bottom is smaller than that at
the top, and conversely for $\delta < 0$, consistent with results from physical
expectations and edge-wave analysis. However, compared to the equivalent graph
in terms of edge-waves in Fig.~\ref{fig:amp_phase_from_Phi}$a,b$, there is a
wavenumber dependence in $\gamma_{\rm eigen}$, when our physical argument would
suggest that the amplitude ratio should only depend on $\delta$. The observation
of $\gamma_{\rm eigen}$ is consistent if we remember that diagnosed quantities
between two separate edge-wave structures do not necessarily need to be the same
as that diagnosed from their combinations.

For the phase-tilt in the streamfunction, in the standard Eady case with $\delta
= 0$, the maximally growing mode has $\Delta\epsilon_{\rm eigen} = \pi/2$, where
the top signal lags behind the bottom signal by quarter of a wavelength. As the
wavenumber increases, the tilt in the normal-mode reduces, while the opposite
is true as the wavenumber decreases. For $\delta\neq0$, the behavior of
$\Delta\epsilon_{\rm eigen}$ in the streamfunction is asymmetric with the sign
of $\delta$. Notably, the most unstable wavenumber for $\delta < 0$ has
$\Delta\epsilon_{\rm eigen} = \pi/2$, indicating the most unstable mode is still
able to access a standard Eady-like configuration in the normal-mode, albeit
with decreasing growth rate. However, while a phase-shift of $\pi/2$ would be
the optimum configuration for constructive interference in the edge-wave
interaction framework \cite[e.g.,][]{Hoskins-et-al85, DaviesBishop94,
Heifetz-et-al04a}, we should note that the phase-shift $\Delta\epsilon$ need not
coincide with the phase-tilt $\Delta\epsilon_{\rm eigen}$ (cf.
Fig.~\ref{fig:amp_phase_from_Phi}$c,d$ and
Fig.~\ref{fig:amp_phase_from_edge_wave}$c,d$). As was demonstrated in
Fig.~\ref{fig:edge_vs_eigen_shift_l00}, $\Delta\epsilon$ is generally not at the
expected optimal because one needs to take into account of the interaction
function.

One could then wonder whether there is in fact any specific meaning to the value
of the phase-tilt $\Delta\epsilon_{\rm eigen}$. It is certainly true that the
phase-tilt $\Delta\epsilon_{\rm eigen}$ (as well as the PV edge-wave
phase-shift) should relate to an \emph{energetic} interpretation of the
instability \cite[e.g., \S6.7.2 of][]{Vallis-GFD}, in relation to the fact that
perturbations of meridional velocity $v'$ and buoyancy $b'$ should be overall
positively correlated, so that the zonally averaged meridional buoyancy flux
$\overline{v'b'}$ is poleward, leading to a decrease in available potential
energy. While the statement about the energetics is true, the problematic aspect
is linking that to the raw value of $\Delta\epsilon_{\rm eigen}$. Since in the
linear instability analysis we should only be talking about rates and
efficiencies, we might suspect that $\Delta\epsilon_{\rm eigen} = \pi/2$ would
correspond to maximum efficiency in $\overline{v'b'}$ in reducing available
potential energy, i.e. maximum linear correlation. A straightforward linear
regression analysis for the correlation of $v'$ and $b'$ over the whole spatial
domain (not shown) indicates this is simply not true: $v'$ and $b'$ are
maximally correlated in the linear sense at small $k$, and its dependence as a
function of $k$ and fixed $\delta$ bears no resemblance to $\Delta\epsilon_{\rm
eigen}$ or the growth rate plots (the scatter plots of $v'$ against $b'$ becomes
increasingly `circular' with increasing $k$). The behavior of $\overline{v'b'}$
itself (making the choice of normalizing the eigenfunction $\tilde{\psi}(z)$ to
have unit magnitude) also bears no resemblance to $\Delta\epsilon_{\rm eigen}$
or the growth rate: for fixed $\delta$, maximum $\overline{v'b'}$ occurs at a
$k$ larger than the wavenumber at which there is maximum growth (not shown). So
while $\Delta\epsilon_{\rm eigen}$ would in some way be related to the energetic
as well as the kinematic/mechanistic view of the instability problem, it may
perhaps be simpler to not attribute too much meaning to $\Delta\epsilon_{\rm
eigen}$. The fact that $\Delta\epsilon_{\rm eigen} = \pi/2$ for the most
unstable mode is curious and is perhaps worthy of further exploration, but we
argue here that if we are invoking the edge-wave interaction mechanism, it is
perhaps quantitatively misleading or not entirely appropriate to provide
evidence in terms of something that is not in edge-wave form.


\section{Analysis in the GEOMETRIC framework}\label{sec:geometric}

The previous section highlights subtleties when using the streamfunction
eigenfunction with a mechanistic explanation, and argues that it is the
edge-wave basis that are more dynamically relevant. Does that mean the
instability eigenfunctions have little utility relative to the edge-wave basis?
We provide a processing of the eigenfunctions in terms of the GEOMETRIC
framework of \cite{Marshall-et-al12, MaddisonMarshall13} (see also
\cite{Hoskins-et-al83, WatermanHoskins13}) that considers geometric quantities
such as anisotropy factors and angles of eddy variance ellipses, and turns out
to have mechanistic links with baroclinic instability (cf. the barotropic case,
considered in \cite{Tamarin-et-al16}). The quantitative links between energetics
and mechanistic interpretations are demonstrated here for a case where both the
edge-wave basis and eddy fluxes are well-defined. The strong correlation of the
geometric quantities with that diagnosed from the edge-wave framework provides a
suggestion that in cases where the edge-wave basis is less well-defined (e.g.,
the linear Charney--Green problem, or data from the nonlinear evolution of
baroclinic instability), the GEOMETRIC framework may still be utilized and has
energetic and dynamical relevance.

As a recap to the work of \cite{Hoskins-et-al83, Marshall-et-al12,
MaddisonMarshall13}, in the QG limit, it is known that the eddy forcing on to the
mean state is determined by the object
\cite[e.g.,][]{Marshall-et-al12}
\begin{equation}
    \boldsymbol{\mathsf{E}} = \begin{pmatrix}-M + P & N & 0 \\ N & M+P & 0 \\ -S & R & 0 \end{pmatrix},
\end{equation}
with
\begin{equation}\begin{aligned}
  M = \frac{1}{2}\overline{v'^2 - u'^2} = -\gamma_m E \cos 2\phi_m \cos^2\lambda, \qquad N = \overline{u'v'} = \gamma_m E \sin 2\phi_m \cos^2\lambda,\\
  P = \frac{1}{2 N_0}\overline{b'^2} = E \sin^2\lambda, \qquad R = \frac{f_0}{N_0^2}\overline{u'b'} = \gamma_b \frac{f_0}{N_0} E \cos\phi_b \sin2\lambda, \qquad S = \frac{f_0}{N_0^2}\overline{v'b'} = \gamma_b \frac{f_0}{N_0} E \sin\phi_b \sin2\lambda,
\end{aligned}\end{equation}
where $M,N$ denote the eddy momentum fluxes (related to the Reynolds stresses),
$R,S$ denote the eddy buoyancy fluxes (related to the form stresses), $P$ is the
eddy potential energy, and $E = P + K$ is the total eddy energy, with the eddy
kinetic energy $K$ defined in the usual way. In the framework, $E$ becomes the
only dimensional variable, which is arbitrary up to a multiplicative constant
for the linear instability problem. By contrast, the non-dimensional geometric
quantities related to the eddy variance ellipses are independent of the
arbitrary mulitplicative constant, and are given by
\begin{equation}\begin{aligned}
    \gamma_m = \frac{\sqrt{M^2 + N^2}}{K},\qquad \gamma_b = \frac{N_0}{2 f_0} \sqrt{\frac{R^2 + S^2}{KP}},\\
    \sin 2\phi_m = \frac{N}{\sqrt{M^2 + N^2}},\qquad \sin \phi_b = \frac{S}{\sqrt{R^2 + S^2}},\\
    \frac{K}{E} = \cos^2\lambda, \quad \frac{P}{E}=\sin^2\lambda, \quad \tan^2\lambda = \frac{P}{K},
\end{aligned}\end{equation}
where the overbar denotes a mean operator (zonal average for the present work),
$\gamma_{m,b}$ are the momentum and buoyancy anisotropy parameters, $\phi_{m,b}$
are angle parameters related to the eddy momentum and buoyancy ellipses, while
$\lambda$ is an angle relating to the eddy energy partition. Note that there is
a degeneracy in the angle parameters, where for example we could define $\phi_b$
in terms of $\cos\phi_b = R / \sqrt{R^2 + S^2}$ as in \cite{Marshall-et-al12}.

One idea relating to parameterization of eddy fluxes is that the non-dimensional
geometric parameters might be more universal and related to dynamics and/or
instability characteristics, so are perhaps be easier to parameterize. For
example, it is known that in barotropic/horizontal shear instabilities, $\phi_m$
directly relates to the tilt angle of the eddy, where if the eddy tilts into the
shear we have instability, while if eddy tilts with the shear we would have the
converse, with eddies fluxing momentum back into the mean state
\cite[e.g.,][]{Hoskins-et-al83, Marshall-et-al12, WatermanHoskins13,
WatermanLilly15, Tamarin-et-al16}. In the present modified Eady problem, it can
be demonstrated that it is the eddy buoyancy rather than momentum fluxes that
are non-trivial, consistent with the present set up leading to a pure baroclinic
instability. The parameters of interest are then
\begin{equation}
    \tan 2\phi_t = \gamma_b \tan 2\lambda, \quad \gamma_t = \frac{\cos 2\lambda}{\cos 2\phi_t},
\end{equation}
where $\phi_t$ and $\gamma_t$ are the angle and anisotropy parameter of a
vertical eddy in physical space \cite{Hoskins-et-al83, Marshall-et-al12,
Youngs-et-al17}. From this, we note that we can define a non-dimensional
parameter $\alpha$ where
\begin{equation}\label{eq:alpha}
  \alpha = \gamma_b \sin\phi_b \sin 2\lambda = \gamma_t \sin\phi_b \sin 2\phi_t.
\end{equation}
The $\alpha$ parameter is a combination of geometric parameters that closely
relates to the Eady growth rate, is bounded in magnitude by unity in the QG
limit \cite{Marshall-et-al12}, and is one of the tuning parameters that is at
present prescribed in parameterizations of baroclinic restratification effects
\cite[e.g.,][]{Mak-et-al22}. An interest here is on the dependence of $\phi_t$
and $\alpha$ on $\delta$, and what is the dominant contribution to the variation
of $\alpha$, with the possibility to aid/inform our parameterization efforts for
baroclinic eddies and its feedback onto the mean state in theoretical and/or
numerical models \cite[e.g.,][]{Poulsen-et-al19, Mak-et-al22}.

For the case of $l=0$, i.e. no meridional variation, $u' = 0$,
and so $R = N = 0$ while $M^2 = K$, and so
\begin{equation}
    \gamma_m = 1, \quad \phi_m = 0, \quad \phi_b = \pm\frac{\pi}{2}, \quad \alpha = \pm\gamma_b \sin 2\lambda = \gamma_t \sin 2\phi_t.
\end{equation}
For the present set up it is the sign of $\alpha$ that distinguishes whether we
have instability or not, since $\phi_b$ is defined in term of the zonal mean
meridional advection of eddy buoyancy fluxes $S$, which is the principal
interest for baroclinic instability ($S>0$ is poleward flux of buoyancy, and so
$\sin\phi_b$ and $\alpha$ are both positive).

\begin{figure}
  \includegraphics[width=0.6\textwidth]{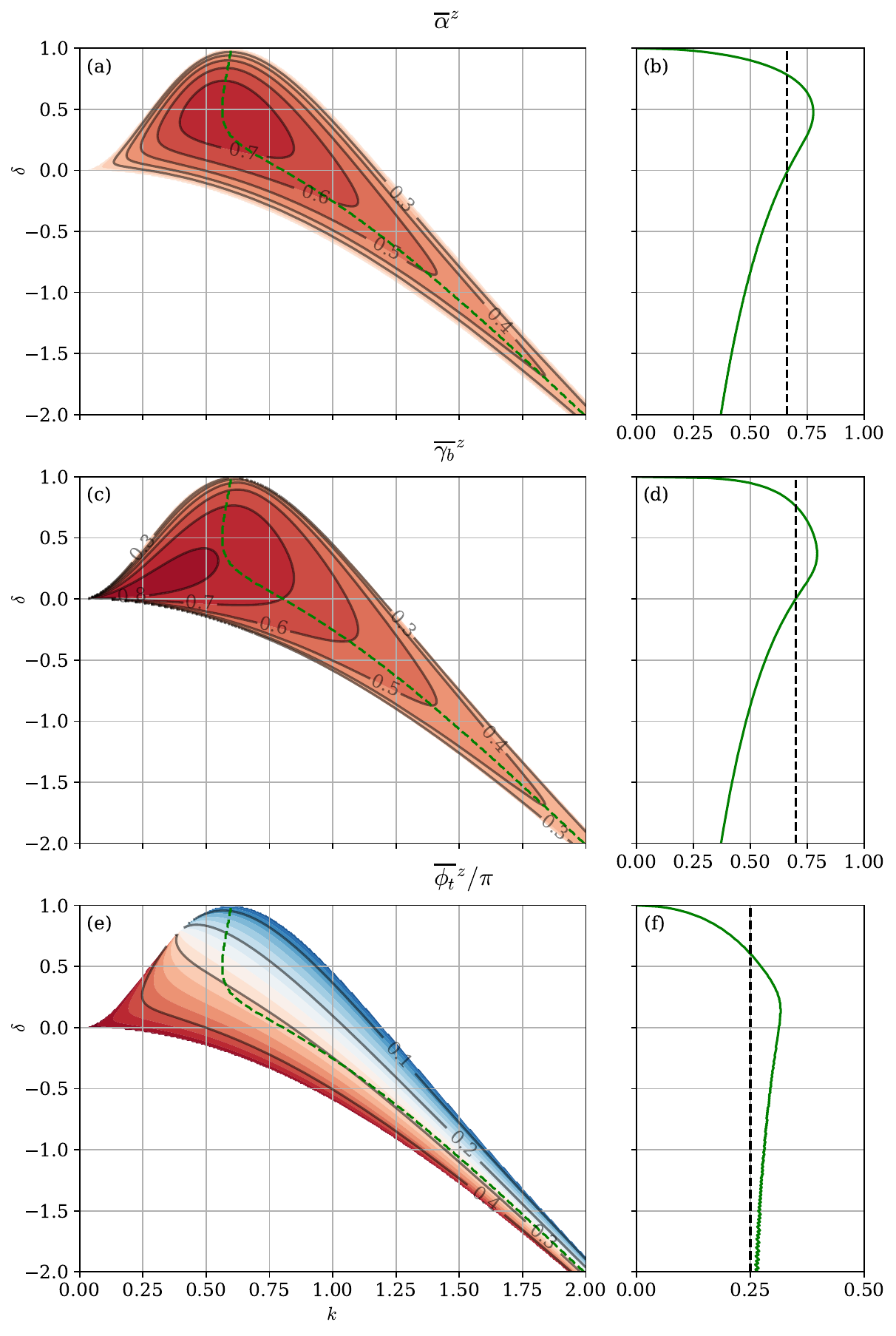}
  \caption{Vertically averaged ($a$) $\overline{\alpha}^z = \overline{\gamma_b
  \sin 2\lambda}^z$ ($c$) buoyancy anisotropy factor $\overline{\gamma_b}^z$ and
  ($e$) vertical tile angle $\overline{\phi_t}^z$ (in multiples of $\pi$) as a
  function of $k$ and $\delta$, and ($b,d,f$) the respective quantities along
  the line of $k_{\max}(\delta)$. The line of $k_{\max}(\delta)$ marked on as
  green has been marked on panel ($a,c,e$) as the green dashed line.}
  \label{fig:eigen_geometric_l00}
\end{figure}

Fig.~\ref{fig:eigen_geometric_l00} shows the value of the vertically averaged
$\alpha$, buoyancy anisotropy $\gamma_b$ and the vertical tilt $\phi_t$ over the
unstable region in parameter space. It may be seen that $\alpha$ correlates
strongly with the growth rate (Fig.~\ref{fig:growth_rates_cr_l00}$a$), in line
with the analysis of \cite{Marshall-et-al12}. Although the values of $\alpha$
reported here are around an order of magnitude larger than what are used in
parameterizations for numerical ocean models \cite[e.g.,][]{Mak-et-al22}, we
note that the diagnosed value here is for the linear instability analysis,
whilst the values used in parameterizations are applied as averages over both
energetic and quiescent regions and to mimic the feedback during the nonlinear
phase. The sensitivity of $\alpha$ to $\delta$ is of most relevance, suggesting
it should be $\alpha$ that is decreased in value in the presence of slopes,
particularly in the $\delta < 0$ scenario (where topographic and
isentropic/isopycnal slopes are opposite in orientation), consistent with a
previous diagnostic result in the nonlinear regime \cite{Wei-et-al22}.

The dominant contribution to the variation of $\alpha$ (in both the average and
pointwise sense) is principally through the buoyancy anisotropy parameter
$\gamma_b$ (see Fig~\ref{fig:eigen_geometric_l00}$a$ and $c$). The eddy energy
partition angle parameter $\lambda$ plays a secondary role, since $\sin2\lambda$
is diagnosed to be close to 1 in value in the pointwise sense (not shown). The
presence of a sloping boundary would naturally be expected to force an
anisotropy, and the feature of $\alpha$ correlating well with $\gamma_b$ and not
$\sin2\lambda$ seems to be consistent with diagnoses from numerical simulations
in a global ocean circulation model in the nonlinear regime
\cite{Poulsen-et-al19}, and diagnoses in the nonlinear regime of an idealized
baroclinic flow over a topographic slope (personal communication with Huaiyu Wei
and Yan Wang). The decreasing values of $\gamma_b$ for fixed $\delta$ as $k$ is
increased corresponds to a statement in the previous section that the scatter
plot of $v'$ against $b'$ becomes increasing `circular' in the same limit
\cite[cf.][]{Marshall-et-al12}. If we consider instead $\alpha$ in terms of
$\phi_t$ and $\gamma_t$, then we see here that $\sin2\phi_t$ would not correlate
well with $\alpha$, and neither would $\gamma_t$ (not shown). So while there is
a flexibility for the form of $\alpha$ used, it would seem that, in the linear
regime at least, it is $\gamma_b$ that is more relevant, and the presence of the
slopes modifies the anisotropy of the state.

The vertical eddy angle $\phi_t$ has the behavior that $\phi_t \searrow 0$ as
$k$ is increased for fixed $\delta$ (Fig.~\ref{fig:eigen_geometric_l00}$e$). One
interpretation of the tilt angle could be that it is related to $\Delta\epsilon$
of two interacting edge-waves \cite[cf.,][for the horizontal
case]{Tamarin-et-al16}, having an optimal phase shift leading to instability
($\phi_t = \pi/4$ might be expected to be analogous to the optimum phase-shift
of $\Delta\epsilon=\pi/2$; cf. \cite{Tamarin-et-al16} for the case where the
velocity shear is purely in the horizontal). Indeed,
Fig.~\ref{fig:eigen_geometric_l00}$e,f$ showing $\phi_t$ over parameter space
resembles that of $\Delta\epsilon$ over parameter space, as shown in
Fig.~\ref{fig:amp_phase_from_edge_wave}$c,d$. In that sense, even though
$\phi_t$ is defined in terms of $S \sim \overline{v'b'}$ (through the definition
of $\gamma_b$) and makes no reference to edge-wave structures whatsoever, there
are apparent mechanistic links of $\phi_t$ with $\Delta\epsilon$. In that
regard, $\phi_t$ could perhaps serve as a possible proxy for edge-wave
phase-shifts that is easier to diagnose in cases where the definition of
edge-wave structures becomes more ambiguous (e.g., dynamics in a nonlinear
setting). Again, the phase-tilt $\Delta\epsilon_{\rm eigen}$
(Fig.~\ref{fig:amp_phase_from_Phi}$c$) bears little resemblance to $\phi_t$
(Fig.~\ref{fig:eigen_geometric_l00}$e$) nor $\Delta\epsilon$
(Fig.~\ref{fig:amp_phase_from_edge_wave}$e$) over parameter space.


The diagnosed results above are presented as vertically averaged quantities,
which does not demonstrate the asymmetry introduced with $\delta$. Examination
of the full vertical profiles does in fact show the quantities to be
increasingly concentrated towards the lower boundary when $\delta <0$ (and
vice-versa), consistent with the known behavior of $\gamma$ in
\eqref{eq:amp_ratio} and Fig.~\ref{fig:amp_phase_from_edge_wave}$a$, the lower
edge-wave being the increasingly dominant contribution (not shown). We close
this section with a note that the above results should be interpreted with the
caveat that the diagnoses are in the linear and nonlinear regime respectively,
and the two are not necessarily directly comparable (e.g., it is not obvious
that the instability has to leave a strong imprint on the nonlinear eddy fluxes
locally).


\section{Closing remarks}\label{sec:conclude}


\subsection{Conclusions}

The present work aims to clarify and point out some links between several
concepts in baroclinic instability, such as the underlying mathematical
symmetries of the governing system, a mechanistic interpretation of the shear
instability problem \cite[e.g.,][]{Bretherton66a, Hoskins-et-al85,
Heifetz-et-al04a, HeifetzGuha19, MengGuo23}, and geometric parameters of the
eddy variance ellipses \cite[e.g.,][]{Hoskins-et-al83, Marshall-et-al12,
Tamarin-et-al16}. We considered the Eady problem in the quasi-geostrophic system
as a working example, modified to include a weak linear bottom slope. Making an
assumption about the magnitude of slopes, the presence of the slope only affects
the dynamics via a bottom boundary contribution to potential vorticity (PV), and
the standard analysis leads to closed form solutions, as is known in the
literature \cite[e.g.,][]{BlumsackGierasch72, Mechoso80, Isachsen11, Brink12,
ChenKamenkovich13, Chen-et-al20}. The resulting modified Eady system was claimed
here to be parity-time ($\mathcal{PT}$) symmetric \cite[cf.][]{Qin-et-al19,
Zhang-et-al20, David-et-al22, MengGuo23}, with consequences for the the solution
spectrum; we refer the reader to Appendix \ref{app:A} for the full details and
other more speculative links.

To clarify aspects of the edge-wave mechanism and the modification by the
presence of a slope, we perform an edge-wave interaction analysis
\cite[cf.,][]{DaviesBishop94, Heifetz-et-al04a, HeifetzGuha19}. For the present
system, where edge-wave locations are well-defined, the edge-wave analysis is
really a rephrasing of the standard instability problem in a different choice of
basis, and the standard modal instability problem is effectively one of finding
fixed points of a two-dimensional dynamical system for the amplitude ratio and
edge-wave phase-shift. Physically, we expect that for $\delta < 0$, where
topographic slope and isentrope/isopycnals have horizontal gradients of opposite
signs, more vertical stretching is allowed (cf. Fig.~\ref{fig:setup}$b$), and
thus is adds to the background PV gradient at the bottom. The bottom wave is
then stronger, and its characteristic should be more apparent in the overall
normal-modes. For $\delta<0$, $c_r^\pm > 0 = c_r^{\pm}(\delta=0)$ with our
choice of basic state set up, since the bottom wave counter-propagates to the
right (or east). The opposite is true for $\delta > 0$. The asymmetry is
supported by the resulting edge-wave analysis, where the amplitude ratios are
simply functions of $\delta$ but not of the wavenumber
(Fig.~\ref{fig:amp_phase_from_edge_wave}$a$); the dependence on $\delta$ only
should be expected from the PV point of view, but such a dependence is not in
fact seen in the analysis of the tilted streamfunction eigenfunction
(Fig.~\ref{fig:amp_phase_from_Phi}$a$).

The phase-shift of the edge-waves $\Delta\epsilon$ associated with the most
unstable mode is not necessarily at the theoretical optimum of $\pi/2$ (actually
slightly larger), and generically differs from the phase-\emph{tilt} in the
unstable streamfunction $\Delta\epsilon_{\rm eigen}$, which does seem to be at
$\pi/2$ for $\delta\leq 0$. We argue that $\Delta\epsilon$ rather than
$\Delta\epsilon_{\rm eigen}$ should be the quantity of interest if a
kinematic/mechanistic interpretation is to be invoked. The unstable
normal-mode is a linear combination of the edge-wave structures (e.g.
Fig.~\ref{fig:edge_vs_eigen_shift_l00}), and the phase-shifts in the untilted
edge-waves structures do not have to correspond to the phase-tilts in the tilted
streamfunction eigenfunctions. We are of the opinion that references to
$\Delta\epsilon_{\rm eigen}$ should generally be avoided (an exception perhaps
for the Phillips problem \cite{Phillips56}, where the entries of the
normal-modes are defined as per-layer quantities and could be argued to
already be in edge-wave form). If some reference is to be made to the energetics
of the instability, the geometric parameters associated with eddy variance
ellipses \cite[e.g.,][]{Hoskins-et-al85, Marshall-et-al12} such as buoyancy
anisotropy $\gamma_b$ serve as better measures of the correlation for the
meridional velocity and buoyancy perturbations $v'$ and $b'$. On the other hand,
the vertical eddy tilt $\phi_t$, although defined with no reference to the
edge-wave structures themselves, display characteristics of the edge-wave
phase-shift $\Delta\epsilon$. The realized eddy efficiency parameter $\alpha$ is
shown to correlate strongly with the growth rate, in line with the definition
given in \cite{Marshall-et-al12}. It is found here that the dominant
contribution to $\alpha$ comes from $\gamma_b$, and so decreases in $\alpha$
with variations in $\delta$ arises the decrease in $\gamma_b$, consistent
somewhat with previous diagnoses from a nonlinear realization of baroclinic
dynamics in a numerical ocean model \cite{Stewart-et-al15}.

Previous works have argued that the changes in instability characteristics
arises from changes in the edge-wave interaction, but only considers the need
for phase-speed matching \cite{Chen-et-al20}. We argue that view is incomplete
as the argument does not extend to all of parameter space, and does not explain
all the instability characteristics. Here we clarify the mechanistic
explanation. For $\delta<0$, the bottom wave increases in strength, and not only
does it intrinsically propagate faster (to the right or east), but leads to a
larger action-at-a-distance. The top wave experiences a larger \emph{hindering},
which can be compensated by reducing the vertical interaction function via
moving to larger wavenumbers (which also reduces the intrinsic propagation speed
of both edge-waves), and the weaker interaction function leads to weaker growth
rates. It seems that it is always possible for such a compensation to be
achieved in the $\delta<0$ case, albeit over a decreasing bandwidth. For
$\delta>0$, the opposite is true, except the reduction in interaction from the
weakened bottom wave cannot be arbitrarily increased by moving to smaller
wavenumbers, since that increases the intrinsic propagation speed of both edge
waves. In this case the top edge-wave will become too fast for the given
background flow and wave interaction for phase-locking to be possible. The
instability bandwidth decreases in width and bounded away from the small
wavenumber as $\delta$ is increased towards 1. At $\delta\geq1$ there is no
instability whatsoever, as counter-propagation of the bottom wave is no longer
possible since the PV gradient vanishes or reverses sign.

\begin{figure}
  \includegraphics[width=\textwidth]{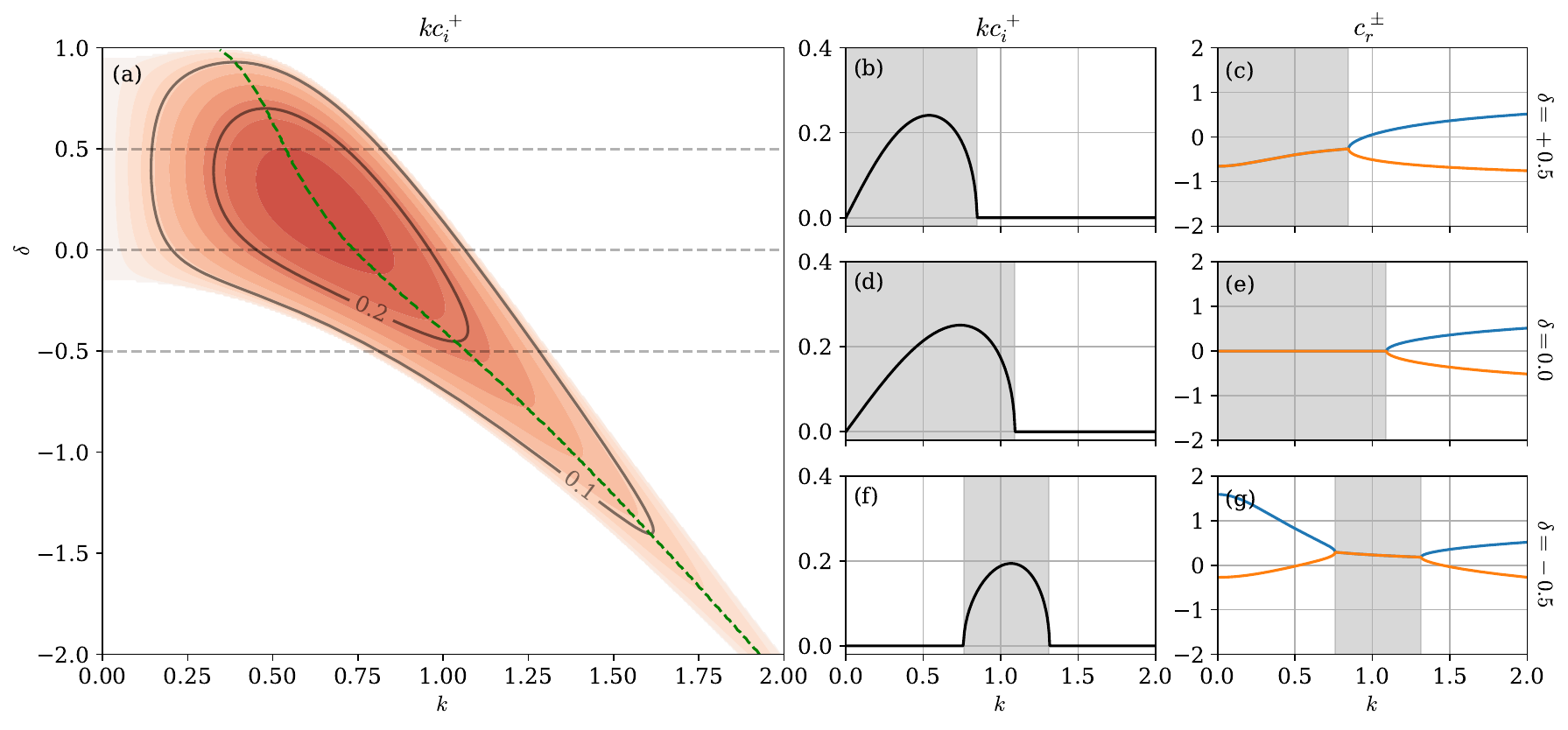}
  \caption{Sample instability characteristics for $l=0.5$, with the non-trivial
  meridional structure function $g(y)= \sin(y/2)$. ($a$) Growth rate as a
  function of the non-dimensional wavenumber $k$ and $\delta$ parameter, with
  darker shadings denoting higher values (strength of shading is the same used
  in Fig.~\ref{fig:growth_rates_cr_l00}), sample contours of growth rates, and
  the green dashed line denoting $k_{\max}(\delta)$ where the growth rate is
  maximized. Also shown are the growth rates $kc^+_i$ and phase-speed $c_r^\pm$
  for ($b,c$) $\delta = +0.5$, ($d,e$) the standard Eady problem $\delta = 0.0$,
  and ($f,g$) $\delta = -0.5$. The shaded regions in panels $b$-$g$ denote the
  regions where there the growth rates are non-zero.}
  \label{fig:growth_rates_cr_l05}
\end{figure}

Although the present work focuses largely on the zero meridional wavenumber case
$l=0$, the results and observations apply to the $l\neq0$ case. As a
demonstration, the instability characteristics of $l=0.5$ with the meridional
structure function $g(y) = \sin(ly)$ is shown in
Fig.~\ref{fig:growth_rates_cr_l05}. Here the instabilities generally possess
weaker growth rates, with a change in the domain of instability in the
$\delta>0$ region, but otherwise the general behavior of the growth rates and
phase-speed $c_r^\pm$ are similar to what has been reported (since the system is
still $\mathcal{PT}$ symmetric). The physical rationalization in terms of edge
waves is still applicable. The reduction in growth rates can be attributed to
weakened interaction from $l\neq0$, since that increases the value of $\mu =
\mbox{Bu}\sqrt{k^2 + l^2}$. The expanded domain of instability for $\delta > 0$
can be rationalized as edge-waves having smaller intrinsic propagation speed in
general (because of the increased $\mu$), and so the top edge-wave can be held
in a phase-locking configuration for the given basic flow, and the weakened
interaction associated with weaker bottom edge-wave can in fact be compensated
by increasing the interaction via decreasing the zonal wavenumber $k$. Edge-wave
amplitude ratios $\gamma$ is as given by \eqref{eq:amp_ratio}, and the edge-wave
phase-shift $\Delta\epsilon$ has similar behavior to that observed in
Fig.~\ref{fig:amp_phase_from_edge_wave}$c$ (not shown). While the computation of
the geometric parameters is slightly more complex as the relevant geometric
variables are now two-dimensional, the conclusions and interpretations are
essentially the same, except that the dominant contribution to $\alpha$ now
comes from $\gamma_b\sin\phi_b$ (since $R$ is no longer trivial and so $\phi_b
\neq \pm \pi/2$), but the energy partition contribution $\sin2\lambda$ is still
close to 1 and largely constant over the domain. In particular, the vertical
eddy tilt $\phi_t$ follows the phase-shift of the edge-waves more so than the
normal-mode phase-tilts (in terms of spatial distribution, average values down
the center line, or domain-averaged values; not shown).

\subsection{Discussions and outlooks}

The present work explores the behavior of the $\alpha$ parameter from a linear
instability point of view, to supplement parameterization efforts of particular
relevance to ocean modeling. We find that there is indeed a suppression of
$\alpha$ in the presence of slopes, and the dominant contribution comes
principally from changes in the buoyancy anisotropy parameter (as well as the
eddy buoyancy angle where it is present). The result here would be consistent
with previous observations that buoyancy fluxes are suppressed over sloped
regions \cite[e.g.,][]{Isachsen11, Brink12, Isachsen15, Manucharyan-et-al17,
Hetland17, TrodahlIsachsen18, WangStewart18, ManucharyanIsachsen19,
Chen-et-al20, WeiWang21}, with previous diagnostic results but in the nonlinear
regime in a realistic global ocean model \cite{Poulsen-et-al19}, and is
consistent with a parameterization that $\alpha$ be suppressed over slopes in a
way that is dependent on the slope Burger number \cite{Wei-et-al22}, related to
the $\delta$ utilized in the present work. This work thus provides a consistency
rationalization for the proposal of \cite{Wei-et-al22}, which was empirical in
nature. While the present work is for the linear regime, previous works have
suggested that the linear instability characteristics can be useful in informing
parameterizations that are invoked for the nonlinear regime
\cite[e.g.,][]{Green70, Stone72, Killworth97, Killworth98, Eden11}.

\begin{figure}
  \includegraphics[width=\textwidth]{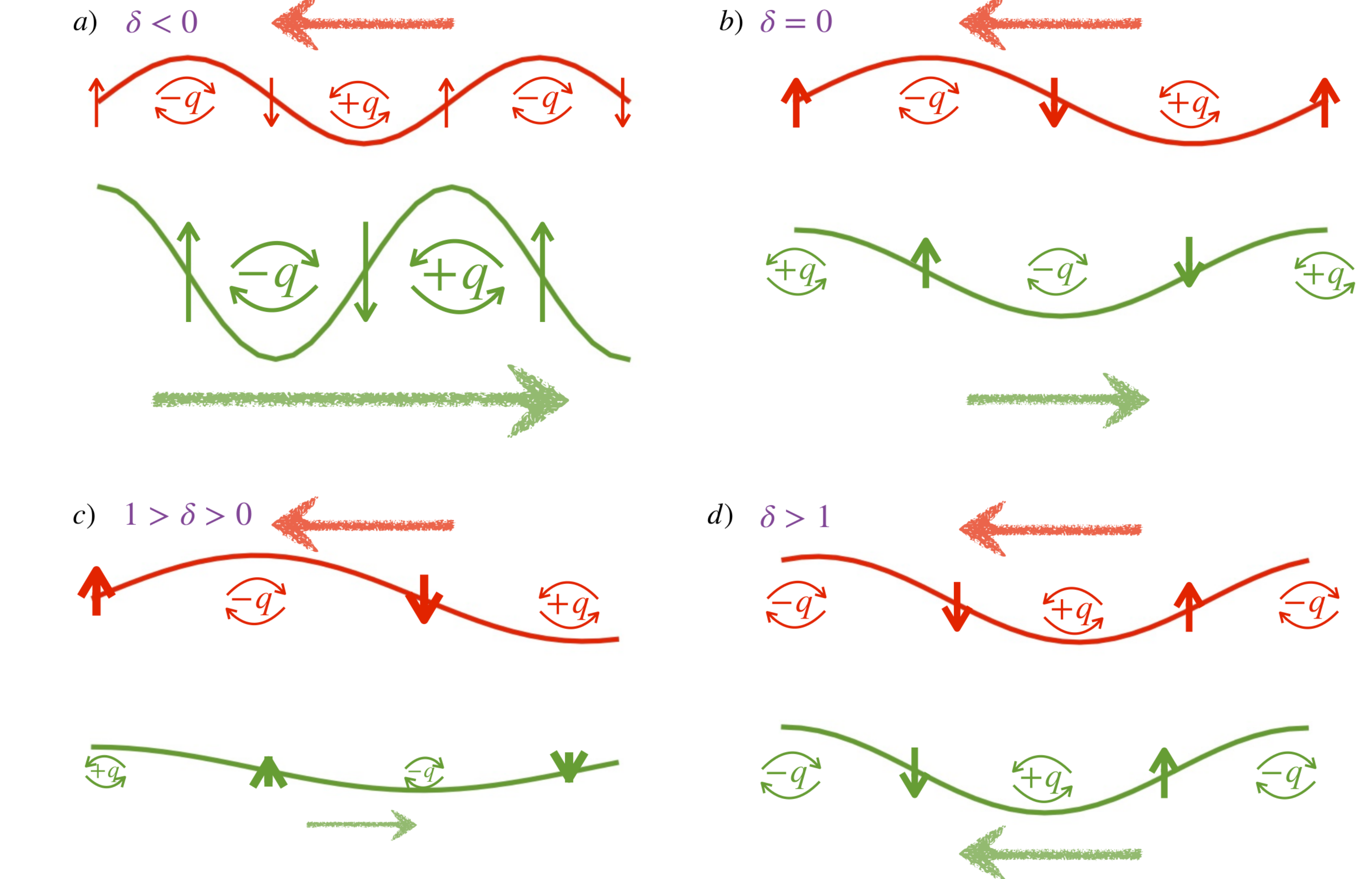}
  \caption{Pictorial schematic for the change in instability characteristics
  over parameter space in terms of interacting Rossby edge-waves. ($a$) For the
  $\delta<0$ case, the bottom PV gradient is stronger, leader to a stronger
  bottom edge-wave that strengthens the magnitude of interaction and edge-wave
  propagation speed (represented by longer arrows), but this effect can be
  compensated by going to shorter waves (represented here by the thinner
  arrows). ($b$) The standard Eady case $\delta = 0$, where the interaction is
  symmetric. ($c$) For $1> \delta > 0$, the bottom PV gradient weakens, leading
  to a weaker bottom edge-wave and a weaker interaction (represented by the
  shorter arrows), but this can be offset somewhat by going to longer waves
  (represented by the thicker arrows). ($d$) For $\delta > 1$, the PV gradient
  switches sign and the bottom edge-wave no longer counter-propagates against
  the background mean flow, no phase-locking configuration is possible, and
  there is no instability.}
  \label{fig:schematic2}
\end{figure}

The stabilization of baroclinic instability in the presence of small slopes
(related to the suppression of $\alpha$) is rationalized in the PV point of
view, where the presence of a slope modifies the background PV gradient and
leads to modifications in the interaction of the Rossby edge-waves. The
rationalization is given in Sec.~\ref{sec:rationalise}, but is also summarized
in the schematic given in Fig.~\ref{fig:schematic2}. The point we clarify here
is that it is the asymmetry in the wave amplitudes that lead to changes in the
mutual interaction, that in turn modifies the phase-locking configuration, and
the end result dictates a change in the phase-shift. Arguments based solely on
phase-shifts and phase-speed matching is missing a key link in the interaction,
and does not fully explain the instability characteristics over the whole
parameter space. Further, we highlight that, from a mechanistic point of view,
it is the untilted edge-wave basis and its phase-shift that is of relevance, and
the phase-tilt in the tilted normal-modes is generally to be avoided as it can
be misleading. The physical rationalization applies generically to baroclinic
instability over slopes; sample analysis on the analogous Phillips problem
\cite[cf.,][]{ChenKamenkovich13} shows similar results and interpretations to
here (not shown). It would further be interesting to see how some of the
analysis and points of view (edge-waves and GEOMETRIC framework) here carry over
to the case of transient / optimal growth \cite[e.g.,][]{HeifetzMethven05}, but
we leave this for a follow up study.

It was found that the vertical tilt angle $\phi_t$ (rather than the buoyancy
angle $\phi_b$) possesses qualitative similarities to the edge-wave phase
shifts, even though the former is defined with no reference to the edge-waves
whatsoever. The observation is in line with the results in
\cite{Tamarin-et-al16}, who find that in the barotropic instability of the
Rayleigh sheet that there is a relation between the horizontal tilt angles
$\phi_m$ (the momentum flux angles) and the edge-wave phase-shift. Then perhaps
a parameterization of baroclinic instability could be based on the geometric
parameters such as $\phi_t$, which informs the spatial structure of the eddy
fluxes as well as an efficiency via $\alpha$, with a realized magnitude based on
a parameterized eddy energy \cite[cf.][]{Marshall-et-al12, Mak-et-al22}. For
diagnostic purposes, in cases where the edge-waves are not so well-defined
(e.g., nonlinear cases, although see next paragraph), diagnoses of $\phi_t$
could be done instead as a proxy for the phase-shift, since the geometric
parameters considered here applies generically to cases where eddy fluxes can be
diagnosed \cite{Marshall-et-al12}.

For the present linear problem the edge-waves are well-defined, and
Sec.~\ref{sec:edgewave} provides a manual construction of the edge-wave basis
(e.g., where one part of the edge-wave couplet has an untilted structure and no
PV signal at the location of the other edge-wave). There are problems where edge
wave structures are harder to define (e.g., instability in the presence of
planetary $\beta$ \cite{Charney47, Green60, Tamarin-et-al16}, the Rayleigh sheet
problem in the presence of a magnetic field \cite{Heifetz-et-al15}), or cases
where one might naturally expect linear theory to play no longer play a dominant
role (e.g., during a nonlinear evolution). However, there are still ways to
define an edge-wave basis, for example making use of orthogonality in the wave
activity quantities such as pseudomomentum and pseudoenergy
\cite[e.g.,][]{Held85, Shepherd90, Heifetz-et-al04a, HeifetzGuha19}. It is
possible to extend the present analysis to some of the other models, such as the
Charney--Green model, which has been pointed to possibly be more relevant than
the Eady model in certain oceanic settings \cite[e.g.,][]{Isachsen11,
Isachsen15, TrodahlIsachsen18, ManucharyanIsachsen19}, problems with smooth
profiles and/or profiles containing multiple PV gradients (e.g. jet profiles),
and mixed barotropic-baroclinic problems \cite[e.g.,][]{James87, Nakamura93a}.
In particular, if one believes that there is some sort of quasi-linear control
on the linear instability on the nonlinear evolution, then perhaps the procedure
will be illuminating for an analogous analysis for simulations of the nonlinear
phase of the shear instabilities \cite[e.g.,][]{Harnik-et-al14}. These are left
as possible future investigations to be pursued.





%
%


%


\begin{acknowledgments}


This work was supported by the the RGC General Research Fund 11308021 and the
Center for Ocean Research in Hong Kong and Macau, a joint research center
between the Laoshan Laboratory and Hong Kong University of Science and
Technology. The Python Jupyter notebook used to generate and analyze data for
the problem may be found at
\url{https://github.com/julianmak/julianmak.github.io/blob/master/files/Eady/Eady_analysis.ipynb}.
GK was supported by a post-doctoral funding through the Center for Ocean
Research in Hong Kong and Macau. Some of the work was initiated while EQYO was a
visiting student at the Hong Kong University of Science and Technology. EQYO is
supported by the Australian Government Research Training Program 678 Scholarship
(RTP) and the Australian Research Council Special 676 Research Initiative,
Australian Centre for Excellence in Antarctic Science (ARC project number 677
SR200100008). JM would like to thank Liiyung Yeow and the hospitality of the
staff at the Edison Hotel at Penang, Malaysia that lead to the discussion of
$\mathcal{PT}$-symmetry in classical systems and the writing of the first draft
of the present article.

Authorship is alphabetical after the first author. Resources, Supervision,
Project administration, Funding acquisition: JM. Conceptualization,
Visualization: JM, NH, EH. Methodology: JM, EH, GK. Software: JM, GK. Formal
analysis: JM, NH, EH, GK. Writing - Original Draft: JM, NH, EH. Validation,
Writing - Review \& Editing: everyone.


\end{acknowledgments}


\appendix

\section{Parity-Time $\mathcal{PT}$ symmetry of the modified Eady problem}\label{app:A}

Note that in the text we made the observation that Eq.~\eqref{eq:lin-nondim} is
invariant under the transformation
\begin{equation}
  \mathcal{P}: (x,y,z) \mapsto (-x,-y,-z), \qquad \mathcal{T}: (t, \psi) \mapsto (-t, -\psi).
\end{equation}
More formally, an operator $\mathcal{H}$ governing a system is Parity-Time
($\mathcal{PT}$) symmetric if it satisfies
\begin{equation}
  (\mathcal{PT})\mathcal{H}^*(\mathcal{PT})^{-1} = \mathcal{H},
\end{equation}
where $\mathcal{H}^*$ denotes the complex conjugate (rather than the Hermitian
conjugate $\mathcal{H}^\dagger$, which for a matrix representation involves a
transpose on top of the element-wise complex conjugate).

$\mathcal{PT}$ symmetry is a concept that originates from quantum physics
\cite[e.g.,][]{Bender-PT}, but has found recent interest in classical fluid
systems also \cite[e.g.,][]{Qin-et-al19, Zhang-et-al20, David-et-al22}. Here we
show explicitly that the present modified Eady problem is $\mathcal{PT}$
symmetric, which leads to certain features in the solution spectrum that was
previously highlighted in Sec.~\ref{sec:eady}. The ideas and tools were previous
given in the work of \cite{David-et-al22} for the Phillips-like problem (two
layer QG with uniform flow in each layer, cf. \cite{Phillips56, Vallis-GFD}).
The following exposition is largely given for completeness, but also serves to
highlight very suggestive links between $\mathcal{PT}$ symmetry, shear
instability and the edge-wave formalism, possibly enabling tools to be borrowed
from quantum physics to further our understanding of classical fluid systems
(e.g. nonlinear shear instability), or providing classical analogues with
well-understood physics to complement the mathematical analysis of quantum
systems.

Note that we can write the linearized system \eqref{eq:lin-nondim} in the form
\begin{equation}\label{eq:eig1}
  \frac{\partial}{\partial t} \mathcal{L} \varphi = \mathcal{M} \varphi,
\end{equation}
which is a generalized eigenvalue problem for the relevant operators
$\mathcal{L}$ and $\mathcal{M}$ acting on an eigenfunction $\varphi$. If we are
taking modal solutions as in \eqref{eq:eigenfunc}, then $\mathcal{L}$ has an
explicit representation whose inverse that can in principle be computed for, and
we can define $M = \mathcal{L}^{-1}\mathcal{M}$ where
\begin{equation}\label{eq:eigenvalue-eq}
  c\phi = M\phi,
\end{equation}
with the eigenvector $\phi = (a,b)$ in this case. For the system here, it can be
shown that (using again the notation $C = \cosh \mu$ and $S = \sinh \mu$)
\begin{equation}\label{eq:M}
  L = \mu\begin{pmatrix}-S & C \\ S & C\end{pmatrix}, \quad L^{-1} = \frac{1}{2\mu CS} \begin{pmatrix}-C & C \\ S & S\end{pmatrix}, \quad M = \frac{-1}{SC}\begin{pmatrix} \cfrac{\delta}{2\mu} C^2 & \left(1-\cfrac{\delta}{2}\right)\cfrac{CS}{\mu} - C^2 \\ \left(1-\cfrac{\delta}{2}\right)\cfrac{CS}{\mu} - S^2 & \cfrac{\delta}{2\mu} S^2 \end{pmatrix},
\end{equation}
where $L$ denotes the representation of $\mathcal{L}$ when modal solutions
\eqref{eq:eigenfunc} are taken.

In the present work, $M$ is already real, $\mathcal{PT}$ happens to be the
negative identity in the matrix representation relevant for the present system
\cite{David-et-al22}, so $M$ is $\mathcal{PT}$-symmetric. Another way to see
that $M$ is $\mathcal{PT}$-symmetric is to note that, for $P$ and $T$ denoting
the matrix representations of $\mathcal{P}$ and $\mathcal{T}$, $\mu$ is
invariant under $P : (k,l) \mapsto (-k, -l)$ (interpreting $\mu = \mbox{Bu}|k|$
if $l=0$), while $T$ introduces a minus sign through its action on the
eigenvector $\phi$, but this is done twice, so since $M$ is real, the above
equality holds.

A particular consequence for $M$ being $\mathcal{PT}$ symmetric is that the
entries $M$ are all real (which we have manually demonstrated here in
\eqref{eq:M}, but is in fact a general feature). Further, the dispersion
relation \eqref{eq:eigenvalue-eq}, for a two component system, satisfies
\begin{equation}\label{eq:c}
  c^2 - \mbox{Tr}(M) + \mbox{Det}(M) = 0, \qquad c = \frac{1}{2}\left(\mbox{Tr}(M) \pm \sqrt{\mbox{Tr}(M)^2 - 4\mbox{Det}(M)}\right),
\end{equation}
where the trace and determinant of $M$ are given by
\begin{equation*}
  M = \begin{pmatrix}a_1 & a_2 \\ a_3 & a_4\end{pmatrix}, \qquad \mbox{Tr}(M) = a_1 + a_3, \qquad \mbox{Det}(M) = a_1 a_4 - a_2 a_3.
\end{equation*}
The entries of $M$ are given in \eqref{eq:M}, and the resulting dispersion
relation and be shown to coincide exactly with that given \eqref{eq:c-equ}
obtained from standard means. Further, the eigenfunction of the system (defined
up to some arbitrary constant) can be written as
\begin{equation}\label{eq:eig2}
  \phi = \begin{pmatrix}a \\ b \end{pmatrix} = \begin{pmatrix}a_2 \\ -a_1 + \cfrac{1}{2}\left(\mbox{Tr}(M) \pm \sqrt{\mbox{Tr}(M)^2 - 4\mbox{Det}(M)}\right) \end{pmatrix},
\end{equation}
and the eigenfunction $\phi$ is $\mathcal{PT}$-symmetric if $\phi$ is real.
However, for the eigenvector of an unstable mode in the present system,
$\mbox{Tr}(M)^2 - 4\mbox{Det}(M) < 0$, and the resulting eigenvector becomes
complex and ceases being an eigenfunction of the $\mathcal{PT}$ operator
(because there is an extra minus introduced under a complex conjugate of the
eigenvector), and we have what is termed \emph{spontaneous breaking of
$\mathcal{PT}$-symmetry}. The boundary between the region with and without
spontaneous $\mathcal{PT}$-symmetry breaking are called \emph{exceptional
points}, and these correspond precisely to the locations of marginal stability.
The collision of eigenvalues $c^\pm$ on the real axis into complex conjugates is
related to what are known as \emph{Krein collisions} \cite[e.g.,][]{Bender-PT,
Zhang-et-al20}.

It is perhaps easy to see that, using the same framework of
\cite{David-et-al22}, other shear flow instability problems should also be
$\mathcal{PT}$ symmetric, such as Phillips-like problems
\cite[cf.,][]{ChenKamenkovich13, David-et-al22}, Charney--Green-like problems
\cite[cf.][]{Charney47, Green60, Vallis-GFD}, the standard Rayleigh sheet
problem \cite{MengGuo23} in hydrodynamics, the Rayleigh sheet problem in
magnetohydrodynamics with a uniform background magnetic field
\cite[e.g.,][]{Heifetz-et-al15}, and more general shear flow problems
presumably. Additionally, given the link between shear instabilities and its
physical interpretation as a pair of interacting edge-waves, one is left to
wonder on the exact links between $\mathcal{PT}$ symmetry and interacting edge
waves. Suggestive links include that spontaneous breaking of $\mathcal{PT}$
symmetry seems to correspond exactly to when phase-locking occurs
\cite[cf.,][]{David-et-al22}. A Krein collision of the eigenvalue occurs at
exception points requires opposite signed Krein signatures, reminiscent of the
requirement that shear instabilities require modes of opposite signed wave
activity to collide (as positive/negative energy modes \cite{Cairns79} or
pseudomomentum \cite[e.g.,][]{Held85}), which is sometimes interpreted as a
necessary (but not sufficient) condition for instability is the need for
counter-propagation of edge-waves \cite[e.g.,][]{Heifetz-et-al04a}. 

While an explicit link to the edge-wave system with $\mathcal{PT}$ symmetry has
been highlighted in the recent work of \cite{MengGuo23}, and low-dimensional
edge-wave systems (low-dimensional in the sense of dynamical systems) could be
considered as a rephrasing of certain shear instability problem
\cite[e.g.,][]{Heifetz-et-al04a}, the links with the general shear instability
problem for general shear flows remain to be explored (since generic shear flow
instability problems should be regarded an infinite dimensional dynamical
system). For example, what is the analogue of the Krein signature of the modes
of the governing operator in the more standard fluid dynamics context? (Likely
something related to pseudomomentum or pseudoenergy?) How are the standard shear
flow linear stability conditions related to the properties of the governing
operator, and in terms of interacting edge-waves? (Related to proofs of a purely
real spectrum in $\mathcal{PT}$ symmetric systems \cite[e.g.,][]{Dorey-et-al01,
Mostafazadeh02}?) Are there conditions beyond the standard necessary but not
sufficient conditions of shear instability derivable from the $\mathcal{PT}$
symmetry property, using techniques drawn from quantum physics? Most references
of $\mathcal{PT}$ symmetry in relation to shear instability systems so far is on
the linear problem, but are there nonlinear analogues of the stability
conditions such as the Arnol'd conditions \cite[e.g.,][]{Shepherd90} derivable
from a similar approach? Such links could in principle provide a
mechanistic/physical interpretation to a theoretical properties related to
$\mathcal{PT}$ symmetry, and techniques in relation to analyzing $\mathcal{PT}$
symmetric systems from quantum physics could provide new approaches for
improving our understanding of classical fluid/plasma systems. The role of
$\mathcal{PT}$ symmetry for fluid/plasma systems remains to be fully explored,
and research in this direction is a subject of future work.





\begin{thebibliography}{89}%
\makeatletter
\providecommand \@ifxundefined [1]{%
 \@ifx{#1\undefined}
}%
\providecommand \@ifnum [1]{%
 \ifnum #1\expandafter \@firstoftwo
 \else \expandafter \@secondoftwo
 \fi
}%
\providecommand \@ifx [1]{%
 \ifx #1\expandafter \@firstoftwo
 \else \expandafter \@secondoftwo
 \fi
}%
\providecommand \natexlab [1]{#1}%
\providecommand \enquote  [1]{``#1''}%
\providecommand \bibnamefont  [1]{#1}%
\providecommand \bibfnamefont [1]{#1}%
\providecommand \citenamefont [1]{#1}%
\providecommand \href@noop [0]{\@secondoftwo}%
\providecommand \href [0]{\begingroup \@sanitize@url \@href}%
\providecommand \@href[1]{\@@startlink{#1}\@@href}%
\providecommand \@@href[1]{\endgroup#1\@@endlink}%
\providecommand \@sanitize@url [0]{\catcode `\\12\catcode `\$12\catcode
  `\&12\catcode `\#12\catcode `\^12\catcode `\_12\catcode `\%12\relax}%
\providecommand \@@startlink[1]{}%
\providecommand \@@endlink[0]{}%
\providecommand \url  [0]{\begingroup\@sanitize@url \@url }%
\providecommand \@url [1]{\endgroup\@href {#1}{\urlprefix }}%
\providecommand \urlprefix  [0]{URL }%
\providecommand \Eprint [0]{\href }%
\providecommand \doibase [0]{https://doi.org/}%
\providecommand \selectlanguage [0]{\@gobble}%
\providecommand \bibinfo  [0]{\@secondoftwo}%
\providecommand \bibfield  [0]{\@secondoftwo}%
\providecommand \translation [1]{[#1]}%
\providecommand \BibitemOpen [0]{}%
\providecommand \bibitemStop [0]{}%
\providecommand \bibitemNoStop [0]{.\EOS\space}%
\providecommand \EOS [0]{\spacefactor3000\relax}%
\providecommand \BibitemShut  [1]{\csname bibitem#1\endcsname}%
\let\auto@bib@innerbib\@empty
\bibitem [{\citenamefont {Vallis}(2006)}]{Vallis-GFD}%
  \BibitemOpen
  \bibfield  {author} {\bibinfo {author} {\bibfnamefont {G.~K.}\ \bibnamefont
  {Vallis}},\ }\href@noop {} {\emph {\bibinfo {title} {{Atmospheric and Oceanic
  Fluid Dynamics}}}}\ (\bibinfo  {publisher} {Cambridge University Press},\
  \bibinfo {year} {2006})\BibitemShut {NoStop}%
\bibitem [{\citenamefont {Lovelace}\ \emph {et~al.}(1999)\citenamefont
  {Lovelace}, \citenamefont {Li}, \citenamefont {Colgate},\ and\ \citenamefont
  {Nelson}}]{Lovelace-et-al99}%
  \BibitemOpen
  \bibfield  {author} {\bibinfo {author} {\bibfnamefont {R.~V.~E.}\
  \bibnamefont {Lovelace}}, \bibinfo {author} {\bibfnamefont {H.}~\bibnamefont
  {Li}}, \bibinfo {author} {\bibfnamefont {S.~A.}\ \bibnamefont {Colgate}},\
  and\ \bibinfo {author} {\bibfnamefont {A.~F.}\ \bibnamefont {Nelson}},\
  }\bibfield  {title} {\bibinfo {title} {Rossby wave instability of {K}eplerian
  accretion disks},\ }\href@noop {} {\bibfield  {journal} {\bibinfo  {journal}
  {Astrophys. J.}\ }\textbf {\bibinfo {volume} {513}},\ \bibinfo {pages} {805}
  (\bibinfo {year} {1999})}\BibitemShut {NoStop}%
\bibitem [{\citenamefont {Kaspi}\ and\ \citenamefont
  {Flierl}(2007)}]{KaspiFlierl07}%
  \BibitemOpen
  \bibfield  {author} {\bibinfo {author} {\bibfnamefont {Y.}~\bibnamefont
  {Kaspi}}\ and\ \bibinfo {author} {\bibfnamefont {G.~R.}\ \bibnamefont
  {Flierl}},\ }\bibfield  {title} {\bibinfo {title} {{Formation of jets by
  baroclinic instability on gas planet atmospheres}},\ }\href
  {https://doi.org/10.1175/JAS4009.1} {\bibfield  {journal} {\bibinfo
  {journal} {J. Atmos. Sci.}\ }\textbf {\bibinfo {volume} {64}},\ \bibinfo
  {pages} {3177} (\bibinfo {year} {2007})}\BibitemShut {NoStop}%
\bibitem [{\citenamefont {Hughes}\ \emph {et~al.}(2007)\citenamefont {Hughes},
  \citenamefont {Rosner},\ and\ \citenamefont
  {Weiss}}]{Hughes-et-al-Tachocline}%
  \BibitemOpen
  \bibfield  {author} {\bibinfo {author} {\bibfnamefont {D.~W.}\ \bibnamefont
  {Hughes}}, \bibinfo {author} {\bibfnamefont {R.}~\bibnamefont {Rosner}},\
  and\ \bibinfo {author} {\bibfnamefont {N.~O.}\ \bibnamefont {Weiss}},\
  }\href@noop {} {\emph {\bibinfo {title} {The solar tachocline}}}\ (\bibinfo
  {publisher} {Cambridge {U}niversity {P}ress},\ \bibinfo {year}
  {2007})\BibitemShut {NoStop}%
\bibitem [{\citenamefont {Teed}\ \emph {et~al.}(2010)\citenamefont {Teed},
  \citenamefont {Jones},\ and\ \citenamefont {Hollerbach}}]{Teed-et-al10}%
  \BibitemOpen
  \bibfield  {author} {\bibinfo {author} {\bibfnamefont {R.~J.}\ \bibnamefont
  {Teed}}, \bibinfo {author} {\bibfnamefont {C.~A.}\ \bibnamefont {Jones}},\
  and\ \bibinfo {author} {\bibfnamefont {R.}~\bibnamefont {Hollerbach}},\
  }\bibfield  {title} {\bibinfo {title} {{Rapidly rotating plane layer
  convection with zonal flow}},\ }\href
  {https://doi.org/10.1080/03091929.2010.512290} {\bibfield  {journal}
  {\bibinfo  {journal} {Geophys. Astrophys. Fluid Dyn.}\ }\textbf {\bibinfo
  {volume} {104}},\ \bibinfo {pages} {457} (\bibinfo {year}
  {2010})}\BibitemShut {NoStop}%
\bibitem [{\citenamefont {Polichtchouk}\ and\ \citenamefont
  {Cho}(2012)}]{PolichtchoukCho12}%
  \BibitemOpen
  \bibfield  {author} {\bibinfo {author} {\bibfnamefont {I.}~\bibnamefont
  {Polichtchouk}}\ and\ \bibinfo {author} {\bibfnamefont {J.~Y.}\ \bibnamefont
  {Cho}},\ }\bibfield  {title} {\bibinfo {title} {{Baroclinic instability on
  hot extrasolar planets}},\ }\href
  {https://doi.org/10.1111/j.1365-2966.2012.21312.x} {\bibfield  {journal}
  {\bibinfo  {journal} {Mon. Not. R. Astron. Soc.}\ }\textbf {\bibinfo {volume}
  {424}},\ \bibinfo {pages} {1307} (\bibinfo {year} {2012})}\BibitemShut
  {NoStop}%
\bibitem [{\citenamefont {Gilman}\ and\ \citenamefont
  {Dikpati}(1402)}]{GilmanDikpati14}%
  \BibitemOpen
  \bibfield  {author} {\bibinfo {author} {\bibfnamefont {P.~A.}\ \bibnamefont
  {Gilman}}\ and\ \bibinfo {author} {\bibfnamefont {M.}~\bibnamefont
  {Dikpati}},\ }\bibfield  {title} {\bibinfo {title} {{Baroclinic instability
  in the solar tachocline}},\ }\href
  {https://doi.org/10.1088/0004-637X/787/1/60} {\bibfield  {journal} {\bibinfo
  {journal} {Astrophys. J.}\ }\textbf {\bibinfo {volume} {787}},\ \bibinfo
  {pages} {60} (\bibinfo {year} {1402})}\BibitemShut {NoStop}%
\bibitem [{\citenamefont {Gilman}(2015)}]{Gilman15}%
  \BibitemOpen
  \bibfield  {author} {\bibinfo {author} {\bibfnamefont {P.~A.}\ \bibnamefont
  {Gilman}},\ }\bibfield  {title} {\bibinfo {title} {{Effect of toroidal fields
  on baroclinic instability in the solar tachocline}},\ }\href
  {https://doi.org/10.1088/0004-637X/801/1/22} {\bibfield  {journal} {\bibinfo
  {journal} {Astrophys. J.}\ }\textbf {\bibinfo {volume} {801}},\ \bibinfo
  {pages} {22} (\bibinfo {year} {2015})}\BibitemShut {NoStop}%
\bibitem [{\citenamefont {Read}\ \emph {et~al.}(2020)\citenamefont {Read},
  \citenamefont {Kennedy}, \citenamefont {Lewis}, \citenamefont {Scolan},
  \citenamefont {{Tabataba-Vakili}}, \citenamefont {Wang}, \citenamefont
  {Wright},\ and\ \citenamefont {Young}}]{Read-et-al20}%
  \BibitemOpen
  \bibfield  {author} {\bibinfo {author} {\bibfnamefont {P.}~\bibnamefont
  {Read}}, \bibinfo {author} {\bibfnamefont {D.}~\bibnamefont {Kennedy}},
  \bibinfo {author} {\bibfnamefont {N.}~\bibnamefont {Lewis}}, \bibinfo
  {author} {\bibfnamefont {H.}~\bibnamefont {Scolan}}, \bibinfo {author}
  {\bibfnamefont {F.}~\bibnamefont {{Tabataba-Vakili}}}, \bibinfo {author}
  {\bibfnamefont {Y.}~\bibnamefont {Wang}}, \bibinfo {author} {\bibfnamefont
  {S.}~\bibnamefont {Wright}},\ and\ \bibinfo {author} {\bibfnamefont
  {R.}~\bibnamefont {Young}},\ }\bibfield  {title} {\bibinfo {title}
  {{Baroclinic and barotropic instabilities in planetary atmospheres:
  energetics, equilibration and adjustment}},\ }\href
  {https://doi.org/10.5194/npg-27-147-2020} {\bibfield  {journal} {\bibinfo
  {journal} {Nonlin. Processes Geophys.}\ }\textbf {\bibinfo {volume} {27}},\
  \bibinfo {pages} {147} (\bibinfo {year} {2020})}\BibitemShut {NoStop}%
\bibitem [{\citenamefont {{Yellin-Bergovoy}}\ \emph {et~al.}(2021)\citenamefont
  {{Yellin-Bergovoy}}, \citenamefont {Umurhan},\ and\ \citenamefont
  {Heifetz}}]{YellinBergovoy-et-al21}%
  \BibitemOpen
  \bibfield  {author} {\bibinfo {author} {\bibfnamefont {R.}~\bibnamefont
  {{Yellin-Bergovoy}}}, \bibinfo {author} {\bibfnamefont {O.~M.}\ \bibnamefont
  {Umurhan}},\ and\ \bibinfo {author} {\bibfnamefont {E.}~\bibnamefont
  {Heifetz}},\ }\bibfield  {title} {\bibinfo {title} {{A minimal model for
  vertical shear instability in protoplanetary accretion disks}},\ }\href
  {https://doi.org/10.1080/03091929.2021.1941921} {\bibfield  {journal}
  {\bibinfo  {journal} {Geophys. Astrophys. Fluid Dyn.}\ }\textbf {\bibinfo
  {volume} {115}},\ \bibinfo {pages} {674} (\bibinfo {year}
  {2021})}\BibitemShut {NoStop}%
\bibitem [{\citenamefont {Charney}(1947)}]{Charney47}%
  \BibitemOpen
  \bibfield  {author} {\bibinfo {author} {\bibfnamefont {J.~G.}\ \bibnamefont
  {Charney}},\ }\bibfield  {title} {\bibinfo {title} {Dynamics of long waves in
  a baroclinic westerly current},\ }\href@noop {} {\bibfield  {journal}
  {\bibinfo  {journal} {J. Meteor.}\ }\textbf {\bibinfo {volume} {4}},\
  \bibinfo {pages} {135} (\bibinfo {year} {1947})}\BibitemShut {NoStop}%
\bibitem [{\citenamefont {Eady}(1949)}]{Eady49}%
  \BibitemOpen
  \bibfield  {author} {\bibinfo {author} {\bibfnamefont {E.~T.}\ \bibnamefont
  {Eady}},\ }\bibfield  {title} {\bibinfo {title} {{Long waves and cyclone
  waves}},\ }\href {https://doi.org/10.1111/j.2153-3490.1949.tb01265.x}
  {\bibfield  {journal} {\bibinfo  {journal} {Tellus}\ }\textbf {\bibinfo
  {volume} {1}},\ \bibinfo {pages} {33} (\bibinfo {year} {1949})}\BibitemShut
  {NoStop}%
\bibitem [{\citenamefont {Phillips}(1956)}]{Phillips56}%
  \BibitemOpen
  \bibfield  {author} {\bibinfo {author} {\bibfnamefont {N.~A.}\ \bibnamefont
  {Phillips}},\ }\bibfield  {title} {\bibinfo {title} {{The general circulation
  of the atmosphere: a numerical experiment}},\ }\href@noop {} {\bibfield
  {journal} {\bibinfo  {journal} {Q. J. Roy. Met. Soc.}\ }\textbf {\bibinfo
  {volume} {82}},\ \bibinfo {pages} {123} (\bibinfo {year} {1956})}\BibitemShut
  {NoStop}%
\bibitem [{\citenamefont {Green}(1960)}]{Green60}%
  \BibitemOpen
  \bibfield  {author} {\bibinfo {author} {\bibfnamefont {J.~S.~A.}\
  \bibnamefont {Green}},\ }\bibfield  {title} {\bibinfo {title} {{A problem in
  baroclinic instability}},\ }\href {https://doi.org/10.1002/qj.49708636813}
  {\bibfield  {journal} {\bibinfo  {journal} {Q. J. Roy. Met. Soc.}\ }\textbf
  {\bibinfo {volume} {86}},\ \bibinfo {pages} {237} (\bibinfo {year}
  {1960})}\BibitemShut {NoStop}%
\bibitem [{\citenamefont {Charney}\ and\ \citenamefont
  {Stern}(1962)}]{CharneyStern62}%
  \BibitemOpen
  \bibfield  {author} {\bibinfo {author} {\bibfnamefont {J.~G.}\ \bibnamefont
  {Charney}}\ and\ \bibinfo {author} {\bibfnamefont {M.~E.}\ \bibnamefont
  {Stern}},\ }\bibfield  {title} {\bibinfo {title} {On the stability of
  internal baroclinic jets in a rotating atmosphere},\ }\href@noop {}
  {\bibfield  {journal} {\bibinfo  {journal} {J. Atmos. Sci.}\ }\textbf
  {\bibinfo {volume} {19}},\ \bibinfo {pages} {159} (\bibinfo {year}
  {1962})}\BibitemShut {NoStop}%
\bibitem [{\citenamefont {Pedlosky}(1964{\natexlab{a}})}]{Pedlosky64a}%
  \BibitemOpen
  \bibfield  {author} {\bibinfo {author} {\bibfnamefont {J.}~\bibnamefont
  {Pedlosky}},\ }\bibfield  {title} {\bibinfo {title} {The stability of
  currents in the atmosphere and the ocean: {P}art {I}},\ }\href@noop {}
  {\bibfield  {journal} {\bibinfo  {journal} {J. Atmos. Sci.}\ }\textbf
  {\bibinfo {volume} {21}},\ \bibinfo {pages} {201} (\bibinfo {year}
  {1964}{\natexlab{a}})}\BibitemShut {NoStop}%
\bibitem [{\citenamefont {Pedlosky}(1964{\natexlab{b}})}]{Pedlosky64b}%
  \BibitemOpen
  \bibfield  {author} {\bibinfo {author} {\bibfnamefont {J.}~\bibnamefont
  {Pedlosky}},\ }\bibfield  {title} {\bibinfo {title} {The stability of
  currents in the atmosphere and the ocean: {P}art {II}},\ }\href@noop {}
  {\bibfield  {journal} {\bibinfo  {journal} {J. Atmos. Sci.}\ }\textbf
  {\bibinfo {volume} {21}},\ \bibinfo {pages} {342} (\bibinfo {year}
  {1964}{\natexlab{b}})}\BibitemShut {NoStop}%
\bibitem [{\citenamefont {Shepherd}(1990{\natexlab{a}})}]{Shepherd90}%
  \BibitemOpen
  \bibfield  {author} {\bibinfo {author} {\bibfnamefont {T.~G.}\ \bibnamefont
  {Shepherd}},\ }\bibfield  {title} {\bibinfo {title} {Symmetries, conservation
  laws, and {H}amiltonian structure in geophysical fluid dynamics},\
  }\href@noop {} {\bibfield  {journal} {\bibinfo  {journal} {Adv. Geophys.}\ ,\
  \bibinfo {pages} {287}} (\bibinfo {year} {1990}{\natexlab{a}})}\BibitemShut
  {NoStop}%
\bibitem [{\citenamefont {Shepherd}(1990{\natexlab{b}})}]{Shepherd88}%
  \BibitemOpen
  \bibfield  {author} {\bibinfo {author} {\bibfnamefont {T.~G.}\ \bibnamefont
  {Shepherd}},\ }\bibfield  {title} {\bibinfo {title} {{Nonlinear saturation of
  baroclinic instability: part I: the two-layer model}},\ }\href@noop {}
  {\bibfield  {journal} {\bibinfo  {journal} {J. Atmos. Sci.}\ }\textbf
  {\bibinfo {volume} {45}},\ \bibinfo {pages} {2014} (\bibinfo {year}
  {1990}{\natexlab{b}})}\BibitemShut {NoStop}%
\bibitem [{\citenamefont {Shepherd}(1990{\natexlab{c}})}]{Shepherd89}%
  \BibitemOpen
  \bibfield  {author} {\bibinfo {author} {\bibfnamefont {T.~G.}\ \bibnamefont
  {Shepherd}},\ }\bibfield  {title} {\bibinfo {title} {{Nonlinear saturation of
  baroclinic instability: part II: continuous stratified fluid}},\ }\href@noop
  {} {\bibfield  {journal} {\bibinfo  {journal} {J. Atmos. Sci.}\ }\textbf
  {\bibinfo {volume} {46}},\ \bibinfo {pages} {888} (\bibinfo {year}
  {1990}{\natexlab{c}})}\BibitemShut {NoStop}%
\bibitem [{\citenamefont {Thorncroft}\ \emph {et~al.}(1993)\citenamefont
  {Thorncroft}, \citenamefont {Hoskins},\ and\ \citenamefont
  {{McIntyre}}}]{Thorncroft-et-al93}%
  \BibitemOpen
  \bibfield  {author} {\bibinfo {author} {\bibfnamefont {C.~D.}\ \bibnamefont
  {Thorncroft}}, \bibinfo {author} {\bibfnamefont {B.~J.}\ \bibnamefont
  {Hoskins}},\ and\ \bibinfo {author} {\bibfnamefont {M.~E.}\ \bibnamefont
  {{McIntyre}}},\ }\bibfield  {title} {\bibinfo {title} {Two paradigms of
  baroclinic-wave life-cycle behaviour},\ }\href
  {https://doi.org/10.1002/qj.49711950903} {\bibfield  {journal} {\bibinfo
  {journal} {Q. J. Roy. Met. Soc.}\ }\textbf {\bibinfo {volume} {119}},\
  \bibinfo {pages} {17} (\bibinfo {year} {1993})}\BibitemShut {NoStop}%
\bibitem [{\citenamefont {Larichev}\ and\ \citenamefont
  {Held}(1995)}]{LarichevHeld95}%
  \BibitemOpen
  \bibfield  {author} {\bibinfo {author} {\bibfnamefont {V.~D.}\ \bibnamefont
  {Larichev}}\ and\ \bibinfo {author} {\bibfnamefont {I.~M.}\ \bibnamefont
  {Held}},\ }\bibfield  {title} {\bibinfo {title} {{Eddy amplitudes and fluxes
  in a homogeneous model of fully developed baroclinic instability}},\ }\href
  {https://doi.org/10.1175/1520-0485(1995)025<2285:EAAFIA>2.0.CO;2} {\bibfield
  {journal} {\bibinfo  {journal} {J. Phys. Oceanogr.}\ }\textbf {\bibinfo
  {volume} {25}},\ \bibinfo {pages} {2285} (\bibinfo {year}
  {1995})}\BibitemShut {NoStop}%
\bibitem [{\citenamefont {Spall}\ and\ \citenamefont
  {Chapman}(1998)}]{SpallChapman98}%
  \BibitemOpen
  \bibfield  {author} {\bibinfo {author} {\bibfnamefont {M.~A.}\ \bibnamefont
  {Spall}}\ and\ \bibinfo {author} {\bibfnamefont {D.~C.}\ \bibnamefont
  {Chapman}},\ }\bibfield  {title} {\bibinfo {title} {{On the efficiency of
  baroclinic eddy heat transport across narrow fronts}},\ }\href
  {https://doi.org/10.1175/1520-0485(1998)028<2275:OTEOBE>2.0.CO;2} {\bibfield
  {journal} {\bibinfo  {journal} {J. Phys. Oceanogr.}\ }\textbf {\bibinfo
  {volume} {28}},\ \bibinfo {pages} {2275} (\bibinfo {year}
  {1998})}\BibitemShut {NoStop}%
\bibitem [{\citenamefont {Thompson}\ and\ \citenamefont
  {Young}(2007)}]{ThompsonYoung07}%
  \BibitemOpen
  \bibfield  {author} {\bibinfo {author} {\bibfnamefont {A.~F.}\ \bibnamefont
  {Thompson}}\ and\ \bibinfo {author} {\bibfnamefont {W.~R.}\ \bibnamefont
  {Young}},\ }\bibfield  {title} {\bibinfo {title} {{Two-layer baroclinic eddy
  heat fluxes: Zonal flows and energy balance}},\ }\href
  {https://doi.org/10.1175/JAS4000.1} {\bibfield  {journal} {\bibinfo
  {journal} {J. Atmos. Sci.}\ }\textbf {\bibinfo {volume} {64}},\ \bibinfo
  {pages} {3214} (\bibinfo {year} {2007})}\BibitemShut {NoStop}%
\bibitem [{\citenamefont {Esler}(2008)}]{Esler08a}%
  \BibitemOpen
  \bibfield  {author} {\bibinfo {author} {\bibfnamefont {J.~G.}\ \bibnamefont
  {Esler}},\ }\bibfield  {title} {\bibinfo {title} {The turbulent equilibration
  of an unstable baroclinic jet},\ }\href@noop {} {\bibfield  {journal}
  {\bibinfo  {journal} {J. Fluid Mech.}\ }\textbf {\bibinfo {volume} {599}},\
  \bibinfo {pages} {241} (\bibinfo {year} {2008})}\BibitemShut {NoStop}%
\bibitem [{\citenamefont {Bachman}\ and\ \citenamefont
  {Fox-Kemper}(2013)}]{BachmanFoxKemper13}%
  \BibitemOpen
  \bibfield  {author} {\bibinfo {author} {\bibfnamefont {S.~D.}\ \bibnamefont
  {Bachman}}\ and\ \bibinfo {author} {\bibfnamefont {B.}~\bibnamefont
  {Fox-Kemper}},\ }\bibfield  {title} {\bibinfo {title} {{Eddy parametrization
  challenge suite I: Eady spindown}},\ }\href
  {https://doi.org/10.1016/j.ocemod.2012.12.003} {\bibfield  {journal}
  {\bibinfo  {journal} {Ocean Modell.}\ }\textbf {\bibinfo {volume} {64}},\
  \bibinfo {pages} {12} (\bibinfo {year} {2013})}\BibitemShut {NoStop}%
\bibitem [{\citenamefont {Bachman}\ \emph {et~al.}(2017)\citenamefont
  {Bachman}, \citenamefont {Marshall}, \citenamefont {Maddison},\ and\
  \citenamefont {Mak}}]{Bachman-et-al17}%
  \BibitemOpen
  \bibfield  {author} {\bibinfo {author} {\bibfnamefont {S.~D.}\ \bibnamefont
  {Bachman}}, \bibinfo {author} {\bibfnamefont {D.~P.}\ \bibnamefont
  {Marshall}}, \bibinfo {author} {\bibfnamefont {J.~R.}\ \bibnamefont
  {Maddison}},\ and\ \bibinfo {author} {\bibfnamefont {J.}~\bibnamefont
  {Mak}},\ }\bibfield  {title} {\bibinfo {title} {{Evaluation of a scalar
  transport coefficient based on geometric constraints}},\ }\href
  {https://doi.org/10.1016/j.ocemod.2016.12.004} {\bibfield  {journal}
  {\bibinfo  {journal} {Ocean Modell.}\ }\textbf {\bibinfo {volume} {109}},\
  \bibinfo {pages} {44} (\bibinfo {year} {2017})}\BibitemShut {NoStop}%
\bibitem [{\citenamefont {Chang}\ and\ \citenamefont
  {Held}(2022)}]{ChangHeld22}%
  \BibitemOpen
  \bibfield  {author} {\bibinfo {author} {\bibfnamefont {C.}~\bibnamefont
  {Chang}}\ and\ \bibinfo {author} {\bibfnamefont {I.~M.}\ \bibnamefont
  {Held}},\ }\bibfield  {title} {\bibinfo {title} {{A scaling theory for the
  diffusivity of poleward eddy heat transport based on Rhines scaling and the
  global entropy budget}},\ }\href {https://doi.org/10.1175/JAS-D-21-0242.1}
  {\bibfield  {journal} {\bibinfo  {journal} {J. Atmos. Sci.}\ }\textbf
  {\bibinfo {volume} {79}},\ \bibinfo {pages} {1743} (\bibinfo {year}
  {2022})}\BibitemShut {NoStop}%
\bibitem [{\citenamefont {Green}(1970)}]{Green70}%
  \BibitemOpen
  \bibfield  {author} {\bibinfo {author} {\bibfnamefont {J.~S.~A.}\
  \bibnamefont {Green}},\ }\bibfield  {title} {\bibinfo {title} {{Transfer
  properties of the large-scale eddies and the general circulation of the
  atmosphere}},\ }\href {https://doi.org/10.1002/qj.49709640802} {\bibfield
  {journal} {\bibinfo  {journal} {Q. J. Roy. Met. Soc.}\ }\textbf {\bibinfo
  {volume} {96}},\ \bibinfo {pages} {157} (\bibinfo {year} {1970})}\BibitemShut
  {NoStop}%
\bibitem [{\citenamefont {Stone}(1972)}]{Stone72}%
  \BibitemOpen
  \bibfield  {author} {\bibinfo {author} {\bibfnamefont {P.~H.}\ \bibnamefont
  {Stone}},\ }\bibfield  {title} {\bibinfo {title} {{A simplified
  radiative-dynamical model for the static stability of rotating
  atmospheres}},\ }\href@noop {} {\bibfield  {journal} {\bibinfo  {journal} {J.
  Atmos. Sci.}\ }\textbf {\bibinfo {volume} {29}},\ \bibinfo {pages} {405}
  (\bibinfo {year} {1972})}\BibitemShut {NoStop}%
\bibitem [{\citenamefont {Killworth}(1997)}]{Killworth97}%
  \BibitemOpen
  \bibfield  {author} {\bibinfo {author} {\bibfnamefont {P.~D.}\ \bibnamefont
  {Killworth}},\ }\bibfield  {title} {\bibinfo {title} {{On the
  parameterization of eddy transfer, Part I. Theory}},\ }\href@noop {}
  {\bibfield  {journal} {\bibinfo  {journal} {J. Mar. Res.}\ }\textbf {\bibinfo
  {volume} {55}},\ \bibinfo {pages} {1171} (\bibinfo {year}
  {1997})}\BibitemShut {NoStop}%
\bibitem [{\citenamefont {Killworth}(1998)}]{Killworth98}%
  \BibitemOpen
  \bibfield  {author} {\bibinfo {author} {\bibfnamefont {P.~D.}\ \bibnamefont
  {Killworth}},\ }\bibfield  {title} {\bibinfo {title} {{On the
  parameterization of eddy transfer, Part II. Tests with a channel model}},\
  }\href@noop {} {\bibfield  {journal} {\bibinfo  {journal} {J. Mar. Res.}\
  }\textbf {\bibinfo {volume} {56}},\ \bibinfo {pages} {349} (\bibinfo {year}
  {1998})}\BibitemShut {NoStop}%
\bibitem [{\citenamefont {Eden}(2011)}]{Eden11}%
  \BibitemOpen
  \bibfield  {author} {\bibinfo {author} {\bibfnamefont {C.}~\bibnamefont
  {Eden}},\ }\bibfield  {title} {\bibinfo {title} {A closure for meso-scale
  eddy fluxes based on linear instability theory},\ }\href
  {https://doi.org/10.1016/j.ocemod.2011.05.009} {\bibfield  {journal}
  {\bibinfo  {journal} {Ocean Modell.}\ }\textbf {\bibinfo {volume} {39}},\
  \bibinfo {pages} {362} (\bibinfo {year} {2011})}\BibitemShut {NoStop}%
\bibitem [{\citenamefont {Marshall}\ \emph {et~al.}(2012)\citenamefont
  {Marshall}, \citenamefont {Maddison},\ and\ \citenamefont
  {Berloff}}]{Marshall-et-al12}%
  \BibitemOpen
  \bibfield  {author} {\bibinfo {author} {\bibfnamefont {D.~P.}\ \bibnamefont
  {Marshall}}, \bibinfo {author} {\bibfnamefont {J.~R.}\ \bibnamefont
  {Maddison}},\ and\ \bibinfo {author} {\bibfnamefont {P.~S.}\ \bibnamefont
  {Berloff}},\ }\bibfield  {title} {\bibinfo {title} {{A framework for
  parameterizing eddy potential vorticity fluxes}},\ }\href
  {https://doi.org/10.1175/JPO-D-11-048.1} {\bibfield  {journal} {\bibinfo
  {journal} {J. Phys. Oceanogr.}\ }\textbf {\bibinfo {volume} {42}},\ \bibinfo
  {pages} {539} (\bibinfo {year} {2012})}\BibitemShut {NoStop}%
\bibitem [{\citenamefont {Maddison}\ and\ \citenamefont
  {Marshall}(2013)}]{MaddisonMarshall13}%
  \BibitemOpen
  \bibfield  {author} {\bibinfo {author} {\bibfnamefont {J.~R.}\ \bibnamefont
  {Maddison}}\ and\ \bibinfo {author} {\bibfnamefont {D.~P.}\ \bibnamefont
  {Marshall}},\ }\bibfield  {title} {\bibinfo {title} {{The Eliassen--Palm flux
  tensor}},\ }\href {https://doi.org/10.1017/jfm.2013.259} {\bibfield
  {journal} {\bibinfo  {journal} {J. Fluid Mech.}\ }\textbf {\bibinfo {volume}
  {729}},\ \bibinfo {pages} {69} (\bibinfo {year} {2013})}\BibitemShut
  {NoStop}%
\bibitem [{\citenamefont {Hoskins}\ \emph
  {et~al.}(1985{\natexlab{a}})\citenamefont {Hoskins}, \citenamefont {James},\
  and\ \citenamefont {White}}]{Hoskins-et-al83}%
  \BibitemOpen
  \bibfield  {author} {\bibinfo {author} {\bibfnamefont {B.~J.}\ \bibnamefont
  {Hoskins}}, \bibinfo {author} {\bibfnamefont {I.~N.}\ \bibnamefont {James}},\
  and\ \bibinfo {author} {\bibfnamefont {G.~H.}\ \bibnamefont {White}},\
  }\bibfield  {title} {\bibinfo {title} {{The shape, propagation and mean-flow
  interaction of large-scale weather systems}},\ }\href
  {https://doi.org/10.1175/1520-0469(1983)040<1595:TSPAMF>2.0.CO;2} {\bibfield
  {journal} {\bibinfo  {journal} {J. Atmos. Sci.}\ }\textbf {\bibinfo {volume}
  {40}},\ \bibinfo {pages} {1595} (\bibinfo {year}
  {1985}{\natexlab{a}})}\BibitemShut {NoStop}%
\bibitem [{\citenamefont {Waterman}\ and\ \citenamefont
  {Hoskins}(2013)}]{WatermanHoskins13}%
  \BibitemOpen
  \bibfield  {author} {\bibinfo {author} {\bibfnamefont {S.}~\bibnamefont
  {Waterman}}\ and\ \bibinfo {author} {\bibfnamefont {B.~J.}\ \bibnamefont
  {Hoskins}},\ }\bibfield  {title} {\bibinfo {title} {Eddy shape, orientation,
  propagation, and mean flow feedback in western boundary current jets},\
  }\href {https://doi.org/10.1175/JPO-D-12-0152.1} {\bibfield  {journal}
  {\bibinfo  {journal} {J. Phys. Oceanogr.}\ }\textbf {\bibinfo {volume}
  {43}},\ \bibinfo {pages} {1666} (\bibinfo {year} {2013})}\BibitemShut
  {NoStop}%
\bibitem [{\citenamefont {Mak}\ \emph {et~al.}(2018)\citenamefont {Mak},
  \citenamefont {Maddison}, \citenamefont {Marshall},\ and\ \citenamefont
  {Munday}}]{Mak-et-al18}%
  \BibitemOpen
  \bibfield  {author} {\bibinfo {author} {\bibfnamefont {J.}~\bibnamefont
  {Mak}}, \bibinfo {author} {\bibfnamefont {J.~R.}\ \bibnamefont {Maddison}},
  \bibinfo {author} {\bibfnamefont {D.~P.}\ \bibnamefont {Marshall}},\ and\
  \bibinfo {author} {\bibfnamefont {D.~R.}\ \bibnamefont {Munday}},\ }\bibfield
   {title} {\bibinfo {title} {{Implementation of a geometrically informed and
  energetically constrained mesoscale eddy parameterization in an ocean
  circulation model}},\ }\href {https://doi.org/10.1175/JPO-D-18-0017.1}
  {\bibfield  {journal} {\bibinfo  {journal} {J. Phys. Oceanogr.}\ }\textbf
  {\bibinfo {volume} {48}},\ \bibinfo {pages} {2363} (\bibinfo {year}
  {2018})}\BibitemShut {NoStop}%
\bibitem [{\citenamefont {Mak}\ \emph {et~al.}(2022)\citenamefont {Mak},
  \citenamefont {Marshall}, \citenamefont {Madec},\ and\ \citenamefont
  {Maddison}}]{Mak-et-al22}%
  \BibitemOpen
  \bibfield  {author} {\bibinfo {author} {\bibfnamefont {J.}~\bibnamefont
  {Mak}}, \bibinfo {author} {\bibfnamefont {D.~P.}\ \bibnamefont {Marshall}},
  \bibinfo {author} {\bibfnamefont {G.}~\bibnamefont {Madec}},\ and\ \bibinfo
  {author} {\bibfnamefont {J.~R.}\ \bibnamefont {Maddison}},\ }\bibfield
  {title} {\bibinfo {title} {{Acute sensitivity of global ocean circulation and
  heat content to eddy energy dissipation time-scale}},\ }\href
  {https://doi.org/10.1029/2021GL097259} {\bibfield  {journal} {\bibinfo
  {journal} {Geophys. Res. Lett.}\ }\textbf {\bibinfo {volume} {49}},\ \bibinfo
  {pages} {e2021GL097259} (\bibinfo {year} {2022})}\BibitemShut {NoStop}%
\bibitem [{\citenamefont {Mak}\ \emph {et~al.}(2023)\citenamefont {Mak},
  \citenamefont {Maddison}, \citenamefont {Marshall}, \citenamefont {Ruan},\
  and\ \citenamefont {Wang}}]{Mak-et-al23}%
  \BibitemOpen
  \bibfield  {author} {\bibinfo {author} {\bibfnamefont {J.}~\bibnamefont
  {Mak}}, \bibinfo {author} {\bibfnamefont {J.~R.}\ \bibnamefont {Maddison}},
  \bibinfo {author} {\bibfnamefont {D.~P.}\ \bibnamefont {Marshall}}, \bibinfo
  {author} {\bibfnamefont {X.}~\bibnamefont {Ruan}},\ and\ \bibinfo {author}
  {\bibfnamefont {Y.}~\bibnamefont {Wang}},\ }\bibfield  {title} {\bibinfo
  {title} {{Scale-awareness in an eddy energy constrained mesoscale eddy
  parameterization}},\ }\href {https://doi.org/10.1029/2023MS003886} {\bibfield
   {journal} {\bibinfo  {journal} {J. Adv. Model. Earth. Syst.}\ }\textbf
  {\bibinfo {volume} {15}},\ \bibinfo {pages} {e2023MS003886} (\bibinfo {year}
  {2023})}\BibitemShut {NoStop}%
\bibitem [{\citenamefont {Blumsack}\ and\ \citenamefont
  {Gierasch}(1972)}]{BlumsackGierasch72}%
  \BibitemOpen
  \bibfield  {author} {\bibinfo {author} {\bibfnamefont {S.~K.}\ \bibnamefont
  {Blumsack}}\ and\ \bibinfo {author} {\bibfnamefont {P.}~\bibnamefont
  {Gierasch}},\ }\bibfield  {title} {\bibinfo {title} {{Mars: The effects of
  topography on baroclinic instability}},\ }\href
  {https://doi.org/10.1175/1520-0469(1972)029<1081:mteoto>2.0.co;2} {\bibfield
  {journal} {\bibinfo  {journal} {J. Atmos. Sci.}\ }\textbf {\bibinfo {volume}
  {29}},\ \bibinfo {pages} {1081} (\bibinfo {year} {1972})}\BibitemShut
  {NoStop}%
\bibitem [{\citenamefont {Mechoso}(1980)}]{Mechoso80}%
  \BibitemOpen
  \bibfield  {author} {\bibinfo {author} {\bibfnamefont {C.~R.}\ \bibnamefont
  {Mechoso}},\ }\bibfield  {title} {\bibinfo {title} {{Baroclinic instability
  of flows along sloping boundaries}},\ }\href
  {https://doi.org/10.1175/1520-0469(1980)037<1393:biofas>2.0.co;22} {\bibfield
   {journal} {\bibinfo  {journal} {J. Atmos. Sci.}\ }\textbf {\bibinfo {volume}
  {37}},\ \bibinfo {pages} {1393} (\bibinfo {year} {1980})}\BibitemShut
  {NoStop}%
\bibitem [{\citenamefont {Isachsen}(2011)}]{Isachsen11}%
  \BibitemOpen
  \bibfield  {author} {\bibinfo {author} {\bibfnamefont {P.~E.}\ \bibnamefont
  {Isachsen}},\ }\bibfield  {title} {\bibinfo {title} {{Baroclinic instability
  and eddy tracer transport across sloping bottom topography: How well does a
  modified Eady model do in primitive equation simulations?}},\ }\href
  {https://doi.org/10.1016/j.ocemod.2010.09.007} {\bibfield  {journal}
  {\bibinfo  {journal} {Ocean Modell.}\ }\textbf {\bibinfo {volume} {39}},\
  \bibinfo {pages} {183} (\bibinfo {year} {2011})}\BibitemShut {NoStop}%
\bibitem [{\citenamefont {Brink}(2012)}]{Brink12}%
  \BibitemOpen
  \bibfield  {author} {\bibinfo {author} {\bibfnamefont {K.}~\bibnamefont
  {Brink}},\ }\bibfield  {title} {\bibinfo {title} {{Baroclinic instability of
  an idealized tidal mixing front}},\ }\href
  {https://doi.org/10.1357/002224012805262716} {\bibfield  {journal} {\bibinfo
  {journal} {J. Mar. Res.}\ }\textbf {\bibinfo {volume} {70}},\ \bibinfo
  {pages} {661} (\bibinfo {year} {2012})}\BibitemShut {NoStop}%
\bibitem [{\citenamefont {Chen}\ and\ \citenamefont
  {Kamenkovich}(2013)}]{ChenKamenkovich13}%
  \BibitemOpen
  \bibfield  {author} {\bibinfo {author} {\bibfnamefont {C.}~\bibnamefont
  {Chen}}\ and\ \bibinfo {author} {\bibfnamefont {I.}~\bibnamefont
  {Kamenkovich}},\ }\bibfield  {title} {\bibinfo {title} {{Effects of
  topography on baroclinic instability}},\ }\href
  {https://doi.org/10.1175/JPO-D-12-0145.1} {\bibfield  {journal} {\bibinfo
  {journal} {J. Phys. Oceanogr.}\ }\textbf {\bibinfo {volume} {43}},\ \bibinfo
  {pages} {790} (\bibinfo {year} {2013})}\BibitemShut {NoStop}%
\bibitem [{\citenamefont {Isachsen}(2015)}]{Isachsen15}%
  \BibitemOpen
  \bibfield  {author} {\bibinfo {author} {\bibfnamefont {P.~E.}\ \bibnamefont
  {Isachsen}},\ }\bibfield  {title} {\bibinfo {title} {{Baroclinic instability
  and the mesoscale eddy field around the Lofoten Basin}},\ }\href
  {https://doi.org/10.1002/2014JC010448} {\bibfield  {journal} {\bibinfo
  {journal} {J. Geophys. Res. Oceans}\ }\textbf {\bibinfo {volume} {120}},\
  \bibinfo {pages} {2884} (\bibinfo {year} {2015})}\BibitemShut {NoStop}%
\bibitem [{\citenamefont {Pedlosky}(2016)}]{Pedlosky16}%
  \BibitemOpen
  \bibfield  {author} {\bibinfo {author} {\bibfnamefont {J.}~\bibnamefont
  {Pedlosky}},\ }\bibfield  {title} {\bibinfo {title} {{Baroclinic instability
  over topography: Unstable at any wave number}},\ }\href@noop {} {\bibfield
  {journal} {\bibinfo  {journal} {J. Mar. Res.}\ }\textbf {\bibinfo {volume}
  {74}},\ \bibinfo {pages} {1} (\bibinfo {year} {2016})}\BibitemShut {NoStop}%
\bibitem [{\citenamefont {Manucharyan}\ \emph {et~al.}(2017)\citenamefont
  {Manucharyan}, \citenamefont {Thompson},\ and\ \citenamefont
  {Spall}}]{Manucharyan-et-al17}%
  \BibitemOpen
  \bibfield  {author} {\bibinfo {author} {\bibfnamefont {G.~E.}\ \bibnamefont
  {Manucharyan}}, \bibinfo {author} {\bibfnamefont {A.~F.}\ \bibnamefont
  {Thompson}},\ and\ \bibinfo {author} {\bibfnamefont {M.~A.}\ \bibnamefont
  {Spall}},\ }\bibfield  {title} {\bibinfo {title} {{Eddy-Memory mode of
  multi-decadal variability in residual-mean ocean circulations with
  application to the Beaufort Gyre}},\ }\href
  {https://doi.org/10.1175/JPO-D-16-0194.1} {\bibfield  {journal} {\bibinfo
  {journal} {J. Phys. Oceanogr.}\ }\textbf {\bibinfo {volume} {47}},\ \bibinfo
  {pages} {855} (\bibinfo {year} {2017})}\BibitemShut {NoStop}%
\bibitem [{\citenamefont {Hetland}(2017)}]{Hetland17}%
  \BibitemOpen
  \bibfield  {author} {\bibinfo {author} {\bibfnamefont {R.~D.}\ \bibnamefont
  {Hetland}},\ }\bibfield  {title} {\bibinfo {title} {{Suppression of
  baroclinic instabilities in buoyancy-driven flow over sloping bathymetry}},\
  }\href {https://doi.org/10.1175/jpo-d-15-0240.1} {\bibfield  {journal}
  {\bibinfo  {journal} {J. Phys. Oceanogr.}\ }\textbf {\bibinfo {volume}
  {47}},\ \bibinfo {pages} {49} (\bibinfo {year} {2017})}\BibitemShut {NoStop}%
\bibitem [{\citenamefont {Trodahl}\ and\ \citenamefont
  {Isachsen}(2018)}]{TrodahlIsachsen18}%
  \BibitemOpen
  \bibfield  {author} {\bibinfo {author} {\bibfnamefont {M.}~\bibnamefont
  {Trodahl}}\ and\ \bibinfo {author} {\bibfnamefont {P.~E.}\ \bibnamefont
  {Isachsen}},\ }\bibfield  {title} {\bibinfo {title} {{Topographic influence
  on baroclinic instability and the mesoscale eddy field in the northern North
  Atlantic Ocean and the Nordic seas}},\ }\href
  {https://doi.org/10.1175/JPO-D-17-0220.1} {\bibfield  {journal} {\bibinfo
  {journal} {J. Phys. Oceanogr.}\ }\textbf {\bibinfo {volume} {48}},\ \bibinfo
  {pages} {2593} (\bibinfo {year} {2018})}\BibitemShut {NoStop}%
\bibitem [{\citenamefont {Wang}\ and\ \citenamefont
  {Stewart}(2018)}]{WangStewart18}%
  \BibitemOpen
  \bibfield  {author} {\bibinfo {author} {\bibfnamefont {Y.}~\bibnamefont
  {Wang}}\ and\ \bibinfo {author} {\bibfnamefont {A.~L.}\ \bibnamefont
  {Stewart}},\ }\bibfield  {title} {\bibinfo {title} {{Eddy dynamics over
  continental slopes under retrograde winds: Insights from a model
  inter-comparison}},\ }\href {https://doi.org/10.1016/j.ocemod.2017.11.006}
  {\bibfield  {journal} {\bibinfo  {journal} {Ocean Modell.}\ }\textbf
  {\bibinfo {volume} {121}},\ \bibinfo {pages} {1} (\bibinfo {year}
  {2018})}\BibitemShut {NoStop}%
\bibitem [{\citenamefont {Manucharyan}\ and\ \citenamefont
  {Isachsen}(2019)}]{ManucharyanIsachsen19}%
  \BibitemOpen
  \bibfield  {author} {\bibinfo {author} {\bibfnamefont {G.~E.}\ \bibnamefont
  {Manucharyan}}\ and\ \bibinfo {author} {\bibfnamefont {P.~E.}\ \bibnamefont
  {Isachsen}},\ }\bibfield  {title} {\bibinfo {title} {{Critical role of
  continental slopes in halocline and eddy dynamics of the Ekman-driven
  Beaufort Gyre}},\ }\href {https://doi.org/10.1029/2018JC014624} {\bibfield
  {journal} {\bibinfo  {journal} {J. Geophys. Res. Oceans}\ }\textbf {\bibinfo
  {volume} {124}},\ \bibinfo {pages} {2679} (\bibinfo {year}
  {2019})}\BibitemShut {NoStop}%
\bibitem [{\citenamefont {Chen}\ \emph {et~al.}(2020)\citenamefont {Chen},
  \citenamefont {Chen},\ and\ \citenamefont {Lerczak}}]{Chen-et-al20}%
  \BibitemOpen
  \bibfield  {author} {\bibinfo {author} {\bibfnamefont {S.}~\bibnamefont
  {Chen}}, \bibinfo {author} {\bibfnamefont {C.}~\bibnamefont {Chen}},\ and\
  \bibinfo {author} {\bibfnamefont {J.~A.}\ \bibnamefont {Lerczak}},\
  }\bibfield  {title} {\bibinfo {title} {{On baroclinic instability over
  continental shelves: Testing the utility of Eady-type models}},\ }\href
  {https://doi.org/10.1175/jpo-d-19-0175.1} {\bibfield  {journal} {\bibinfo
  {journal} {J. Phys. Oceanogr.}\ }\textbf {\bibinfo {volume} {50}},\ \bibinfo
  {pages} {3} (\bibinfo {year} {2020})}\BibitemShut {NoStop}%
\bibitem [{\citenamefont {Tanaka}(2021)}]{Tanaka21}%
  \BibitemOpen
  \bibfield  {author} {\bibinfo {author} {\bibfnamefont {Y.}~\bibnamefont
  {Tanaka}},\ }\bibfield  {title} {\bibinfo {title} {{Stability of a flow over
  bottom topography: A general condition and a linear analysis in a two-layer
  quasi-geostrophic model with a possible application to a Kuroshio meander}},\
  }\href {https://doi.org/10.1029/2021JC017849} {\bibfield  {journal} {\bibinfo
   {journal} {J. Geophys. Res. Oceans.}\ }\textbf {\bibinfo {volume} {126}},\
  \bibinfo {pages} {e2021JC017849} (\bibinfo {year} {2021})}\BibitemShut
  {NoStop}%
\bibitem [{\citenamefont {Wei}\ and\ \citenamefont {Wang}(2021)}]{WeiWang21}%
  \BibitemOpen
  \bibfield  {author} {\bibinfo {author} {\bibfnamefont {H.}~\bibnamefont
  {Wei}}\ and\ \bibinfo {author} {\bibfnamefont {Y.}~\bibnamefont {Wang}},\
  }\bibfield  {title} {\bibinfo {title} {{Full-depth scalings for isopycnal
  eddy mixing across continental slopes under upwelling-favorable winds}},\
  }\href {https://doi.org/10.1029/2021MS002498} {\bibfield  {journal} {\bibinfo
   {journal} {J. Adv. Model. Earth. Syst.}\ }\textbf {\bibinfo {volume} {13}},\
  \bibinfo {pages} {e2021MS002498} (\bibinfo {year} {2021})}\BibitemShut
  {NoStop}%
\bibitem [{\citenamefont {Wei}\ \emph {et~al.}(2022)\citenamefont {Wei},
  \citenamefont {Wang}, \citenamefont {Stewart},\ and\ \citenamefont
  {Mak}}]{Wei-et-al22}%
  \BibitemOpen
  \bibfield  {author} {\bibinfo {author} {\bibfnamefont {H.}~\bibnamefont
  {Wei}}, \bibinfo {author} {\bibfnamefont {Y.}~\bibnamefont {Wang}}, \bibinfo
  {author} {\bibfnamefont {A.~L.}\ \bibnamefont {Stewart}},\ and\ \bibinfo
  {author} {\bibfnamefont {J.}~\bibnamefont {Mak}},\ }\bibfield  {title}
  {\bibinfo {title} {{Scalings for eddy buoyancy fluxes across prograde
  shelf/slope fronts}},\ }\href {https://doi.org/10.1029/2022MS003229}
  {\bibfield  {journal} {\bibinfo  {journal} {J. Adv. Model. Earth. Syst.}\
  }\textbf {\bibinfo {volume} {14}},\ \bibinfo {pages} {e2022MS003229}
  (\bibinfo {year} {2022})}\BibitemShut {NoStop}%
\bibitem [{\citenamefont {Wei}\ \emph {et~al.}(2024)\citenamefont {Wei},
  \citenamefont {Wang},\ and\ \citenamefont {Mak}}]{Wei-et-al24}%
  \BibitemOpen
  \bibfield  {author} {\bibinfo {author} {\bibfnamefont {H.}~\bibnamefont
  {Wei}}, \bibinfo {author} {\bibfnamefont {Y.}~\bibnamefont {Wang}},\ and\
  \bibinfo {author} {\bibfnamefont {J.}~\bibnamefont {Mak}},\ }\bibfield
  {title} {\bibinfo {title} {{Parameterizing eddy buoyancy fluxes across
  prograde shelf/slope fronts using a slope-aware GEOMETRIC closure}},\ }\href
  {https://doi.org/10.1175/JPO-D-23-0152.1} {\bibfield  {journal} {\bibinfo
  {journal} {J. Phys. Oceanogr.}\ }\textbf {\bibinfo {volume} {54}},\ \bibinfo
  {pages} {359} (\bibinfo {year} {2024})}\BibitemShut {NoStop}%
\bibitem [{\citenamefont {Nummelin}\ and\ \citenamefont
  {Isachsen}(2024)}]{NummelinIsachsen24}%
  \BibitemOpen
  \bibfield  {author} {\bibinfo {author} {\bibfnamefont {A.}~\bibnamefont
  {Nummelin}}\ and\ \bibinfo {author} {\bibfnamefont {P.~E.}\ \bibnamefont
  {Isachsen}},\ }\bibfield  {title} {\bibinfo {title} {{Parameterizing
  mesoscale eddy buoyancy transport over sloping topography}},\ }\bibfield
  {journal} {\bibinfo  {journal} {ESS Open Archive}\ }\href
  {https://doi.org/10.22541/essoar.168394750.04852652/v2}
  {10.22541/essoar.168394750.04852652/v2} (\bibinfo {year} {2024})\BibitemShut
  {NoStop}%
\bibitem [{\citenamefont {Bretherton}(1966)}]{Bretherton66a}%
  \BibitemOpen
  \bibfield  {author} {\bibinfo {author} {\bibfnamefont {F.~P.}\ \bibnamefont
  {Bretherton}},\ }\bibfield  {title} {\bibinfo {title} {Baroclinic instability
  and the short wavelength cut-off in terms of potential vorticity},\
  }\href@noop {} {\bibfield  {journal} {\bibinfo  {journal} {Q. J. Roy. Met.
  Soc.}\ }\textbf {\bibinfo {volume} {92}},\ \bibinfo {pages} {335} (\bibinfo
  {year} {1966})}\BibitemShut {NoStop}%
\bibitem [{\citenamefont {Hoskins}\ \emph
  {et~al.}(1985{\natexlab{b}})\citenamefont {Hoskins}, \citenamefont
  {McIntyre},\ and\ \citenamefont {Robertson}}]{Hoskins-et-al85}%
  \BibitemOpen
  \bibfield  {author} {\bibinfo {author} {\bibfnamefont {B.~J.}\ \bibnamefont
  {Hoskins}}, \bibinfo {author} {\bibfnamefont {M.~E.}\ \bibnamefont
  {McIntyre}},\ and\ \bibinfo {author} {\bibfnamefont {A.~W.}\ \bibnamefont
  {Robertson}},\ }\bibfield  {title} {\bibinfo {title} {{On the use and
  significance of isentropic potential vorticity maps}},\ }\href
  {https://doi.org/10.1002/qj.49711147002} {\bibfield  {journal} {\bibinfo
  {journal} {Q. J. Roy. Met. Soc.}\ }\textbf {\bibinfo {volume} {111}},\
  \bibinfo {pages} {877} (\bibinfo {year} {1985}{\natexlab{b}})}\BibitemShut
  {NoStop}%
\bibitem [{\citenamefont {Heifetz}\ \emph {et~al.}(2004)\citenamefont
  {Heifetz}, \citenamefont {Bishop}, \citenamefont {Hoskins},\ and\
  \citenamefont {Methven}}]{Heifetz-et-al04a}%
  \BibitemOpen
  \bibfield  {author} {\bibinfo {author} {\bibfnamefont {E.}~\bibnamefont
  {Heifetz}}, \bibinfo {author} {\bibfnamefont {C.~H.}\ \bibnamefont {Bishop}},
  \bibinfo {author} {\bibfnamefont {B.~J.}\ \bibnamefont {Hoskins}},\ and\
  \bibinfo {author} {\bibfnamefont {J.}~\bibnamefont {Methven}},\ }\bibfield
  {title} {\bibinfo {title} {The counter-propagating {R}ossby-wave perspective
  on baroclinic instability. {I}: {M}athematical basis},\ }\href@noop {}
  {\bibfield  {journal} {\bibinfo  {journal} {Q. J. Roy. Met. Soc.}\ }\textbf
  {\bibinfo {volume} {130}},\ \bibinfo {pages} {211} (\bibinfo {year}
  {2004})}\BibitemShut {NoStop}%
\bibitem [{\citenamefont {Heifetz}\ \emph {et~al.}(1999)\citenamefont
  {Heifetz}, \citenamefont {Bishop},\ and\ \citenamefont
  {Alpert}}]{Heifetz-et-al99}%
  \BibitemOpen
  \bibfield  {author} {\bibinfo {author} {\bibfnamefont {E.}~\bibnamefont
  {Heifetz}}, \bibinfo {author} {\bibfnamefont {C.~H.}\ \bibnamefont
  {Bishop}},\ and\ \bibinfo {author} {\bibfnamefont {P.}~\bibnamefont
  {Alpert}},\ }\bibfield  {title} {\bibinfo {title} {{Counter-propagating
  Rossby waves in the barotropic Rayleigh model of shear instability}},\
  }\href@noop {} {\bibfield  {journal} {\bibinfo  {journal} {Q. J. Roy. Met.
  Soc.}\ }\textbf {\bibinfo {volume} {125}},\ \bibinfo {pages} {2835} (\bibinfo
  {year} {1999})}\BibitemShut {NoStop}%
\bibitem [{\citenamefont {Harnik}\ and\ \citenamefont
  {Heifetz}(2007)}]{HarnikHeifetz07}%
  \BibitemOpen
  \bibfield  {author} {\bibinfo {author} {\bibfnamefont {N.}~\bibnamefont
  {Harnik}}\ and\ \bibinfo {author} {\bibfnamefont {E.}~\bibnamefont
  {Heifetz}},\ }\bibfield  {title} {\bibinfo {title} {{Relating overreflection
  and wave geometry to the counter-propagating Rossby wave perspective: Toward
  a deeper mechanistic understanding of shear instability}},\ }\href@noop {}
  {\bibfield  {journal} {\bibinfo  {journal} {J. Atmos. Sci.}\ }\textbf
  {\bibinfo {volume} {64}},\ \bibinfo {pages} {2238} (\bibinfo {year}
  {2007})}\BibitemShut {NoStop}%
\bibitem [{\citenamefont {Rabinovich}\ \emph {et~al.}(2011)\citenamefont
  {Rabinovich}, \citenamefont {Umurhan}, \citenamefont {Harnik}, \citenamefont
  {Lott},\ and\ \citenamefont {Heifetz}}]{Rabinovich-et-al11}%
  \BibitemOpen
  \bibfield  {author} {\bibinfo {author} {\bibfnamefont {A.}~\bibnamefont
  {Rabinovich}}, \bibinfo {author} {\bibfnamefont {O.~M.}\ \bibnamefont
  {Umurhan}}, \bibinfo {author} {\bibfnamefont {N.}~\bibnamefont {Harnik}},
  \bibinfo {author} {\bibfnamefont {F.}~\bibnamefont {Lott}},\ and\ \bibinfo
  {author} {\bibfnamefont {E.}~\bibnamefont {Heifetz}},\ }\bibfield  {title}
  {\bibinfo {title} {Vorticity inversion and action-at-a-distance instability
  in stably stratified shear flow},\ }\href
  {https://doi.org/10.1017/S002211201000529X} {\bibfield  {journal} {\bibinfo
  {journal} {J. Fluid Mech.}\ }\textbf {\bibinfo {volume} {670}},\ \bibinfo
  {pages} {301} (\bibinfo {year} {2011})}\BibitemShut {NoStop}%
\bibitem [{\citenamefont {Heifetz}\ \emph {et~al.}(2015)\citenamefont
  {Heifetz}, \citenamefont {Mak}, \citenamefont {Nycander},\ and\ \citenamefont
  {Umurhan}}]{Heifetz-et-al15}%
  \BibitemOpen
  \bibfield  {author} {\bibinfo {author} {\bibfnamefont {E.}~\bibnamefont
  {Heifetz}}, \bibinfo {author} {\bibfnamefont {J.}~\bibnamefont {Mak}},
  \bibinfo {author} {\bibfnamefont {J.}~\bibnamefont {Nycander}},\ and\
  \bibinfo {author} {\bibfnamefont {O.~M.}\ \bibnamefont {Umurhan}},\
  }\bibfield  {title} {\bibinfo {title} {Interacting vorticity waves as an
  instability mechanism for magnetohydrodynamic shear instabilities},\ }\href
  {https://doi.org/10.1017/jfm.2015.47} {\bibfield  {journal} {\bibinfo
  {journal} {J. Fluid Mech.}\ }\textbf {\bibinfo {volume} {767}},\ \bibinfo
  {pages} {199} (\bibinfo {year} {2015})}\BibitemShut {NoStop}%
\bibitem [{\citenamefont {Heifetz}\ and\ \citenamefont
  {Guha}(2019)}]{HeifetzGuha19}%
  \BibitemOpen
  \bibfield  {author} {\bibinfo {author} {\bibfnamefont {E.}~\bibnamefont
  {Heifetz}}\ and\ \bibinfo {author} {\bibfnamefont {A.}~\bibnamefont {Guha}},\
  }\bibfield  {title} {\bibinfo {title} {{Normal form of synchronization and
  resonance between vorticity waves in shear flow instability}},\ }\href
  {https://doi.org/10.1103/PhysRevE.100.043105} {\bibfield  {journal} {\bibinfo
   {journal} {Phys. Rev. E}\ }\textbf {\bibinfo {volume} {100}},\ \bibinfo
  {pages} {043105} (\bibinfo {year} {2019})}\BibitemShut {NoStop}%
\bibitem [{\citenamefont {Bender}(2018)}]{Bender-PT}%
  \BibitemOpen
  \bibfield  {author} {\bibinfo {author} {\bibfnamefont {C.~M.}\ \bibnamefont
  {Bender}},\ }\href@noop {} {\emph {\bibinfo {title} {PT Symmetry: In Quantum
  And Classical Physics}}},\ \bibinfo {edition} {1st}\ ed.\ (\bibinfo
  {publisher} {World Scientific},\ \bibinfo {year} {2018})\BibitemShut
  {NoStop}%
\bibitem [{\citenamefont {Qin}\ \emph {et~al.}(2019)\citenamefont {Qin},
  \citenamefont {Zhang}, \citenamefont {Glasser},\ and\ \citenamefont
  {Xiao}}]{Qin-et-al19}%
  \BibitemOpen
  \bibfield  {author} {\bibinfo {author} {\bibfnamefont {H.}~\bibnamefont
  {Qin}}, \bibinfo {author} {\bibfnamefont {R.}~\bibnamefont {Zhang}}, \bibinfo
  {author} {\bibfnamefont {A.~S.}\ \bibnamefont {Glasser}},\ and\ \bibinfo
  {author} {\bibfnamefont {J.}~\bibnamefont {Xiao}},\ }\bibfield  {title}
  {\bibinfo {title} {{Kelvin-Helmholtz instability is the result of parity-time
  symmetry breaking}},\ }\href {https://doi.org/10.1063/1.5088498} {\bibfield
  {journal} {\bibinfo  {journal} {Phys. Plasma}\ }\textbf {\bibinfo {volume}
  {26}},\ \bibinfo {pages} {032102} (\bibinfo {year} {2019})}\BibitemShut
  {NoStop}%
\bibitem [{\citenamefont {Zhang}\ \emph {et~al.}(2020)\citenamefont {Zhang},
  \citenamefont {Qin},\ and\ \citenamefont {Xiao}}]{Zhang-et-al20}%
  \BibitemOpen
  \bibfield  {author} {\bibinfo {author} {\bibfnamefont {R.}~\bibnamefont
  {Zhang}}, \bibinfo {author} {\bibfnamefont {H.}~\bibnamefont {Qin}},\ and\
  \bibinfo {author} {\bibfnamefont {J.}~\bibnamefont {Xiao}},\ }\bibfield
  {title} {\bibinfo {title} {{PT-symmetry entails pseudo-Hermiticity regardless
  of diagonalizability}},\ }\href {https://doi.org/10.1063/1.5117211}
  {\bibfield  {journal} {\bibinfo  {journal} {J. Math. Phys.}\ }\textbf
  {\bibinfo {volume} {61}},\ \bibinfo {pages} {012101} (\bibinfo {year}
  {2020})}\BibitemShut {NoStop}%
\bibitem [{\citenamefont {David}\ \emph {et~al.}(2022)\citenamefont {David},
  \citenamefont {Delplace},\ and\ \citenamefont {Venaille}}]{David-et-al22}%
  \BibitemOpen
  \bibfield  {author} {\bibinfo {author} {\bibfnamefont {T.~W.}\ \bibnamefont
  {David}}, \bibinfo {author} {\bibfnamefont {P.}~\bibnamefont {Delplace}},\
  and\ \bibinfo {author} {\bibfnamefont {A.}~\bibnamefont {Venaille}},\
  }\bibfield  {title} {\bibinfo {title} {{How do discrete symmetries shape the
  stability of geophysical flows?}},\ }\href
  {https://doi.org/10.1063/5.0088936} {\bibfield  {journal} {\bibinfo
  {journal} {Phys. Fluids}\ }\textbf {\bibinfo {volume} {34}},\ \bibinfo
  {pages} {056605} (\bibinfo {year} {2022})}\BibitemShut {NoStop}%
\bibitem [{\citenamefont {Davies}\ and\ \citenamefont
  {Bishop}(1994)}]{DaviesBishop94}%
  \BibitemOpen
  \bibfield  {author} {\bibinfo {author} {\bibfnamefont {H.~C.}\ \bibnamefont
  {Davies}}\ and\ \bibinfo {author} {\bibfnamefont {C.~H.}\ \bibnamefont
  {Bishop}},\ }\bibfield  {title} {\bibinfo {title} {{Eady edge waves and rapid
  development}},\ }\href@noop {} {\bibfield  {journal} {\bibinfo  {journal} {J.
  Atmos. Sci.}\ }\textbf {\bibinfo {volume} {51}},\ \bibinfo {pages} {1930}
  (\bibinfo {year} {1994})}\BibitemShut {NoStop}%
\bibitem [{\citenamefont {Tamarin}\ \emph {et~al.}(2016)\citenamefont
  {Tamarin}, \citenamefont {Maddison}, \citenamefont {Heifetz},\ and\
  \citenamefont {Marshall}}]{Tamarin-et-al16}%
  \BibitemOpen
  \bibfield  {author} {\bibinfo {author} {\bibfnamefont {T.}~\bibnamefont
  {Tamarin}}, \bibinfo {author} {\bibfnamefont {J.~R.}\ \bibnamefont
  {Maddison}}, \bibinfo {author} {\bibfnamefont {E.}~\bibnamefont {Heifetz}},\
  and\ \bibinfo {author} {\bibfnamefont {D.~P.}\ \bibnamefont {Marshall}},\
  }\bibfield  {title} {\bibinfo {title} {{{A geometric interpretation of eddy
  Reynolds stresses in barotropic ocean jets}}},\ }\href
  {https://doi.org/10.1175/JPO-D-15-0139.1} {\bibfield  {journal} {\bibinfo
  {journal} {J. Phys. Oceanogr.}\ }\textbf {\bibinfo {volume} {46}},\ \bibinfo
  {pages} {2285} (\bibinfo {year} {2016})}\BibitemShut {NoStop}%
\bibitem [{\citenamefont {Held}(1985)}]{Held85}%
  \BibitemOpen
  \bibfield  {author} {\bibinfo {author} {\bibfnamefont {I.~M.}\ \bibnamefont
  {Held}},\ }\bibfield  {title} {\bibinfo {title} {{Pseudomomentum and the
  orthogonality of modes in shear flows}},\ }\href
  {https://doi.org/10.1175/1520-0469(1985)042<2280:PATOOM>2.0.CO;2} {\bibfield
  {journal} {\bibinfo  {journal} {J. Atmos. Sci.}\ }\textbf {\bibinfo {volume}
  {42}},\ \bibinfo {pages} {2280} (\bibinfo {year} {1985})}\BibitemShut
  {NoStop}%
\bibitem [{\citenamefont {Heifetz}\ \emph {et~al.}(2009)\citenamefont
  {Heifetz}, \citenamefont {Harnik},\ and\ \citenamefont
  {Tamarin}}]{Heifetz-et-al09}%
  \BibitemOpen
  \bibfield  {author} {\bibinfo {author} {\bibfnamefont {E.}~\bibnamefont
  {Heifetz}}, \bibinfo {author} {\bibfnamefont {N.}~\bibnamefont {Harnik}},\
  and\ \bibinfo {author} {\bibfnamefont {T.}~\bibnamefont {Tamarin}},\
  }\bibfield  {title} {\bibinfo {title} {Canonical hamiltonian representation
  of pseudoenergy in shear flows using counter-propagating rossby waves},\
  }\href@noop {} {\bibfield  {journal} {\bibinfo  {journal} {Q. J. Roy. Met.
  Soc.}\ }\textbf {\bibinfo {volume} {135}},\ \bibinfo {pages} {2161} (\bibinfo
  {year} {2009})}\BibitemShut {NoStop}%
\bibitem [{\citenamefont {Strogatz}(2000)}]{Strogatz-Dynamical}%
  \BibitemOpen
  \bibfield  {author} {\bibinfo {author} {\bibfnamefont {S.~H.}\ \bibnamefont
  {Strogatz}},\ }\href@noop {} {\emph {\bibinfo {title} {{Nonlinear Dynamics
  And Chaos: With Applications To Physics, Biology, Chemistry, And Engineering
  (Studies in Nonlinearity)}}}},\ \bibinfo {edition} {1st}\ ed.\ (\bibinfo
  {publisher} {CRC Press},\ \bibinfo {year} {2000})\BibitemShut {NoStop}%
\bibitem [{\citenamefont {Heifetz}\ and\ \citenamefont
  {Guha}(2017)}]{HeifetzGuha17}%
  \BibitemOpen
  \bibfield  {author} {\bibinfo {author} {\bibfnamefont {E.}~\bibnamefont
  {Heifetz}}\ and\ \bibinfo {author} {\bibfnamefont {A.}~\bibnamefont {Guha}},\
  }\bibfield  {title} {\bibinfo {title} {{A generalized action-angle
  representation of wave interaction in stratified shear flows}},\ }\href
  {https://doi.org/10.1017/jfm.2017.719} {\bibfield  {journal} {\bibinfo
  {journal} {J. Fluid Mech.}\ }\textbf {\bibinfo {volume} {834}},\ \bibinfo
  {pages} {220} (\bibinfo {year} {2017})}\BibitemShut {NoStop}%
\bibitem [{\citenamefont {Nakamura}(1993{\natexlab{a}})}]{Nakamura93b}%
  \BibitemOpen
  \bibfield  {author} {\bibinfo {author} {\bibfnamefont {N.}~\bibnamefont
  {Nakamura}},\ }\bibfield  {title} {\bibinfo {title} {{Momentum flux, flow
  symmetry, and the nonlinear barotropic governor}},\ }\href
  {https://doi.org/10.1175/1520-0469(1993)050<2159:MFFSAT>2.0.CO;2} {\bibfield
  {journal} {\bibinfo  {journal} {J. Atmos. Sci.}\ }\textbf {\bibinfo {volume}
  {50}},\ \bibinfo {pages} {2159} (\bibinfo {year}
  {1993}{\natexlab{a}})}\BibitemShut {NoStop}%
\bibitem [{\citenamefont {Waterman}\ and\ \citenamefont
  {Lilly}(2015)}]{WatermanLilly15}%
  \BibitemOpen
  \bibfield  {author} {\bibinfo {author} {\bibfnamefont {S.}~\bibnamefont
  {Waterman}}\ and\ \bibinfo {author} {\bibfnamefont {J.~M.}\ \bibnamefont
  {Lilly}},\ }\bibfield  {title} {\bibinfo {title} {{Geometric decomposition of
  eddy feedbacks in barotropic systems}},\ }\href
  {https://doi.org/10.1175/JPO-D-14-0177.1} {\bibfield  {journal} {\bibinfo
  {journal} {J. Phys. Oceanogr.}\ }\textbf {\bibinfo {volume} {45}},\ \bibinfo
  {pages} {1009} (\bibinfo {year} {2015})}\BibitemShut {NoStop}%
\bibitem [{\citenamefont {Youngs}\ \emph {et~al.}(2017)\citenamefont {Youngs},
  \citenamefont {Thompson}, \citenamefont {Lazar},\ and\ \citenamefont
  {Richards}}]{Youngs-et-al17}%
  \BibitemOpen
  \bibfield  {author} {\bibinfo {author} {\bibfnamefont {M.~K.}\ \bibnamefont
  {Youngs}}, \bibinfo {author} {\bibfnamefont {A.~F.}\ \bibnamefont
  {Thompson}}, \bibinfo {author} {\bibfnamefont {A.}~\bibnamefont {Lazar}},\
  and\ \bibinfo {author} {\bibfnamefont {K.~J.}\ \bibnamefont {Richards}},\
  }\bibfield  {title} {\bibinfo {title} {{ACC meanders, energy transfer, and
  mixed barotropic--baroclinic instability}},\ }\href
  {https://doi.org/10.1175/JPO-D-10-0160.1} {\bibfield  {journal} {\bibinfo
  {journal} {J. Phys. Oceanogr.}\ }\textbf {\bibinfo {volume} {47}},\ \bibinfo
  {pages} {1291} (\bibinfo {year} {2017})}\BibitemShut {NoStop}%
\bibitem [{\citenamefont {Poulsen}\ \emph {et~al.}(2019)\citenamefont
  {Poulsen}, \citenamefont {Jochum}, \citenamefont {Maddison}, \citenamefont
  {Marshall},\ and\ \citenamefont {Nuterman}}]{Poulsen-et-al19}%
  \BibitemOpen
  \bibfield  {author} {\bibinfo {author} {\bibfnamefont {M.~B.}\ \bibnamefont
  {Poulsen}}, \bibinfo {author} {\bibfnamefont {M.}~\bibnamefont {Jochum}},
  \bibinfo {author} {\bibfnamefont {J.~R.}\ \bibnamefont {Maddison}}, \bibinfo
  {author} {\bibfnamefont {D.~P.}\ \bibnamefont {Marshall}},\ and\ \bibinfo
  {author} {\bibfnamefont {R.}~\bibnamefont {Nuterman}},\ }\bibfield  {title}
  {\bibinfo {title} {{A geometric interpretation of Southern Ocean eddy form
  stress}},\ }\href {https://doi.org/10.1175/JPO-D-18-0220.1} {\bibfield
  {journal} {\bibinfo  {journal} {J. Phys. Oceanogr.}\ }\textbf {\bibinfo
  {volume} {49}},\ \bibinfo {pages} {2553} (\bibinfo {year}
  {2019})}\BibitemShut {NoStop}%
\bibitem [{\citenamefont {Meng}\ and\ \citenamefont {Guo}(2023)}]{MengGuo23}%
  \BibitemOpen
  \bibfield  {author} {\bibinfo {author} {\bibfnamefont {C.}~\bibnamefont
  {Meng}}\ and\ \bibinfo {author} {\bibfnamefont {Z.}~\bibnamefont {Guo}},\
  }\href@noop {} {\bibinfo {title} {Vorticity wave interaction and exceptional
  points in shear flow instabilities}} (\bibinfo {year} {2023}),\ \Eprint
  {https://arxiv.org/abs/2303.11842} {arXiv:2303.11842 [physics.flu-dyn]}
  \BibitemShut {NoStop}%
\bibitem [{\citenamefont {Stewart}\ \emph {et~al.}(2015)\citenamefont
  {Stewart}, \citenamefont {Spence}, \citenamefont {Waterman}, \citenamefont
  {{Le Sommer}}, \citenamefont {Molines}, \citenamefont {Lilly},\ and\
  \citenamefont {England}}]{Stewart-et-al15}%
  \BibitemOpen
  \bibfield  {author} {\bibinfo {author} {\bibfnamefont {K.~D.}\ \bibnamefont
  {Stewart}}, \bibinfo {author} {\bibfnamefont {P.}~\bibnamefont {Spence}},
  \bibinfo {author} {\bibfnamefont {S.}~\bibnamefont {Waterman}}, \bibinfo
  {author} {\bibfnamefont {J.}~\bibnamefont {{Le Sommer}}}, \bibinfo {author}
  {\bibfnamefont {J.-M.}\ \bibnamefont {Molines}}, \bibinfo {author}
  {\bibfnamefont {J.~M.}\ \bibnamefont {Lilly}},\ and\ \bibinfo {author}
  {\bibfnamefont {M.~H.}\ \bibnamefont {England}},\ }\bibfield  {title}
  {\bibinfo {title} {{Anisotropy of eddy variability in the global ocean}},\
  }\href {https://doi.org/10.1016/j.ocemod.2015.09.005} {\bibfield  {journal}
  {\bibinfo  {journal} {Ocean Modell.}\ }\textbf {\bibinfo {volume} {95}},\
  \bibinfo {pages} {53} (\bibinfo {year} {2015})}\BibitemShut {NoStop}%
\bibitem [{\citenamefont {Heifetz}\ and\ \citenamefont
  {Methven}(2005)}]{HeifetzMethven05}%
  \BibitemOpen
  \bibfield  {author} {\bibinfo {author} {\bibfnamefont {E.}~\bibnamefont
  {Heifetz}}\ and\ \bibinfo {author} {\bibfnamefont {J.}~\bibnamefont
  {Methven}},\ }\bibfield  {title} {\bibinfo {title} {{Relating optimal growth
  to counter-propagating Rossby Waves in shear instability}},\ }\href
  {https://doi.org/10.1063/1.1937064} {\bibfield  {journal} {\bibinfo
  {journal} {Phys. Fluids}\ }\textbf {\bibinfo {volume} {17}},\ \bibinfo
  {pages} {064107} (\bibinfo {year} {2005})}\BibitemShut {NoStop}%
\bibitem [{\citenamefont {James}(1987)}]{James87}%
  \BibitemOpen
  \bibfield  {author} {\bibinfo {author} {\bibfnamefont {I.~N.}\ \bibnamefont
  {James}},\ }\bibfield  {title} {\bibinfo {title} {{Suppression of baroclinic
  instability in horizontally sheared flows}},\ }\href
  {https://doi.org/10.1175/1520-0469(1987)044<3710:SOBIIH>2.0.CO;2} {\bibfield
  {journal} {\bibinfo  {journal} {J. Atmos. Sci.}\ }\textbf {\bibinfo {volume}
  {44}},\ \bibinfo {pages} {3710} (\bibinfo {year} {1987})}\BibitemShut
  {NoStop}%
\bibitem [{\citenamefont {Nakamura}(1993{\natexlab{b}})}]{Nakamura93a}%
  \BibitemOpen
  \bibfield  {author} {\bibinfo {author} {\bibfnamefont {N.}~\bibnamefont
  {Nakamura}},\ }\bibfield  {title} {\bibinfo {title} {{An illustrative model
  of instabilities in meridionally and vertically sheared flows}},\ }\href
  {https://doi.org/10.1175/1520-0469(1993)050<0357:AIMOII>2.0.CO;2} {\bibfield
  {journal} {\bibinfo  {journal} {J. Atmos. Sci.}\ }\textbf {\bibinfo {volume}
  {50}},\ \bibinfo {pages} {357} (\bibinfo {year}
  {1993}{\natexlab{b}})}\BibitemShut {NoStop}%
\bibitem [{\citenamefont {Harnik}\ \emph {et~al.}(2014)\citenamefont {Harnik},
  \citenamefont {Dritschel},\ and\ \citenamefont {Heifetz}}]{Harnik-et-al14}%
  \BibitemOpen
  \bibfield  {author} {\bibinfo {author} {\bibfnamefont {N.}~\bibnamefont
  {Harnik}}, \bibinfo {author} {\bibfnamefont {D.~G.}\ \bibnamefont
  {Dritschel}},\ and\ \bibinfo {author} {\bibfnamefont {E.}~\bibnamefont
  {Heifetz}},\ }\bibfield  {title} {\bibinfo {title} {{On the equilibration of
  asymmetric barotropic instability}},\ }\href@noop {} {\bibfield  {journal}
  {\bibinfo  {journal} {Q. J. Roy. Met. Soc.}\ }\textbf {\bibinfo {volume}
  {140}},\ \bibinfo {pages} {2444} (\bibinfo {year} {2014})}\BibitemShut
  {NoStop}%
\bibitem [{\citenamefont {Cairns}(1979)}]{Cairns79}%
  \BibitemOpen
  \bibfield  {author} {\bibinfo {author} {\bibfnamefont {R.~A.}\ \bibnamefont
  {Cairns}},\ }\bibfield  {title} {\bibinfo {title} {The role of negative
  energy waves in some instabilities of parallel flows},\ }\href@noop {}
  {\bibfield  {journal} {\bibinfo  {journal} {J. Fluid Mech.}\ }\textbf
  {\bibinfo {volume} {92}},\ \bibinfo {pages} {1} (\bibinfo {year}
  {1979})}\BibitemShut {NoStop}%
\bibitem [{\citenamefont {Dorey}\ \emph {et~al.}(2001)\citenamefont {Dorey},
  \citenamefont {Dunning},\ and\ \citenamefont {Tateo}}]{Dorey-et-al01}%
  \BibitemOpen
  \bibfield  {author} {\bibinfo {author} {\bibfnamefont {P.}~\bibnamefont
  {Dorey}}, \bibinfo {author} {\bibfnamefont {C.}~\bibnamefont {Dunning}},\
  and\ \bibinfo {author} {\bibfnamefont {R.}~\bibnamefont {Tateo}},\ }\bibfield
   {title} {\bibinfo {title} {{Spectral equivalences, Bethe ansatz equations,
  and reality properties in PT-symmetric quantum mechanics}},\ }\href
  {https://doi.org/10.1088/0305-4470/34/28/305} {\bibfield  {journal} {\bibinfo
   {journal} {J. Phys. A Math.}\ }\textbf {\bibinfo {volume} {34}},\ \bibinfo
  {pages} {5679} (\bibinfo {year} {2001})}\BibitemShut {NoStop}%
\bibitem [{\citenamefont {Mostafazadeh}(2002)}]{Mostafazadeh02}%
  \BibitemOpen
  \bibfield  {author} {\bibinfo {author} {\bibfnamefont {A.}~\bibnamefont
  {Mostafazadeh}},\ }\bibfield  {title} {\bibinfo {title} {{Pseudo-Hermiticity
  versus PT symmetry: The necessary condition for the reality of the spectrum
  of a non-Hermitian Hamiltonian}},\ }\href {https://doi.org/10.1063/1.1418246}
  {\bibfield  {journal} {\bibinfo  {journal} {J. Math. Phys.}\ }\textbf
  {\bibinfo {volume} {43}},\ \bibinfo {pages} {205} (\bibinfo {year}
  {2002})}\BibitemShut {NoStop}%
\end{thebibliography}

%

\end{document}